\documentclass[reprint,twocolumn,showpacs,superscriptaddress,longbibliography,email,floatfix,aps,prl]{revtex4-2}

\usepackage{ifthen,amsthm,amsmath,amsxtra,amsfonts,dsfont,graphicx,bm,tikz,scalerel,wasysym,bbm,graphicx,amsthm,braket,physics,empheq}
\usepackage{tikz}
\usepackage[inline]{enumitem}
\usepackage{MyCircuitTikz}
\usepackage{hyperref}
\hypersetup{
     colorlinks=true,
     linkcolor=red,
     filecolor=blue,
     citecolor=blue,      
     urlcolor=blue,
     }
\usepackage{color}
\usepackage{MnSymbol}%
\usepackage{wasysym}%

\newcommand{\ttilde}{~}
\newcommand{\dket}[1]{|#1\rangle\!\rangle}
\newcommand{\dbra}[1]{\langle\!\langle#1|}
\newcommand{\dbraket}[2]{\langle\!\langle#1|#2\rangle\!\rangle}

\begin{document}
 \raggedbottom
\title{Exact first-detection probability in a locally monitored solvable quantum circuit}

\author{Cecilia De Fazio}
\email{cecilia.de-fazio@uni-tuebingen.de}
\affiliation{Institut f\"ur Theoretische Physik and Center for Integrated Quantum Science and Technology, Universit\"at T\"ubingen, Auf der Morgenstelle 14, 72076 T\"ubingen, Germany}
\author{Igor Lesanovsky}
\affiliation{Institut f\"ur Theoretische Physik and Center for Integrated Quantum Science and Technology, Universit\"at T\"ubingen, Auf der Morgenstelle 14, 72076 T\"ubingen, Germany}
\affiliation{School of Physics and Astronomy and Centre for the Mathematics and Theoretical Physics of Quantum Non-Equilibrium Systems, The University of Nottingham, Nottingham, NG7 2RD, United Kingdom}
\author{Gabriele Perfetto}
\affiliation{Institut für Theoretische Physik, ETH Zürich, Wolfgang-Pauli-Str. 27, 8093 Zürich, Switzerland}

\begin{abstract}
We study the interplay between local stroboscopic measurements and the unitary quantum circuit dynamics of the deterministic Floquet-quantum East model. We derive exact analytical results for the probability of first detecting a subset of qubits back in their initial state. This is achieved via a renewal equation linking the first-detection probability to the subsystem Loschmidt echo, which encodes the effective open dynamics of the monitored qubits. The resulting statistics exhibit two regimes set by the competition between the measurement probing time and the relaxation timescale of the unitary circuit dynamics: while infrequent monitoring leads to geometric first-detection statistics, frequent monitoring induces correlations that manifest as deviations from this geometric behavior. These results provide a rare exact characterization of how local measurement backaction reshapes detection statistics in an interacting quantum many-body system. In our local protocol, return probabilities are controlled by the size of the monitored subregion and can therefore be accessed on digital quantum simulators via local mid-circuit measurements.
\end{abstract}

\maketitle
\textbf{Introduction.--- } 
\begin{figure}[t]
    \centering
\includegraphics[width=\columnwidth]{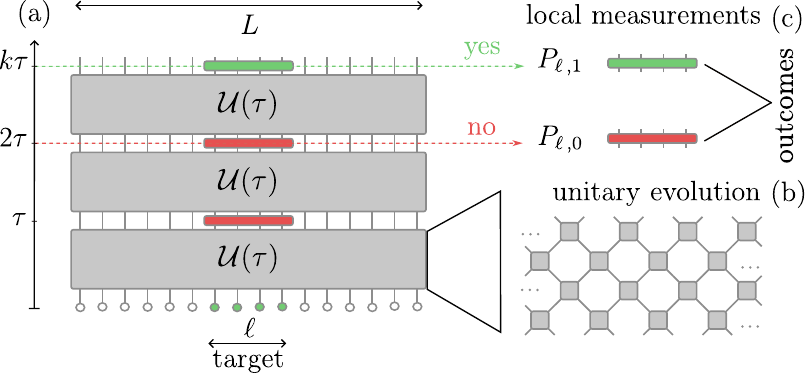}
   \caption{ \textbf{Measurement protocol.} (a) Circuit representation of the state in Eq.\ttilde\eqref{eq:statek} for a system of size $L$. A monitored subregion of size $\ell\ll L$ is initialized in the target product state $\ket{\mathbf{0}_\ell}$ (green circles). The unitary circuit dynamics is interspersed with local stroboscopic measurements on the monitored subregion at times $\tau,2\tau,\dots,k\tau$, where $\tau$ is the probing time.
(b) The brickwork operator $\mathcal{U}(\tau)=(\mathcal{U}_i\,\mathcal{U}_h)^\tau$, built from alternating layers of two-site gates applied at half-integer and integer sites, as given in Eq.\ttilde\eqref{eq:brickwork0}.
(c) Local projective measurements are performed on the monitored subsystem only. Each outcome answers the question ``is the subsystem in the target state?'', projecting the local state onto the target subspace if the answer is yes ($P_{\ell,1}$, green), or onto its complement ($P_{\ell,0}$, red) otherwise, as expressed by Eq.\ttilde\eqref{eq:localmeasurement}.
}
  \label{fig:fig1}
\end{figure}
Monitored quantum circuits, in which unitary dynamics is interspersed with local measurements \cite{FisherReview2023}, have attracted growing attention for a twofold reason. On the one hand, quantum circuits constitute the basic architecture of quantum computation \cite{Pino2021,Corcoles2021,Deist2022,Koh2023,Graham2023}, making it essential to understand and control the role of measurements in information extraction. On the other hand, they provide a versatile setting for modeling many-body dynamics \cite{Chan2018a,Khemani2018}: their locality, local unitarity, discrete-time evolution, together with specific gate properties, such as randomness \cite{Nahum2017,vonKeyserlingk2018,Nahum2018}, space–time symmetries \cite{Bertini2019dual,piroli2020exact,klobas2021exact,yu2024hierarchical,Foligno2024,Bertini2025review}, or kinetic constraints \cite{horssen2015dynamics,pancotti2020quantum,Bertini2024b,Bertini2024}, can introduce structure and simplifications that provide insight into the resulting dissipative dynamics induced by measurements \cite{Claeys2022a,Cotler2023,Li2026}. This has led to the discovery of a rich set of collective out-of-equilibrium phenomena ranging from measurement-induced phase transitions \cite{Li2018,Skinner2019,Chan2019} to active–inactive dynamical phase coexistence revealed through ancilla-assisted monitoring \cite{Cech2025a,Cech2025b,Cech2026,Cea2026,Cech2026b}.

A complementary diagnostic of monitored quantum dynamics is the first-detection return probability (FDRP)~\cite{grunbaum_recurrence_2013,Dhar2015b,Dhar2015QuantumModel,Barkai2016, Friedman2017QuantumProblem}. In quantum walks, the FDRP addresses when a walker is first detected back at its initial location under stroboscopic measurements, and naturally relates to spatial search in quantum algorithms \cite{Aaronson:2005, Kendon2010, Childs:2010_review,quantum_walk_grover,Farhi_Gutmann98,quantum_walk_spatial_search,tornow2023,yin2025restart,yin2025resonance,liu2026fractionally,King2025_optimalsearch,king2026timecomplexity} and network dynamics \cite{Barabasi:2016_review, santhanam2026,Folz:2023}. Beyond single-particle quantum walks \cite{QuantumWalksTodd2006,grunbaum_recurrence_2013,Dhar2015QuantumModel,Dhar2015b,Barkai2016,Friedman2017QuantumProblem,Barkai2018,Barkai2020,Thiel_spectral,majumdar2023,roy_causality_2025,Vecchio_2026,roy2026phase}, however, studies of the FDRP are limited. The main results \cite{Dittle_fdt_many_body,imparato_many_body_fdt,Walter2025,yin2025preparation,liu2026fractionally} in this direction have been obtained through the quantum renewal equation, which expresses the first-detection amplitude, whose modulus squared gives the FDRP, 
in terms of the Loschmidt amplitude \cite{Gorin2006}. In all these studies, however, the FDRP becomes extremely small already after the first few attempts. This is a consequence of the \textit{global}, spatially-resolved monitoring of the many-body wavefunction, which resides in a large Hilbert space whose dimension grows exponentially in the system size $L$. For \textit{local} monitoring of a few degrees of freedom, instead, exact results are only available for mean values \cite{liu2026fractionally}, while no predictions exist for the full FDRP. Local monitoring is, in fact, fundamentally harder to treat than global monitoring, as it requires an open-system description: although the measurements provide information only about the monitored region, their backaction affects the rest of the system through the correlations and entanglement 
between the two.

In this manuscript we solve the local monitoring problem exactly for an interacting quantum circuit with kinetic constraints, the deterministic Floquet-quantum East model (DFQEM) \cite{Bertini2024}, computing the FDRP of a small but finite subregion of size $\ell \ll L$, see Fig.~\ref{fig:fig1}. 
This model exhibits additional properties \cite{yu2024hierarchical}, similar in spirit to dual unitarity \cite{Bertini2019dual}, enabling access to the local dynamics for suitable choices of initial states \cite{Bertini2024} and making the DFQEM a rare example of an analytically tractable problem in interacting many-body dynamics. To compute the FDRP, we derive a renewal equation in terms of the \textit{subsystem} Loschmidt echo, which captures the effective open dynamics of the monitored subregion, with the remaining unmonitored qubits acting as an external bath. Importantly, this bath is coupled to the subregion through the same gates that generate its internal dynamics, so that no weak-coupling separation between system and environment is assumed. Nevertheless, the structure of the circuit allows us to compute the local FDRP exactly. The results show a transition from uncorrelated to correlated measurement outcomes as the probing time is decreased below the relaxation timescale of the monitored region \cite{Bertini2024}. Moreover, the FDRP remains at accessible magnitudes over many attempts, with a decay rate set by $\ell$ and the probing time. These features make the local stroboscopic protocol amenable to implementation in digital quantum simulators with mid-circuit measurements \cite{midcircuit_1,midcircuit_2,midcircuit_3,midcircuit_4,midcircuit_5}.
 
\textbf{Local measurement protocol.--- } We consider the setting illustrated in Fig.~\ref{fig:fig1}(a): a chain of length $L$ undergoing unitary evolution interspersed with local projective measurements on a subsystem of size $\ell \ll L$. We first describe the unitary dynamics in the absence of measurements. On two qubits, it is implemented through a CNOT gate \cite{nielsen2010quantum},
\begin{equation} \label{eq:localgate}
u = \begin{tikzpicture}[baseline={([yshift=-0.6ex]current bounding box.center)},scale=0.5]
\gate{0}{0}
\end{tikzpicture} =  \openone \otimes \mathbb{P} + \mathbb{X} \otimes (\openone-\mathbb{P})\,, 
\end{equation}
where  $\mathbb{P} = (\openone+\mathbb{Z})/2$ is the projector onto the state zero $\ket{0}=(1 \,\,\, 0)^T$ of a single qubit, $\openone$ is the single-site identity, and $\{\mathbb{X},\mathbb{Y},\mathbb{Z}\}$ are Pauli matrices. This gate enforces the kinetic constraint characteristic of the East model: a qubit can flip only if its right neighbor is in state $\ket{1}=(0 \, \,\, 1)^T$. The many-qubit dynamics is implemented as in Fig.~\ref{fig:fig1}(b), where the brickwork circuit  corresponds to the DFQEM, that is,  the deterministic limit of the ``trotterized" quantum East model \cite{horssen2015dynamics,pancotti2020quantum, Bertini2024b} and the quantum generalization of the classical Floquet-East model \cite{Klobas2024,DeFazio2024}. Here, the total system evolves over a time $\tau$ via the brickwork-structured operator $ \mathcal{U}(\tau)= ( \mathcal{U}_i\, \mathcal{U}_h )^\tau$, where $\mathcal{U}_h$ and $\mathcal{U}_i$ act on alternating (half-integer and integer, respectively) subsets of qubits. Note, indeed, that qubits and time are labeled by integers and half-integers, so that a region of length $x$ contains $2x$ qubits, as also detailed in the Supplemental Material \cite{SM}. The two operators are constructed from the local gate~\eqref{eq:localgate} and related by the one-site shift operator $\mathcal{S}$,
\begin{equation}
\label{eq:brickwork0}
    \mathcal{U}_h = \bigotimes_{j \in \mathcal{N}_{h}} u_{j,j+1/2}\,, 
    \qquad 
    \mathcal{U}_i = \mathcal{S}^\dagger \mathcal{U}_h \mathcal{S},
\end{equation}
where $u_{j,j+1/2}$ denotes the two-site gate acting on a half-integer site $j$ and its right neighbor and $\mathcal{N}_{h}$ is the subset of half-integers between $0$ and $L$. Above, we also assumed periodic boundary conditions, i.e., $u_{j+L,\,j+L+1/2}=u_{j,\,j+1/2}$ for $j$ either integer or half-integer.

We now monitor a \textit{subregion} of size $\ell$ at stroboscopic times $\tau,2\tau \dots k\tau$ and ask whether it returns (locally) to the initial state $\ket{\mathbf{{0}}_\ell} = \ket{0}^{\otimes 2\ell}$,  which we identify as the target state. Here, measurement outcomes are labeled by $m=1,0$, corresponding to projection onto the target subspace or its complement, depending on whether the detection has been successful or not, see Fig.~\ref{fig:fig1}(c). We accordingly define the local operator $P_{\ell,m}$ as
\begin{equation}
\label{eq:localmeasurement}
{P}_{\ell\,,m}= m\, \ketbra{\mathbf{{0}}_\ell} + (1-m)  (\openone_{\ell}  -  \ketbra{\mathbf{{0}}_\ell}\,)\,\,,
\end{equation}
where $\openone_{x} = \openone^{\otimes 2x}$ for a region of generic length $x$. The rest of the system is initially in the bi-translationally invariant product state $\ket{\Phi_{L-\ell}}=(\ket{\phi_1}\otimes\ket{\phi_2})^{\otimes L - \ell}$, where $\ket{\phi_i}$ are \textit{solvable} quantum states, selected to ensure tractable local dynamics, see Refs.~\cite{Bertini2024,SM}. 

The protocol halts at the first successful detection ($m=1$). If this occurs at the $k$-th attempt, the unnormalized state $\ket{\Psi_k}$ of the total system at time $k\tau$ (circuit representation in Fig.\ttilde\ref{fig:fig1}(a)) reads
\begin{equation}
\label{eq:statek}
    \ket{\Psi_k} = \mathcal{P}_{1} \, \mathcal{U}(\tau) \, \left[ \,  \mathcal{P}_{0} \, \mathcal{U}(\tau) \right]^{k-1}    \ket{\Phi_{L-\ell}} \otimes \ket{\mathbf{{0}}_\ell} \,,
\end{equation}
where $\mathcal{P}_{m} = \openone_{L-\ell} \otimes {P}_{\ell,m}$ acts non-trivially only on the monitored region and $\ket{\Phi_{L-\ell}} \otimes \ket{\mathbf{{0}}_\ell}$ is the initial state of the full system. The FDRP $F(k)$ is given by the squared norm of the state~\eqref{eq:statek}, and can equivalently be written as the correlation function:
\begin{equation}
    \label{eq:FDP}
    F(k) = \braket{\Psi_k} = 
    \Big\langle 
        \prod_{j=1}^{k-1} \mathcal{P}_0(j)\,
        \mathcal{P}_1(k)\,
        \prod_{j=1}^{k-1} \mathcal{P}_0(k-j)
    \Big\rangle_0 ,
\end{equation}
\begin{figure}[t]
    \centering
 \includegraphics[width=\columnwidth]{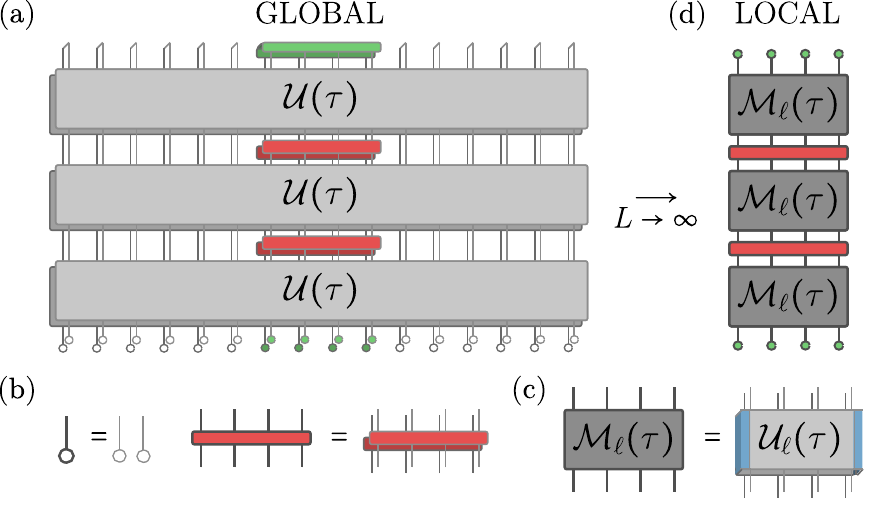}
\caption{  \textbf{Folded space representation.}  (a) Two-layer circuit representing the probability $F(k)$ in Eq.\ttilde\eqref{eq:FDP} for the exemplary value $k=3$ in the folded space, with folding performed at the level of the projector $P_{\ell, 1}$. 
(b) A single-qudit propagator in the folded space (left-hand side)—indicated by darker and thicker lines—acts so that each site corresponds to the tensor product of two qubits. The folded-space projector (left-hand side) likewise factorizes into the tensor product of two layers $P_{\ell,0} \otimes P_{\ell,0}$ (right-hand side). 
(c) The operator $\mathcal{M}_\ell(\tau)$, which implements the evolution of the monitored subregion for $L\to\infty$, is obtained from the brickwork circuit $\mathcal{U}_\ell(\tau)$ acting on the region of size $\ell$ only, with boundary operators (in blue) that encode the effect of the rest of the system. The latter couple the two layers of the folded space, so $\mathcal{M}_\ell(\tau)$ does not factorize into back and front layers.
(d) Local circuit representing the first-detection probability of the subsystem $\ell$ at large $L$, using the compact notation introduced in panels (b) and (c).  }
    \label{fig:fig2}
\end{figure}%
where $\mathcal{P}_m(j)= \mathcal{U}^\dagger (j\tau)\, \mathcal{P}_m \, \mathcal{U} (j\tau)$ are the time-evolved projectors and $\langle\, \cdot\, \rangle_0$ denotes the expectation value in the initial state.  
Eq.\ttilde\eqref{eq:FDP} allows for the exact calculation of $F(k)$, as we explain in detail below.
\begin{figure*}[t!]
    \centering
   \includegraphics[width=1.0\textwidth]{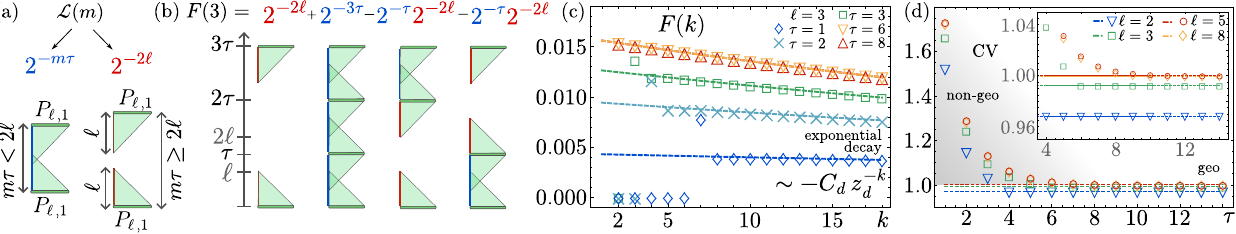}
\caption{  \textbf{Results.} 
(a) Geometrical interpretation of the Loschmidt echo~\eqref{eq:Lamplitudedef}. Each folded-space projector $P_{\ell, 1}$ enforces a larger space-time region of the circuit in  Eq.~\eqref{eq:Lamplitudedef}, forming an isosceles right triangle of base $\ell$ (green), to be in the zero state. For $m\tau < 2\ell$, the two projectors are causally connected, and the exponent of $\mathcal{L} (m)$ scales with the temporal distance $m\tau$ (blue). For $m\tau \ge 2\ell$, $\mathcal{L}(m)$ factorizes into two independent contributions, each scaling exponentially with the triangle base (red).
(b) Example of a partition decomposition of the first-detection probability $F(k)$ for $k=3$ and $\ell < \tau < 2\ell$. Here, Loschmidt echoes contribute differently (red and blue) depending on the temporal distances between consecutive projectors, based on the geometrical argument in panel (a).
(c) Exact results (markers) versus asymptotic behavior (dashed lines) of the first-detection probability $F(k)$ for $\ell=3$ as a function of the number of attempts $k$ for several choices of $\tau$ (colors in the legend) and $k\geq 2$. 
(d)  The coefficient of variation $ \mathrm{CV}$ (markers), defined in Eq.~\eqref{eq:CM}, compared with the geometric distribution prediction (dashed lines) as a function of $\tau$ and for several choices of $\ell$. Deviation of the markers from the dashed lines implies a non-geometric (non-geo) behavior, distinct from the geometric (geo) behavior.
}
\label{fig:fig3}
\end{figure*}

\textbf{Local temporal partitioning.---} We compute the FDRP~\eqref{eq:FDP} by employing the \textit{space-evolution approach}. Originally developed for tensor-network simulations \cite{muellerhermes2012tensor,Hastings2015}, this method was later extended to random unitary circuits \cite{bertini2018exact} and dual-unitary circuits \cite{piroli2020exact}, and has since been applied to a broad range of problems, including the evolution of local observables \cite{piroli2020exact,lerose2021influence,klobas2021exact}, entanglement dynamics \cite{bertini2019entanglement,Bertini2024}, and spectral properties \cite{bertini2018exact}. 

 The computation proceeds in two steps. As shown in Fig.~\ref{fig:fig2}(a), we first map the circuit into a \textit{folded space} by folding the portion of the circuit above the projector $P_{\ell,1}$ onto the lower part of the circuit. Under this folding, the projector $P_{\ell, 1}$ is mapped to the folded-space state $\dbra{\mathbf{{0}}_\ell}= \bra{\mathbf{{0}}_\ell} \otimes \bra{\mathbf{{0}}_\ell}$. More precisely, as depicted in Fig.~\ref{fig:fig2}(b), each site of the folded circuit is associated with a four-dimensional qudit, corresponding to the tensor product of two qubits in the back and front layers. The same construction applies to the folded-space projector $P_{\ell,0}\otimes P_{\ell,0}$, as illustrated in the same panel. The second step exploits the equilibrium properties of the spatial dynamics obtained by exchanging the roles of space and time \cite{SM}. For a fixed time, a spatial transfer matrix evolves temporal states (i.e., states extended in time) along the spatial direction. In the limit $L \to \infty$, the evolution of the monitored region is described by the non-unitary operator $\mathcal{M}_\ell(\tau)$ in Fig.~\ref{fig:fig2}(c), where non-unitarity arises from boundary operators encoding the influence of the rest of the system. Since these boundary operators factorize in time (see Ref.\ttilde\cite{Bertini2024} and \cite{SM} for details), the rest of the system acts as a Markovian bath~\cite{lerose2021influence}. The resulting circuit in Fig.~\ref{fig:fig2}(d) can be expressed as a finite expansion of exactly computable diagrams by decomposing each occurrence of $P_{\ell, 0} \otimes P_{\ell, 0}$ arising from unsuccessful detection into the four terms $\openone_\ell \otimes \openone_\ell$, $P_{\ell, 1} \otimes P_{\ell, 1}$, $-\openone_\ell \otimes P_{\ell, 1}$, and $-P_{\ell, 1} \otimes \openone_\ell$, where the left (right) operator in the tensor product acts on the back (front) layers of the folded space, respectively. 
A key property of the DFQEM is that a local projector such as $P_{\ell, 1}$ on one layer automatically enforces the same projector on the other layer when evaluated in the setting of Fig.~\ref{fig:fig2}(d) \cite{SM}. As a result, each $P_{\ell, 0} \otimes P_{\ell, 0}$ effectively reduces to only two terms, namely $\openone_\ell \otimes \openone_\ell$ and $-P_{\ell, 1} \otimes P_{\ell, 1}$. 

Repeating this step $k-1$ times leads to the expansion
\begin{equation}
\label{eq:expansion}
    F (k) = \sum_{r=1}^\infty (-1)^{r+1} \sum_{m_1,\dots,m_r=1}^\infty \left( \prod_{j=1}^r \mathcal{L} (m_j) \right) \delta_{\sum_i m_i, k}\,,
\end{equation}
where, for a positive integer $m$,  $\mathcal{L} (m)$ is the exact subsystem Loschmidt echo for the DFQEM, defined as the folded-space overlap between the locally evolved state of the $2\ell$ monitored qubits at time $m\tau$ and the folded-space target state $\dket{\mathbf{{0}}_\ell}$ (represented by the green circles):
\begin{equation}
\label{eq:Lamplitudedef}
    \mathcal{L} (m) = 
       \begin{tikzpicture}[baseline={([yshift=-0.6ex]current bounding box.center)},scale=1.0]
       \Mtau{0}{0}{\myell}{\mytau}[\normalsize $\mathcal{M}_\ell (m\tau)$]
       \freelegslocalqubit{0}{0}{\myspacing}{-\dy}{\myell}
       \freelegslocalqubit{0}{\mytau}{\myspacing}{\dy}{\myell}
      \end{tikzpicture}
    = \begin{cases}
2^{-m\tau}, & m\tau < 2\ell, \\
2^{-2\ell}, & m\tau \geq 2\ell.
\end{cases}
\end{equation}
Together, Eqs.\ttilde\eqref{eq:expansion} and\ttilde\eqref{eq:Lamplitudedef} establish the central insight of this work: the FDRP of the monitored subregion is fully determined by its measurement-free local dynamics  (see \cite{SM} for the derivation). Indeed, $\mathcal{L}(m)$ is the probability of finding the subregion in the target state at time $m\tau$ in the absence of intermediate measurements and, while typically defined in terms of the global unitary evolution, here involves only the effective non-unitary evolution $\mathcal{M}_\ell$ of the $2\ell$ monitored qubits, illustrated in Fig.~\ref{fig:fig2}(c). 

Moreover, Eq.\ttilde\eqref{eq:Lamplitudedef} admits a simple geometrical interpretation, illustrated in Fig.~\ref{fig:fig3}(a): each folded-space target state $\dket{\mathbf{{0}}_\ell}$ acts as a projector that forces a larger space-time (isosceles right) triangle to be in the zero state. This is reminiscent of the inactive-region probabilities in the classical Floquet-East model~\cite{Klobas2024,DeFazio2024}, which are obtained by conditioning a local space-time region to be in the zero state and scale with its perimeter. In the quantum setting, however, $\mathcal{L}(m)$ scales with the length of one side of the rectangle enclosing the two green operators for $m\tau <2\ell$ (in blue) or with the base lengths of the two triangles for $m\tau \geq 2\ell$ (in red), rather than with the entire perimeter of these regions.
Similarly, this geometric argument extends to the expansion~\eqref{eq:expansion}, which can be interpreted as a partition function obtained by partitioning the total time $k\tau$ into integer multiples of $\tau$, at the times when the projectors $P_{\ell,1}\otimes P_{\ell,1}$ (represented by the green side of the triangles) are applied in the circuit of Fig.~\ref{fig:fig2}(d). As can be seen from the example in Fig.~\ref{fig:fig3}(b), each term in the expansion corresponds to a temporal partition, where the Loschmidt echoes are evaluated over temporal extents that sum to $k\tau$, but can contribute differently depending on the relative values of each temporal extent and $2\ell$, according to the geometric argument above.

\textbf{Exact subsystem first-detection probability.---} Eq.~\eqref{eq:expansion} also implies that the FDRP $F(k)$ satisfies the following renewal equation, written in terms of the subsystem Loschmidt echo
 \begin{equation}
 F (k) =\mathcal{L} (k)-\sum_{m=1}^{k-1} \mathcal{L} (m) F (k-m).
\label{eq:local_renewal}
 \end{equation}
 We stress that Eq.~\eqref{eq:local_renewal} differs from previously established renewal equations~\cite{Barkai2016,Friedman2017QuantumProblem,Dhar2015QuantumModel,Dhar2015b,Barkai2018,Barkai2020,Thiel_spectral,Walter2025,roy_causality_2025,roy2026phase}, which apply to the first-detection \textit{amplitude}. In contrast, Eqs.~\eqref{eq:expansion} and~\eqref{eq:local_renewal} directly relate the first-detection probability to the Loschmidt echo, owing to the constraint imposed by the DFQEM on the measurement projectors in the folded circuit and the choice of a target state. 

The FDRP is largest at the first attempt, where $F(1)=\mathcal{L}(1)$. For larger $k$, Eq.~\eqref{eq:local_renewal} gives the exact results shown in Fig.~\ref{fig:fig3}(c), which displays the FDRP $F(k)$ as a function of the number of attempts $k$ for several values of $\tau$ and fixed $\ell$. We distinguish the infrequent and frequent measurement regimes $\tau\geq 2\ell$ and $\tau < 2\ell$, respectively, depending on how the probing time $\tau$ relates to the relaxation time $2\ell$ of the DFQEM\ttilde\cite{Bertini2024}. As discussed in~\cite{SM}, a common feature of both regimes is that $\sum_{k=1}^{\infty}F(k)=1$, indicating that the local target state $\ket{\mathbf{0}_\ell}$ is eventually detected with certainty, similarly to recurrent quantum walks \cite{grunbaum_recurrence_2013,Dhar2015b,Dhar2015QuantumModel,Barkai2016,Friedman2017QuantumProblem}. This follows from the saturation of $\mathcal{L}(m)$ to the finite value $2^{-2\ell}$,
set by the finite dimension of the monitored subsystem Hilbert space, even though the full
system becomes infinitely large in the limit $L\to\infty$.
We now analyze the two regimes in detail.

\textbf{Infrequent measurements: relaxation to uncorrelated statistics for $\tau\geq 2\ell$.---} 
In the infrequent-measurement regime, the FDRP obeys a geometric distribution 
\begin{equation}
F(k)=2^{-2\ell} (1-2^{-2\ell})^{k-1},
\label{eq:infrequent_result}
\end{equation}
where the success probability at each measurement is $2^{-2\ell}$. 
This follows from Eqs.\ttilde\eqref{eq:expansion} and\ttilde\eqref{eq:Lamplitudedef},
and is consistent with the known relaxation properties of the model~\cite{Bertini2024}: for $\tau\geq 2\ell$, the monitored subsystem effectively relaxes to the infinite temperature state between consecutive measurements, so the probability of hitting the target state is uniform across all configurations. Indeed, as shown in Fig.~\ref{fig:fig3}(c), in this regime the FDRP is independent of the probing time $\tau$ and depends only on the relaxation time $2\ell$ (see orange and red markers). In this sense, each measurement is an independent event, with detecting the target state equivalent to rolling a $2^{2\ell}$-sided die.

\textbf{Frequent measurements: cross-measurement correlations for $\tau < 2\ell$.--- }
In this regime, we distinguish between early and late detection depending on the integer $n=\lceil 2\ell/\tau\rceil$, which compares the two time scales. For early detection, $1<k< n$, we find $F(k)=0$. This reflects the fact that unsuccessful measurements, while not fully projective, can nevertheless provide information about the state for specific targets such as the empty state $\ket{\mathbf{{0}}_\ell} $. Indeed, failure to detect the target at the first attempt reveals that the state of the system immediately before the measurement has non-zero overlap only with computational-basis states containing at least one qubit in state $\ket{1}$. Such excitations require a finite time to be reabsorbed at the boundaries, during which transitions to $\ket{\mathbf{{0}}_\ell} $ are forbidden. 

In order to describe detection at later attempts $k\gg n$, it is convenient to introduce the generating function $\hat{\mathcal{G}}(z)$, whose derivatives give the FDRPs,
\begin{equation}
\label{eq:FDPgen}
   F (k)=\frac{1}{(k-1)!} \frac{d^{k-1}}{dz^{k-1}} \left. \hat{\mathcal{G}} (z)\right|_{z=0},
\end{equation}
see \cite{SM} for details. This results from the $z$-transform $\hat{\mathcal{L}}(z)=\sum_{k}\mathcal{L}(k)\,z^k$ and $\hat{\mathcal{F}}(z)= z \hat{\mathcal{G}}(z)=\sum_{k} F (k) z^k$ applied to Eq.~\eqref{eq:local_renewal}. We find that $\hat{\mathcal{G}} (z)$ is a rational function, whose order is determined by $n$.  
The late-detection behavior of $F(k)$ is thus determined by the pole $z_d$ of $\hat{\mathcal{G}}(z)$ of smallest modulus, which implies an exponential asymptotic decay for the FDRP
\begin{equation}
F(k) \sim -C_d \,z_d^{-k}, \quad k \gg n,
\label{eq:asymFDP}
\end{equation}
with the coefficient $C_d = \operatorname{Res}_{z=z_d} \hat{\mathcal{G}}(z) < 0$. In Fig.~\ref{fig:fig3}(c), this behavior (dashed lines) shows excellent agreement with the exact numerical results obtained from Eq.~\eqref{eq:local_renewal} (markers).

Finally, we compute the coefficient of variation $\mathrm{CV}$ \cite{SM}, defined as the ratio between the standard deviation $\sigma$ and the mean $\mu$ of the distribution $F(k)$:
\begin{equation}
\label{eq:CM}
    \mathrm{CV} = \frac{\sigma}{\mu} = \frac{\sqrt{\langle k^2 \rangle -\langle k \rangle^2}}{\langle k \rangle }\,,
\end{equation}
with $\langle h(k)\rangle=\sum_k h(k)F(k)$. Fig.\ttilde\ref{fig:fig3}(d) shows that for $\tau<2\ell$, $\mathrm{CV}$ consistently deviates from the geometric distribution prediction $\mathrm{CV}_{\mathrm{geo}}=\sqrt{1-2^{-2\ell}}$, signaling correlations between measurement outcomes; the geometric value is recovered only for $\tau \geq 2\ell$.  

\textbf{Conclusions.---} 
We have derived exact results for the first-detection return probability (FDRP) of a \textit{locally} monitored subregion in the interacting many-body DFQEM. We identified a transition in the FDRP from uncorrelated to correlated measurement records as the probing time $\tau$ is chosen smaller than the relaxation time $2\ell$ of the monitored subregion. Our analysis is based on the renewal equation~\eqref{eq:local_renewal}, which relates the FDRP to the subsystem Loschmidt echo~\eqref{eq:Lamplitudedef}. Beyond the DFQEM, our methodology shows that the FDRP can be viewed as a local multi-time correlation function, making the approach applicable to a broader class of quantum circuits for which techniques to compute correlators are available 
\cite{Vidal2004,Verstraete2004,banuls2009matrix,Schollwoeck2011,Bertini2019dual,Paeckel2019,piroli2020exact,klobas2021exact,Claeys2022a,yu2024hierarchical,Foligno2024,Li2025postselection,Cech2025a,Li2026,Cech2026,Cech2025b,Bertini2025review}. Importantly, our study links directly to the recent experimental measurement of the subsystem-resolved Loschmidt echo \cite{Karch2025}. Similar setups, including quantum simulators with mid-circuit measurements \cite{midcircuit_1,midcircuit_2,midcircuit_3,midcircuit_4,midcircuit_5}, could be used to characterize the scaling of the return probability with the size $\ell$ of the monitored subregion. 
Altogether, the techniques developed here can thus serve as powerful tools for future theoretical and experimental investigations of nonequilibrium locally monitored quantum dynamics.

\let\oldaddcontentsline\addcontentsline
\renewcommand{\addcontentsline}[3]{}
\acknowledgments
\textbf{Acknowledgements. --- } We thank Marcel Cech and Pieter Claeys for useful discussions.  C.D.F.~acknowledges support from the Alexander von Humboldt Foundation through a postdoctoral research fellowship. I.L. acknowledges support from the QuantERA II program (project CoQuaDis, DFG Grant No. 532763411) that has received funding from the EU H2020 research and innovation program under GA No. 101017733. I.L. also acknowledges funding from the Deutsche Forschungsgemeinschaft (DFG, German Research Foundation) through the Research Unit FOR 5413/1, Grant No.~465199066, from Leverhulme Trust (Grant No. RPG-2024-112) and from the European Union through the ERC grant OPEN-2QS (Grant No. 101164443). G.P. acknowledges funding from the Swiss National Science Foundation (SNSF) through Grant No.~10005336. 

\textbf{AI usage. --- } The authors acknowledge using Claude (Anthropic) as a support tool for refining TikZ code for diagrams and for minor language editing. All suggestions were reviewed and approved by the authors.
\let\addcontentsline\oldaddcontentsline

\let\oldaddcontentsline\addcontentsline
\renewcommand{\addcontentsline}[3]{}
\bibliography{biblio.bib}
\let\addcontentsline\oldaddcontentsline

\onecolumngrid
\clearpage

\setcounter{equation}{0}
\setcounter{page}{1}

\setcounter{figure}{0}
\setcounter{table}{0}
\makeatletter
\renewcommand{\theequation}{S\arabic{equation}}
\renewcommand{\thefigure}{S\arabic{figure}}
\renewcommand{\thetable}{S\arabic{table}}
\setcounter{secnumdepth}{2}

\begin{center}
{\Large SUPPLEMENTAL MATERIAL}
\end{center}
\begin{center}
\vspace{0.8cm}
{\Large Exact first-detection probability in a locally monitored solvable quantum circuit}
\end{center}
\begin{center}
Cecilia De Fazio,$^{1}$ Igor Lesanovsky,$^{1,2}$
Gabriele Perfetto$^{3}$ 
\end{center}
\begin{center}
$^1${\em Institut f\"ur Theoretische Physik and Center for Integrated Quantum Science and Technology,\\ Universit\"at T\"ubingen, Auf der Morgenstelle 14, 72076 T\"ubingen, Germany}\\
$^2${\em School of Physics and Astronomy and Centre for the Mathematics\\ and Theoretical Physics of Quantum Non-Equilibrium Systems,\\ The University of Nottingham, Nottingham, NG7 2RD, United Kingdom}\\
$^3${\em Institut für Theoretische Physik, ETH Zürich, Wolfgang-Pauli-Str. 27, 8093 Zürich, Switzerland}
\end{center}

\tableofcontents

\section{Summary of graphic notation and useful relations}
\begin{figure}[h]
    \centering
    \includegraphics[width=0.25\textwidth]{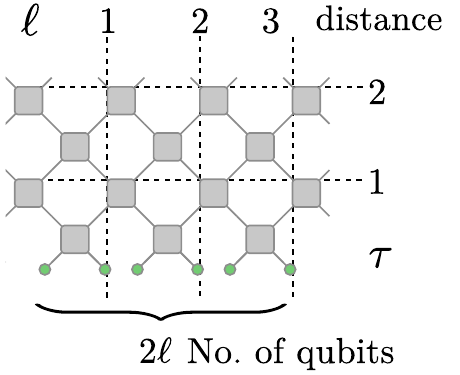}
    \caption{ \textbf{Convention adopted in the circuit representation.} Coordinates defining the length of the monitored region $\ell$, number of qubits $2\ell$ contained in the region, and the probing time $\tau$. In most of the graphics that we show in the Supplemental Material, we fix $\ell=3$ and $\tau=2$. Calculations can be, however, generalized for generic values of $\tau$ and $\ell$. }
    \label{figSM:notation}
\end{figure}
In this work we make extensive use of tensor network notation and circuit techniques, see the website \cite{TNwebsite} for an in-depth review on the subject. Below, we summarize the notation, define the graphical objects, and present the relations and identities used throughout the Supplemental Material. Furthermore, we refer to Fig.\ttilde\ref{figSM:notation} for the metric convention adopted in the letter and the Supplemental Material.
\\

\noindent\textbf{Qubits and gate evolution}: A single-qubit state in two dimensions is represented by a line with an open edge (we refer to it as an open leg):
\begin{equation}
    \ket{\phi}= 
\,,
\end{equation}
where  upward (downward) legs denote kets (bras). The scalar product of two states $\ket{\phi}$ and $\ket{\psi}$ is represented by a diagram in which
the two open legs are joined:
\begin{equation}
\label{eq:scalarprod}
 \newcommand{\dist}{0.3}
    \braket{\psi}{\phi}= 
     %
\,.
\end{equation}
The tensor product of two qubit states is denoted as  
\begin{equation}
    \ket{\phi} \otimes \ket{\psi}= 
     %
\,,
\end{equation}
where the two upward open legs indicate that the resulting state lives in a four-dimensional Hilbert space. Note that the inclination of the legs is introduced only to match the brickwork circuit layout illustrated in Fig.~1(b). 

Two-qubit time-evolution is implemented by the gate $u$ in Eq.~(1) of the main text, whose elements in the computational basis are
\begin{equation}
    \bra{\, k \; l \,} u \ket{ \,i \; j\,} =  %
    \,,
\end{equation}
with $\ket{\, i \; j\,}$ being a tensor product $\ket{i} \otimes \ket{j}\,$ for $i,j=0,1$. \\

\noindent \textbf{Observables:} We consider local observables, primarily expressed in terms of projectors. In particular, for a single qubit, the projector onto the state $\ket{0}$ is 
\begin{equation}
\mathbb{P}=
    %
\,.
\end{equation}
Additionally, we can define projectors onto the state $\ket{1}$ as
\begin{equation}
\openone-\mathbb{P}=
    %
\,.
\end{equation}
Fundamental identities that projectors satisfy can be expressed in the circuit representation as 
\begin{equation}
\label{eq:projrel1}
 \mathbb{P}^2=
     %
=\mathbb{P}(\openone-\mathbb{P})=0\;,\quad
\end{equation}
and those for their action on the computational basis as
\begin{equation}
\label{eq:projrel2}
\mathbb{P} \ket{0} =
     %
    = \mathbb{P}\ket{1} = 0\;.
\end{equation}
Successful measurement outcomes correspond to projectors onto the target state $\ket{\mathbf{{0}}_\ell} = \left(\,
%
 
= \left(\, \mathbb{P}\,\right)^{\otimes 2 \ell}\,.
\end{equation}
In contrast, unsuccessful measurement outcomes are expressed as 
\begin{equation}
\label{eq:getno}
{P}_{\ell\,,0}=
    %
 
= \openone_{\ell} - {P}_{\ell\,,1}
\,,
\end{equation}
with $\openone_{\ell}=\openone^{\otimes2\ell}$ and the single-site identity represented as $\openone = 
%
 \,.
$\\

\noindent\textbf{Folded-space quantities:} In folded space, the scalar product of two single-qubit states in Eq.\ttilde\eqref{eq:scalarprod} takes the form 
\begin{equation}
\label{eq:scalarprodf}
 \newcommand{\dist}{0.3}
         %
\,,
\end{equation}
where the ket (bra) state is in the back (front) layer and we introduce the thick leg $%
$, composed of two layers:
\begin{equation}
\label{eq:fs}
%
=2\,,
\end{equation}
where the last identity represents the norm of the folded-space state $ %
\right)^T$. Throughout the Supplemental Material, we will make use of bent lines to represent the folded-space legs $%
\,
$. This notation serves only to reproduce the geometry of the  brickwork circuit. 

Similarly, we define the unitary two-layer gate as
\begin{equation}
\label{eq:gatef}
    %
    \,,
\end{equation}
where each single-layer gate is given in Eq.~(1) of main text. This gate satisfies generalized dual unitarity  \ttilde\cite{yu2024hierarchical,Bertini2024} implemented through the two-gate relations 
\begin{equation}
\label{eq:generaliseddual}
     %
\;.
\end{equation} 
These relations allow us to contract the left-most (right-most) gate, reducing the two gates to a single one. Similar contractions can be carried out using the relations introduced below. \\

\noindent\textbf{One-layer and two-layer operators:} Operators and qubit states can be expressed in the folded space as illustrated in Fig.~2(b) of main text. In particular, we distinguish between two-layer projectors and one-layer projectors. For a single site, two-layer projectors onto the computational basis are defined as 
\begin{equation}
%
\,.
\end{equation}
Relations in Eqs.\ttilde\eqref{eq:projrel1}-\eqref{eq:projrel2} generalize to two-layer projectors
\begin{equation}
\label{eq:projrel1f}
     %
=0\;.
\end{equation}
Here, we have also defined the two-layer states $%
$. In particular, other identities hold in the folded space, such as these ones defining the norms of these two-layer states
\begin{equation}
\label{eq:normstatezero}
      %
=1\;.
\end{equation}
As a consequence of the East rule satisfied by the gate in Eq.~(1), applying the two-layer gate to a two-layer state zero (or projector in zero) placed on the right implies that the qudit on the left remains unchanged:
\begin{equation}
\label{eq:gatestaterelations}
    %
\;,\quad\quad
\end{equation}
Moreover, pairs of projectors $\projectorf$, as well as single projectors applied to a right leg, can propagate through the gate according to
\begin{equation}
\label{eq:greencircleprop}
  %
  \,.
\end{equation} 
These relations can be used to contract the gate through the identities
\begin{equation}
\label{eq:greencirclegate}
  %
 \;.
\end{equation}
We also consider single-layer operators, such as the zero-state projectors acting on the back and on
the front layer alone, respectively
\begin{equation}
\label{eq:onelayer}
   \;\quad\quad
 %
\end{equation}
The action of two-layer projectors on one of these projectors, when non-zero, gives a two-layer projector
\begin{equation}
\label{eq:onetwolayers}
       %
\end{equation} \\

\noindent\textbf{A special property of the gate:} 
The requirement that the same state be enforced on both layers of the folded circuit can be implemented more generally through the operator $\projectorf[\staropf]$, which we introduce here starting from the following property of the gate\ttilde\eqref{eq:gatef}:
\begin{equation}
\label{eq:relstar}
     %
 = \text{diag} \left( 1, 0 , 0 , 1 \right) \;.
\end{equation}
This operator acts on the single-layer projectors $\projectorf[\backqubitf]$ and $\projectorf[\frontqubitf]$ as 
      \begin{equation}
      \label{eq:actionstarop}
       %
\;,
      \end{equation}
      where the notation $%
$ accounts for the commutativity of the operators $%
\,.
\end{equation}
By the relations in Eq.\ttilde\eqref{eq:projrel1f} one can show that the following identities hold
\begin{equation}
    %
  =2\;.\quad\quad
\end{equation}
Furthermore, the application of the operators $%
$ to the left side of the gate in Eq.\ttilde\eqref{eq:gatef} implies \cite{Bertini2024}
\begin{equation}
\label{eq:staropfgatef}
     %
\;.
\end{equation}
Moreover, two operators $\projectorf[\staropf]$ propagate vertically through the gate 
\begin{equation}
    \label{eq:staropfgateftwo}
       %
 
  \;.
\end{equation}
By combining the first identity in Eq.\ttilde\eqref{eq:staropfgatef} with Eq.\ttilde\eqref{eq:projgaterelations}, we obtain
\begin{equation}
\label{eq:staropfgatefqubitf}
    %
\;.
\end{equation}

\section{Unitary-circuit techniques}
In this section, we present the details of the model considered in the main text. In particular, in Subsec.~\ref{subsec:IIA}, we show an explicit circuit representation of the first-detection probability in Eq.~(5). In Subsec.~\ref{subsec:IIB}, we discuss how to implement the local approach illustrated in panel (d) of Fig.~2. 

\subsection{Brickwork circuits for first-detection probabilities}
\label{subsec:IIA}
We here present the explicit circuits used to compute the first-detection probability in Eq.~(5). At time zero, the system is in a product state: the monitored system is initialized in the target state
$\ket{\mathbf{{0}}_\ell} = \left(\,
\begin{tikzpicture}[baseline={([yshift=-0.6ex]current bounding box.center)},scale=0.5]
     \SingleProp{0}{0}{\hprop}
     \qubit{0}{0}
     \end{tikzpicture} \,\right)^{\otimes 2\ell}$
 and the rest of the system is in a bi-translationally invariant state $   \left(\, \ket{\phi_1} \otimes \ket{\phi_2} \,\right)^{\otimes L-\ell}$ where the two-site units are denoted as
\begin{equation}
\label{eq:lrqubitstates}
 \ket{\phi_1} \otimes \ket{\phi_2} = 
     \begin{tikzpicture}[baseline={([yshift=-0.6ex]current bounding box.center)},scale=0.5]
     \pgfmathparse{0.5}\let\dl\pgfmathresult
      \gridLine{(0.0)-\dl}{-\dl}{(0.0)}{(0.0)}%
       \qubitone{-\dl}{-\dl}
       
        \begin{scope}[shift={(0.3,0)}]
         \gridLine{(0.0}{(0.0)}{+\dl}{(0.0)-\dl} %
       \qubittwo{\dl}{-\dl}
        \end{scope}
      
     \end{tikzpicture}\,,
\end{equation}
with single-qubit states given by $\ket{\phi_\mathrm{j}}= r_j e^{i \alpha_\mathrm{j}} \begin{tikzpicture}[baseline={([yshift=-0.6ex]current bounding box.center)},scale=0.5]
     \SingleProp{0}{0}{\hprop}
     \qubit{0}{0}
     \end{tikzpicture}+\sqrt{1-r_\mathrm{j}^2}\,e^{i \beta_\mathrm{j}}  \begin{tikzpicture}[baseline={([yshift=-0.6ex]current bounding box.center)},scale=0.5]
     \SingleProp{0}{0}{\hprop}
     \qubitr{0}{0}
     \end{tikzpicture}\,,$ for $j=1,2$ and whose parameters satisfy \cite{Bertini2024}
\begin{align}
\label{eq:initialstate}
     0 &=
  r_1r_2\sqrt{1-r_1^2}\sqrt{1-r_2^2} \cos(\alpha_1\!-\!\beta_1)\cos(\alpha_2\!-\!\beta_2) ,
    \nonumber \\
  0 &=
  \left (r^2_1-\frac{1}{2}\right)\left(r^2_2-\frac{1}{2}\right).
\end{align}
 We stress that these relations are crucial for accessing the local dynamics of the monitored region. Moreover, due to the last relation, the two-qubit states are necessarily superpositions of computational-basis states. As explained in the main text, starting from the initial state, the brickwork dynamics alternated with the measurement outcomes~\eqref{eq:getyes} and~\eqref{eq:getno}, resulting from stroboscopic measurements performed with period $\tau$, leads to the state in Eq.~(4), represented here as
\begin{equation}
\label{eq:statekc}
 \ket{\Psi_k} =
  \begin{tikzpicture}[baseline={([yshift=-0.6ex]current bounding box.center)},scale=0.5]
    \pgfmathparse{5}\let\ellpos\pgfmathresult
    \pgfmathparse{2}\let\period\pgfmathresult
    \pgfmathparse{3*\period}\let\tval\pgfmathresult
    \pgfmathparse{1*\period}\let\ttone\pgfmathresult
    \pgfmathparse{2*\period}\let\tttwo\pgfmathresult
    \pgfmathparse{11}\let\totalL\pgfmathresult
    \pgfmathparse{3}\let\elllength\pgfmathresult

    \statePBC{\tval}{\totalL}{\ellpos}{\elllength}
    \getNo{\ellpos}{\ttone}{\elllength}
    \getNo{\ellpos}{\tttwo}{\elllength}
    \getYes{\ellpos}{\tval}{\elllength}

    \pgfmathsetmacro{\dyarrow}{0.6}
    \pgfmathsetmacro{\startgatex}{1 + 2*(\ellpos - 1)}
    \pgfmathsetmacro{\arrowXstart}{\startgatex - \dlonger}
    \pgfmathsetmacro{\arrowXend}{\startgatex + 2*(\elllength-1) + \dlonger}
    \pgfmathsetmacro{\arrowXmid}{0.5*(\arrowXstart+\arrowXend)}

    \pgfmathsetmacro{\arrowYbot}{-\dlonger - \dyarrow}
    \draw[<->, thick] (\arrowXstart, \arrowYbot) -- (\arrowXend, \arrowYbot);
    \node at (\arrowXmid, \arrowYbot-0.4) {$\ell$};

    \pgfmathsetmacro{\arrowYtop}{\yfinal - 0.25 + \dyarrow}
    \pgfmathsetmacro{\arrowXstartL}{0.0 }
    \pgfmathsetmacro{\arrowXendL}{\xfinal + \dlonger}
    \pgfmathsetmacro{\arrowXmidL}{0.5*(\arrowXstartL+\arrowXendL)}
    \draw[<->, thick] (\arrowXstartL, \arrowYtop) -- (\arrowXendL, \arrowYtop);
    \node at (\arrowXmidL, \arrowYtop+0.4) {$L$};
    
    \pgfmathsetmacro{\dxarrow}{1.0}
    \pgfmathsetmacro{\arrowX}{\xfinal + \dxarrow}
    \pgfmathsetmacro{\getNoYtop}{2*\ttone - 0.5 + 0.06}
    \pgfmathsetmacro{\arrowYmid}{0.5*(-\dlonger+\getNoYtop)}
    \draw[<->, thick] (\arrowX, -\dlonger) -- (\arrowX, \getNoYtop);
    \node[anchor=west] at (\arrowX+0.2, \arrowYmid) {$\tau$};

  \end{tikzpicture}
  \,.
\end{equation}
Above, the brickwork unitary operator $U(\tau)$, also illustrated in Fig.~1(b), is implemented as in Eq.~(2), while periodic boundary conditions are enforced by the dashed lines connecting the qubits at the left and right edges of the system. The squared norm of this state gives the first-detection probability in Eq.~(5), which in the folded representation reads
\begin{equation}
\label{eq:fdpfolded}
    F(k) = \braket{\Psi_k} =
     \begin{tikzpicture}[baseline={([yshift=-0.6ex]current bounding box.center)},scale=0.5]
    \pgfmathparse{5}\let\ellpos\pgfmathresult
    \pgfmathparse{2}\let\period\pgfmathresult
    \pgfmathparse{3*\period}\let\tval\pgfmathresult
    \pgfmathparse{1*\period}\let\ttone\pgfmathresult
    \pgfmathparse{2*\period}\let\tttwo\pgfmathresult
    \pgfmathparse{11}\let\totalL\pgfmathresult
    \pgfmathparse{3}\let\elllength\pgfmathresult

    \statePBCf{\tval}{\totalL}{\ellpos}{\elllength}
    \getNof{\ellpos}{\ttone}{\elllength}
    \getNof{\ellpos}{\tttwo}{\elllength}
    
    \pgfmathsetmacro{\startgatex}{1 + 2*(\ellpos - 1)}
    \pgfmathsetmacro{\ddelta}{0.5*\dlonger}
    \foreach \i in {0,...,\numexpr\elllength-1\relax}{%
      \pgfmathsetmacro{\xx}{\startgatex + 2*\i}%
      \pgfmathsetmacro{\oddX}{\xx-\ddelta}%
      \qubitf{\oddX}{\yfinal-\ddelta}%
      \pgfmathsetmacro{\evenX}{\xx+\ddelta}%
      \qubitf{\evenX}{\yfinal-\ddelta}%
    }%
    \foreach \i in {0,...,\numexpr\totalL-1\relax}{%
      \pgfmathtruncatemacro{\skipstart}{\ellpos-1}%
      \pgfmathtruncatemacro{\skipend}{\ellpos+\elllength-2}%
      \ifnum\i<\skipstart
        \pgfmathsetmacro{\xx}{2*\i+2}%
        \pgfmathsetmacro{\oddX}{\xx-1-\ddelta}%
        \dotf{\oddX}{\yfinal-\ddelta}%
        \pgfmathsetmacro{\evenX}{\xx-1+\ddelta}%
        \dotf{\evenX}{\yfinal-\ddelta}%
      \else
        \ifnum\i>\skipend
          \pgfmathsetmacro{\xx}{2*\i+2}%
          \pgfmathsetmacro{\oddX}{\xx-1-\ddelta}%
          \dotf{\oddX}{\yfinal-\ddelta}%
          \pgfmathsetmacro{\evenX}{\xx-1+\ddelta}%
          \dotf{\evenX}{\yfinal-\ddelta}%
        \fi
      \fi
    } 
  \end{tikzpicture}\;,
\end{equation}
This circuit is the explicit representation of the one shown in Fig.~2(a), where the
two layers are encoded in the gate of Eq.\ttilde\eqref{eq:gatef} and in the folded-space states and
operators defined in Eqs.\ttilde\eqref{eq:scalarprodf}--\eqref{eq:gatef} and in
Fig.~2(b).

\subsection{Space-evolution approach}
\label{subsec:IIB}
We now review the ideas developed in~\cite{Bertini2024} to compute the entanglement entropy of the unmonitored DFQEM, revisited here for the computation of the first-detection probability. In particular, we discuss how to pass from the dynamics of the total system, illustrated in Fig.~2(a), to the local dynamics shown in Fig.~2(d), thereby reducing the computation of the first-detection probability to a problem defined solely on the monitored region.

According to Eq.~(5), the
first-detection probability can be interpreted as a multi-time correlation function of local projectors supported on the monitored region of length $\ell$. To keep the discussion general and valid for any number of attempts
$k$, we do not specify the nature of the observables for the moment, and write the first-detection
probability $F(k)$ more generally as a correlator $C_{t,\ell}$ involving observables of finite support,
contained within a box of size $t \times \ell$, where $t = k\tau$ denotes
the total observation time and $\ell$ the spatial support, respectively:
\begin{equation}
\label{eq:correlator}
C_{t,\ell} =     
   \begin{tikzpicture}[baseline={([yshift=-0.6ex]current bounding box.center)},scale=0.5]
    \pgfmathparse{5}\let\ellpos\pgfmathresult
    \pgfmathparse{4}\let\tval\pgfmathresult
    \pgfmathparse{11}\let\totalL\pgfmathresult
    \pgfmathparse{3}\let\elllength\pgfmathresult

      \statef{\tval}{\totalL}{\ellpos}{\elllength}
   
    \pgfmathsetmacro{\startgatex}{1 + 2*(\ellpos - 1)}
    \pgfmathsetmacro{\ddelta}{0.5*\dlonger}
    \foreach \i in {0,...,\numexpr\elllength-1\relax}{%
      \pgfmathsetmacro{\xx}{\startgatex + 2*\i}%
      \pgfmathsetmacro{\oddX}{\xx-\ddelta}%
      \dotf{\oddX}{\yfinal-\ddelta}
      \pgfmathsetmacro{\evenX}{\xx+\ddelta}%
      \dotf{\evenX}{\yfinal-\ddelta}%
    }%
    \foreach \i in {0,...,\numexpr\totalL\relax}{%
      \pgfmathtruncatemacro{\skipstart}{\ellpos-1}%
      \pgfmathtruncatemacro{\skipend}{\ellpos+\elllength-2}%
      \ifnum\i<\skipstart
        \pgfmathsetmacro{\xx}{2*\i+2}%
        \ifnum\i=0\else
          \pgfmathsetmacro{\oddX}{\xx-1-\ddelta}%
          \dotf{\oddX}{\yfinal-\ddelta}%
        \fi
        \ifnum\i=\totalL\else
          \pgfmathsetmacro{\evenX}{\xx-1+\ddelta}%
          \dotf{\evenX}{\yfinal-\ddelta}%
        \fi
      \else
        \ifnum\i>\skipend
          \pgfmathsetmacro{\xx}{2*\i+2}%
          \pgfmathsetmacro{\oddX}{\xx-1-\ddelta}%
          \dotf{\oddX}{\yfinal-\ddelta}%
          \ifnum\i=\totalL\else
            \pgfmathsetmacro{\evenX}{\xx-1+\ddelta}%
            \dotf{\evenX}{\yfinal-\ddelta}%
          \fi
        \fi
      \fi
    }%

    
    \pgfmathsetmacro{\ddelta}{0.5*\dlonger}
    \pgfmathsetmacro{\boxXstart}{\startgatex - 2 - \dlonger}
    \pgfmathsetmacro{\boxXend}{\startgatex + 2*(\elllength-1) + \dlonger}
    \pgfmathsetmacro{\boxYbot}{-\dlonger}
    \pgfmathsetmacro{\boxYtop}{\yfinal - \ddelta}
    \pgfmathparse{1.15}\let\distxl\pgfmathresult
    \pgfmathparse{0.15}\let\distxr\pgfmathresult
    \pgfmathparse{0.20}\let\disty\pgfmathresult
    \draw[line width=0.1pt, myredShadow, rounded corners=3pt, fill=myred, fill opacity=0.15, draw opacity=1.0] (\boxXstart+\distxl, \boxYbot+\disty) rectangle (\boxXend-+\distxr, \boxYtop+\disty);
    
    \pgfmathsetmacro{\dxarrow}{1.5}
    \pgfmathsetmacro{\arrowX}{\xfinal + \dxarrow}
    \pgfmathsetmacro{\arrowYmid}{0.5*(\boxYbot+\boxYtop)}
    \draw[<->, thick] (\arrowX, \boxYbot) -- (\arrowX, \boxYtop);
    \node[anchor=west] at (\arrowX+0.2, \arrowYmid) {$t$};
    
    \pgfmathsetmacro{\dyarrow}{0.6}
    \pgfmathsetmacro{\startgatex}{1 + 2*(\ellpos - 1)}
    \pgfmathsetmacro{\arrowXstart}{\startgatex - \dlonger}
    \pgfmathsetmacro{\arrowXend}{\startgatex + 2*(\elllength-1) + \dlonger}
    \pgfmathsetmacro{\arrowXmid}{0.5*(\arrowXstart+\arrowXend)}
    \pgfmathsetmacro{\arrowYbot}{-\dlonger - \dyarrow}
    \draw[<->, thick] (\arrowXstart, \arrowYbot) -- (\arrowXend, \arrowYbot);
    \node at (\arrowXmid, \arrowYbot-0.4) {$\ell$};
    
    \pgfmathsetmacro{\dotsX}{1.5 - \dlonger - 0.5}%
    \pgfmathtruncatemacro{\leftYend}{2*\tval-1}%
    \foreach \y in {1,3,...,\leftYend}{%
      \node at (\dotsX, \y) {$\dots$};
    }%

    \pgfmathsetmacro{\dotsXend}{2+ 2*(\totalL - 1)+1.5 - \dlonger}%
    \pgfmathtruncatemacro{\rightYend}{2*(\tval-1)}%
    \foreach \y in {0,2,...,\rightYend}{%
      \node at (\dotsXend, \y) {$\dots$};
    }\;.
  \end{tikzpicture}
\end{equation}
The space-evolution approach\ttilde\cite{muellerhermes2012tensor,Hastings2015} is based on exchanging the roles of space and time, exploiting the special space-time geometry of discrete-time circuits such as the brickwork circuit. Its main requirement is translational invariance, which is satisfied by the initial state of the unmonitored qubits, as it consists of repeated two-qubit units defined in Eq.~\eqref{eq:lrqubitstates}.
At a fixed time $t$, we can thus define
the space-transfer matrix, which evolves a temporal state of extent $t$ along the spatial direction, as
\begin{equation}
\label{eq:spacetransfer}
    \mathbb{T}_t = 
\begin{tikzpicture}[baseline={([yshift=-0.6ex]current bounding box.center)},scale=0.5]
  \pgfmathparse{4}\let\tval\pgfmathresult
  \pgfmathparse{1}\let\totalL\pgfmathresult
  \spacetransferf{\tval}{\totalL}
  
  \pgfmathsetmacro{\dxarrow}{1.5}%
  \pgfmathsetmacro{\arrowX}{\xfinal + \dxarrow}%
  \pgfmathsetmacro{\arrowYbot}{-\dlonger}%
  \pgfmathsetmacro{\arrowYtop}{2*\tval-1+\dlonger}%
  \pgfmathsetmacro{\arrowYmid}{0.5*(\arrowYbot+\arrowYtop)}%
  \draw[<->, thick] (\arrowX, \arrowYbot) -- (\arrowX, \arrowYtop);
  \node[anchor=west] at (\arrowX+0.2, \arrowYmid) {$t$};
\end{tikzpicture}\;.
\end{equation}
Note that the space-transfer matrix, unlike the time-evolution operator, depends strongly on the choice of initial state and, in the present case, on the qubit states defined by the relations in Eq.\ttilde\eqref{eq:initialstate}. The stationary state of the space evolution is described by the fixed points, i.e., the left and right eigenstates of the space-transfer matrix with the largest eigenvalue,
here equal to one due to the chosen normalization: 
\begin{equation}
     {}_{t}\dbra{L}  \mathbb{T}_t =  {}_{t}\dbra{L} \,, \quad \quad \mathbb{T}_t \dket{R}_t = \dket{R}_t \,, \quad \quad {}_{t}\dbraket{L}{R}_t =1 \,.
\end{equation}
For the present model and initial state, the fixed points of $\mathbb{T}_t$ can be computed exactly \ttilde\cite{Bertini2024}  
and read:
\begin{equation}
   {}_{t}\dbra{L} = 2^{-t+1}
\begin{tikzpicture}[baseline={([yshift=-0.6ex]current bounding box.center)},scale=0.5]
 \pgfmathparse{4}\let\tval\pgfmathresult
 \leftfpf{0}{0}{\tval}
\end{tikzpicture} \,, \quad \quad \quad \dket{R}_t = 2^{-t}
\begin{tikzpicture}[baseline={([yshift=-0.6ex]current bounding box.center)},scale=0.5]
 \pgfmathparse{4}\let\tval\pgfmathresult
 \rightfpf[0]{0}{0}{\tval}
\end{tikzpicture} \,. 
\end{equation}
Crucially, when the support of the observables is sufficiently small compared to the total system size,
$\ell \ll L$, the spatial stationary state is attained and the correlator in
Eq.\ttilde\eqref{eq:correlator} reduces to
\begin{equation}
\label{eq:fdplocal}
C_{t,\ell} = 2^{-2t+1}
    \begin{tikzpicture}[baseline={([yshift=-0.6ex]current bounding box.center)},scale=0.5]
  \pgfmathparse{4}\let\tval\pgfmathresult
  \pgfmathparse{3}\let\totalL\pgfmathresult
  \localcircuitf{\tval}{\totalL}
  \leftfpf{-1}{0}{\tval}
  \rightfpf[1]{2*\totalL}{1}{\tval}

  \pgfmathsetmacro{\boxXstart}{0.5}
  \pgfmathsetmacro{\boxXend}{2*\totalL + 0.5}
  \pgfmathsetmacro{\boxYbot}{-\dlonger + 0.20}
  \pgfmathsetmacro{\boxYtop}{\myystart + 0.15}
  \pgfmathparse{0.0}\let\distxl\pgfmathresult
  \pgfmathparse{0.0}\let\distxr\pgfmathresult
  \pgfmathparse{0.0}\let\disty\pgfmathresult
  \draw[line width=0.1pt, myredShadow, rounded corners=3pt, fill=myred, fill opacity=0.15, draw opacity=1.0] (\boxXstart-\distxl, \boxYbot-\disty) rectangle (\boxXend+\distxr, \boxYtop+\disty);
\end{tikzpicture}\;.
\end{equation}
This holds for any local correlator and in particular for the first-detection probability $F(k)$ in Eq.\ttilde\eqref{eq:fdpfolded}, which becomes 
\begin{equation}
\label{eq:localFDP}
 F(k) =
  2^{-2k\tau+1}
    \begin{tikzpicture}[baseline={([yshift=-0.6ex]current bounding box.center)},scale=0.5]
   \pgfmathparse{6}\let\tval\pgfmathresult
  \pgfmathparse{3}\let\totalL\pgfmathresult
   \pgfmathparse{1.5}\let\ellpos\pgfmathresult
    \pgfmathparse{3}\let\elllength\pgfmathresult
    \pgfmathparse{2}\let\period\pgfmathresult
    \pgfmathparse{3*\period}\let\tval\pgfmathresult
    \pgfmathparse{1*\period}\let\ttone\pgfmathresult
    \pgfmathparse{2*\period}\let\tttwo\pgfmathresult
  \localcircuitf[0]{\tval}{\totalL}
  \leftfpf{-1}{0}{\tval}
  \rightfpf[0]{2*\totalL}{1}{\tval}
  \pgfmathsetmacro{\myystart}{2*\tval-1+\dlonger}%
  \pgfmathtruncatemacro{\xfinalint}{2*\totalL}%
  \foreach\x in \xlistodd{%
    \ifnum\x=1\else
      \ifnum\x=\xfinalminus\else
        \pgfmathsetmacro{\dotX}{\x+1+2+\dlonger}%
        \qubitf{\dotX}{\myystart}%
      \fi
    \fi
  }%
  \foreach\x in \xlisteven{%
    \ifnum\x=\xfinalint\else
      \pgfmathsetmacro{\dotX}{\x+2-1-\dlonger}%
      \qubitf{\dotX}{\myystart}%
    \fi
  }%
  \pgfmathsetmacro{\dotX}{1+\dlonger}%
  \qubitf{\dotX}{\myystart}%
  \pgfmathsetmacro{\dotXright}{2*\totalL+1-\dlonger}%
  \qubitf{\dotXright}{\myystart}%
  \getNof{\ellpos}{\ttone}{\elllength}
    \getNof{\ellpos}{\tttwo}{\elllength}

\end{tikzpicture}\;
=
 2^{-2k\tau+1}
    \begin{tikzpicture}[baseline={([yshift=-0.6ex]current bounding box.center)},scale=0.5]
  \pgfmathparse{6}\let\tval\pgfmathresult
  \pgfmathparse{3}\let\totalL\pgfmathresult
   \pgfmathparse{1.5}\let\ellpos\pgfmathresult
    \pgfmathparse{3}\let\elllength\pgfmathresult
    \pgfmathparse{2}\let\period\pgfmathresult
    \pgfmathparse{3*\period}\let\tval\pgfmathresult
    \pgfmathparse{1*\period}\let\ttone\pgfmathresult
    \pgfmathparse{2*\period}\let\tttwo\pgfmathresult

  \localcircuitf[0]{\tval}{\totalL}
  \pgfmathsetmacro{\dlonger}{0.65}%
  \pgfmathtruncatemacro{\yfinalminus}{2*\tval-2}%
  \pgfmathsetmacro{\dotXleft}{1-\dlonger}%
  \foreach \y in {0,2,...,\yfinalminus}{%
    \pgfmathsetmacro{\dotYdown}{\y+1-\dlonger}%
    \pgfmathsetmacro{\dotYup}{\y+1+\dlonger}%
    \ifnum\y=0\else
      \dotf{\dotXleft}{\dotYdown}%
    \fi
    \dotf{\dotXleft}{\dotYup}%
  }
  \pgfmathsetmacro{\qbitX}{1-\dlonger}%
  \pgfmathsetmacro{\qbitY}{1-\dlonger}%
  \qubittwodf{\qbitX+0.15}{\qbitY+0.15-1}
\rightfpf[0]{2*\totalL}{1}{\tval}
  \pgfmathsetmacro{\myystart}{2*\tval-1+\dlonger}%
  \pgfmathtruncatemacro{\xfinalint}{2*\totalL}%
  \foreach\x in \xlistodd{%
    \ifnum\x=1\else
      \ifnum\x=\xfinalminus\else
        \pgfmathsetmacro{\dotX}{\x+1+2+\dlonger}%
        \qubitf{\dotX}{\myystart}%
      \fi
    \fi
  }%
  \foreach\x in \xlisteven{%
    \ifnum\x=\xfinalint\else
      \pgfmathsetmacro{\dotX}{\x+2-1-\dlonger}%
      \qubitf{\dotX}{\myystart}%
    \fi
  }%
  \pgfmathsetmacro{\dotX}{1+\dlonger}%
  \qubitf{\dotX}{\myystart}%
  \pgfmathsetmacro{\dotXright}{2*\totalL+1-\dlonger}%
  \qubitf{\dotXright}{\myystart}%
  \nogaterf{0}{0}
  
  \getNof{\ellpos}{\ttone}{\elllength}
    \getNof{\ellpos}{\tttwo}{\elllength}
\end{tikzpicture}\;.
\end{equation}
In the second equality, to contract the leftmost column, we used the first relation in Eq.\ttilde\eqref{eq:generaliseddual} from top to bottom, Eq.\ttilde\eqref{eq:relstar} for the bottom gate, along with the normalization $  \begin{tikzpicture}[baseline={([yshift=-0.6ex]current bounding box.center)},scale=0.5]
 \nogatelf{0}{0}
    \dotf{-0.5}{0.5} 
    \qubitonef{-0.5}{-0.5} 
  \end{tikzpicture}
  = 1$. We also introduced the following state 
\begin{equation}
\label{eq:qubittwod}
 \begin{tikzpicture}[baseline={([yshift=-0.6ex]current bounding box.center)},scale=0.5] 
       \nogaterf{0}{0}
    \qubittwodf{0.5}{-0.5} 
  \end{tikzpicture} =
       \begin{tikzpicture}[baseline={([yshift=-0.6ex]current bounding box.center)},scale=0.5]
    \nogaterf[\staropf]{0}{0}
    \qubittwof{0.5}{-0.5} 
  \end{tikzpicture}
   = 
  r_2^2 \,
    \begin{tikzpicture}[baseline={([yshift=-0.6ex]current bounding box.center)},scale=0.5]
   \nogaterf{0}{0}
    \qubitf{0.5}{-0.5}
  \end{tikzpicture}
  + (1-r_2^2) \,
    \begin{tikzpicture}[baseline={([yshift=-0.6ex]current bounding box.center)},scale=0.5]
    \nogaterf{0}{0}
    \qubitrf{0.5}{-0.5}
    \end{tikzpicture}. 
\end{equation}
Therefore, the first-detection probability in Eq.\ttilde\eqref{eq:localFDP} refers to the dynamics of the monitored region alone. It can be understood as an alternation of measurement outcomes and the non-unitary operator $\mathcal{M}_\ell$, displayed in Fig.~2(c), which we now define for a time interval $t$ as
\begin{equation}
\label{eq:M_ell}
    \mathcal{M}_\ell (t) = 2^{-2t}\,
      \begin{tikzpicture}[baseline={([yshift=1.5ex]current bounding box.center)},scale=0.5]
  \pgfmathparse{2}\let\tval\pgfmathresult
  \pgfmathparse{3}\let\totalL\pgfmathresult
   \pgfmathparse{1.5}\let\ellpos\pgfmathresult
    \pgfmathparse{3}\let\elllength\pgfmathresult

  \localcircuitNOTARGETf[0]{\tval}{\totalL}
  \pgfmathsetmacro{\dlonger}{0.65}%
  \pgfmathtruncatemacro{\yfinalminus}{2*\tval-2}%
  \pgfmathsetmacro{\dotXleft}{1-\dlonger}%
  \foreach \y in {0,2,...,\yfinalminus}{%
    \pgfmathsetmacro{\dotYdown}{\y+1-\dlonger}%
    \pgfmathsetmacro{\dotYup}{\y+1+\dlonger}%
    \ifnum\y=0\else
      \dotf{\dotXleft}{\dotYdown}%
    \fi
    \dotf{\dotXleft}{\dotYup}%
  }
  \pgfmathsetmacro{\qbitX}{1-\dlonger}%
  \pgfmathsetmacro{\qbitY}{1-\dlonger}%
  \dotf{\qbitX}{\qbitY}%
\rightfpf[0]{2*\totalL}{1}{\tval}

  \pgfmathsetmacro{\dlonger}{0.65}%
  \pgfmathsetmacro{\myystart}{2*\tval-1+\dlonger}%
  \pgfmathsetmacro{\dxarrow}{1.5}%
  \pgfmathsetmacro{\arrowX}{2*\totalL+1+\dxarrow}%
  \pgfmathsetmacro{\arrowYbot}{-\dlonger}%
  \pgfmathsetmacro{\arrowYtop}{\myystart}%
  \pgfmathsetmacro{\arrowYmid}{0.5*(\arrowYbot+\arrowYtop)}%
  \draw[<->, thick] (\arrowX, \arrowYbot) -- (\arrowX, \arrowYtop);
  \node[anchor=west] at (\arrowX+0.2, \arrowYmid) {$t$};

  \pgfmathsetmacro{\dyarrow}{0.6}%
  \pgfmathsetmacro{\arrowXstart}{1-\dlonger}%
  \pgfmathsetmacro{\arrowXend}{2*\elllength+1+\dlonger}%
  \pgfmathsetmacro{\arrowXmid}{0.5*(\arrowXstart+\arrowXend)}%
  \pgfmathsetmacro{\arrowYbot}{-\dlonger-\dyarrow}%
  \draw[<->, thick] (\arrowXstart, \arrowYbot) -- (\arrowXend, \arrowYbot);
  \node at (\arrowXmid, \arrowYbot-0.4) {$\ell$};
    \end{tikzpicture}\,,
\end{equation}
In particular, $\mathcal{M}_\ell$ corresponds to the effective evolution operator of the monitored subregion in folded space, with its non-unitarity arising from boundary contributions determined by the fixed points of the space-transfer matrix in Eq.\ttilde\eqref{eq:spacetransfer}. \\

This concludes the overview of the circuit techniques used to implement a local description of the first-detection probability. In the
next section, we exploit these ideas  to compute the building blocks of the
expansion in Eq.~(6).

\section{Computation of the subsystem Loschmidt echo}
\label{sec:compLamplitude}
In this section, we present the derivation of the Loschmidt echo in Eq.~(7). 
We start from the explicit circuit representation, obtained by plugging  Eq.\ttilde\eqref{eq:M_ell} into Eq.~(7):
\begin{equation}
\label{eq:Amplitudecircuit}
 \mathcal{L}(m) =     
\;,
\end{equation}
for a fixed period $m\tau$ and monitored-region length $\ell$; in the diagram above we set $m\tau=4$, and $\ell=3$. 

To compute this quantity, we distinguish two regimes for the total observation time $m\tau$. We first consider the case $m\tau \leq 2\ell$. For concreteness, we illustrate the derivation for the values used in the diagram above, although the same steps apply throughout this regime. In this case, the diagram in
Eq.\ttilde\eqref{eq:Amplitudecircuit} reduces to
\begin{equation}
\label{eq:Amplitudecircuitcomp}
   %
\end{equation}
Here, we exploit the first relation in Eq.\ttilde\eqref{eq:greencircleprop} to propagate the projectors $\projectorf$ at the top and at the bottom through the circuit.  This propagation generates two triangles of green circles that meet, producing a term of the form $ %
$. In the second equality, the
  two projectors on that term propagate diagonally via the same relation. We iterate this last step to all the other terms until we obtain the last equality. Using Eqs.\ttilde\eqref{eq:gatestaterelations} and \ttilde\eqref{eq:greencirclegate}, together with the norms in the last relation of Eq.\ttilde\eqref{eq:fs} and in Eq.\ttilde\eqref{eq:normstatezero}, the resulting diagram in Eq.\ttilde\eqref{eq:Amplitudecircuitcomp} can be contracted to obtain
\begin{equation}
\label{eq:Lmts}
    \mathcal{L}(m) = 2^{-2m\tau}
   %
\; \right)^{m\tau} =2^{-m\tau}\;, \quad \quad \;\text{for}\;\,m\tau\leq 2\ell\,.
\end{equation}

We now turn to the case $m\tau > 2\ell$. For concreteness, we illustrate the derivation for $m\tau=8$ and $\ell=3$, although the same steps apply throughout this regime. In this case, the diagram in Eq.\ttilde\eqref{eq:Amplitudecircuit} gives 
\begin{equation}
    \label{eq:Amplitudecompl}
       %
 
\end{equation}
In more detail: in the first equality above, similarly to the previous case, we use Eq.\ttilde\eqref{eq:greencircleprop} to propagate the projectors $\projectorf$ through the circuit, obtaining two triangles of green circles. In the second equality, we use Eq.\ttilde\eqref{eq:relstar} to introduce the operator $\projectorf[\staropf]$ in the left-most column of the circuit, which we then propagate rightward via the first relation in  Eq.\ttilde\eqref{eq:staropfgatef}, yielding the last diagram. Contracting it further, we obtain
\begin{equation}
\label{eq:Lamplitudelargetau}
    \begin{tikzpicture}[baseline={([yshift=-0.6ex]current bounding box.center)},scale=0.5]
  \pgfmathparse{8}\let\tval\pgfmathresult
  \pgfmathparse{3}\let\totalL\pgfmathresult
   \pgfmathparse{1.5}\let\ellpos\pgfmathresult
    \pgfmathparse{3}\let\elllength\pgfmathresult
    \pgfmathparse{4}\let\period\pgfmathresult
    \pgfmathparse{1*\period}\let\ttone\pgfmathresult

  \localcircuitNOTARGETnogatef{\tval}{\totalL}
  \pgfmathsetmacro{\dlonger}{0.65}%
  \pgfmathtruncatemacro{\yfinalminus}{2*\tval-2}%
  \pgfmathsetmacro{\dotXleft}{1-\dlonger}%
  \foreach \y in {0,2,...,\yfinalminus}{%
    \pgfmathsetmacro{\dotYdown}{\y+1-\dlonger}%
    \pgfmathsetmacro{\dotYup}{\y+1+\dlonger}%
    \ifnum\y=0\else
      \dotf{\dotXleft}{\dotYdown}%
    \fi
    \dotf{\dotXleft}{\dotYup}%
  }
  \pgfmathsetmacro{\qbitX}{1-\dlonger}%
  \pgfmathsetmacro{\qbitY}{1-\dlonger}%
    \dotf{1-\dlonger}{1-\dlonger}
  
\rightfpf[0]{2*\totalL}{1}{\tval}

   \pgfmathsetmacro{\newtval}{2*\tval-1}%
 \hlinequbitf[\staropf]{\ellpos+1-0.5}{2+2*\ttone}{5}
 \hlinequbitf[\staropf]{\ellpos+1-0.5}{1+2*\ttone}{5}
\btrianglequbitf[\staropf]{\ellpos+1-0.5}{3+2*\ttone}{2*\elllength-1}
\ttrianglequbitf[\staropf]{\ellpos+1-0.5}{2*\ttone-2}{2*\elllength-1}
\hlinequbitf[\staropf]{\ellpos+1-0.5}{2*\ttone-1}{5}
  \vlinequbitf[\staropf]{1}{1}{\newtval}
  \btrianglequbitf{1}{1}{6}
  \pgfmathsetmacro{\linelength}{2*\elllength-1}
  \hlinequbitf[\staropf]{\ellpos+1-0.5}{2*\ttone}{\linelength}
  \ttrianglequbitf{1}{\newtval}{5}
  \pgfmathsetmacro{\dlonger}{0.65}%
  \foreach \i in {0,...,\numexpr\totalL-1\relax}{%
    \pgfmathsetmacro{\xxx}{2*(\i)+2}%
    \qubitf{\xxx-\dlonger}{-\dlonger}%
    \qubitf{\xxx+\dlonger}{-\dlonger}%
  }%
  \pgfmathsetmacro{\myystart}{2*\tval-1+\dlonger}%
  \pgfmathtruncatemacro{\xfinalint}{2*\totalL}%
  \foreach\x in \xlistodd{%
    \ifnum\x=1\else
      \ifnum\x=\xfinalminus\else
        \pgfmathsetmacro{\dotX}{\x+1+2+\dlonger}%
        \qubitf{\dotX}{\myystart}%
      \fi
    \fi
  }%
  \foreach\x in \xlisteven{%
    \ifnum\x=\xfinalint\else
      \pgfmathsetmacro{\dotX}{\x+2-1-\dlonger}%
      \qubitf{\dotX}{\myystart}%
    \fi
  }%
  \pgfmathsetmacro{\dotX}{1+\dlonger}%
  \qubitf{\dotX}{\myystart}%
  \pgfmathsetmacro{\dotXright}{2*\totalL+1-\dlonger}%
  \qubitf{\dotXright}{\myystart}%
\end{tikzpicture} =
\begin{tikzpicture}[baseline={([yshift=-0.6ex]current bounding box.center)},scale=0.5]
  \pgfmathparse{8}\let\tval\pgfmathresult
  \pgfmathparse{3}\let\totalL\pgfmathresult
   \pgfmathparse{1.5}\let\ellpos\pgfmathresult
    \pgfmathparse{3}\let\elllength\pgfmathresult
    \pgfmathparse{4}\let\period\pgfmathresult
    \pgfmathparse{1*\period}\let\ttone\pgfmathresult

  \localcircuitNOTARGETnogatef{\tval}{\totalL}
  \pgfmathsetmacro{\dlonger}{0.65}%
  \pgfmathtruncatemacro{\yfinalminus}{2*\tval-2}%
  \pgfmathsetmacro{\dotXleft}{1-\dlonger}%
  \foreach \y in {0,2,...,\yfinalminus}{%
    \pgfmathsetmacro{\dotYdown}{\y+1-\dlonger}%
    \pgfmathsetmacro{\dotYup}{\y+1+\dlonger}%
    \ifnum\y=0\else
      \dotf{\dotXleft}{\dotYdown}%
    \fi
    \dotf{\dotXleft}{\dotYup}%
  }
  \pgfmathsetmacro{\qbitX}{1-\dlonger}%
  \pgfmathsetmacro{\qbitY}{1-\dlonger}%
    \dotf{1-\dlonger}{1-\dlonger}
\rightfpf[0]{2*\totalL}{1}{\tval}

\pgfmathtruncatemacro{\newtval}{2*\tval-1}%
\foreach \tt in {1,...,\numexpr\newtval\relax}{%
  \hlinequbitf[\staropf]{\ellpos-0.5}{\tt}{6}%
}%
  \btrianglequbitf{1}{1}{6}
  \pgfmathsetmacro{\linelength}{2*\elllength-1}
  \ttrianglequbitf{1}{\newtval}{5}
  \pgfmathsetmacro{\dlonger}{0.65}%
  \foreach \i in {0,...,\numexpr\totalL-1\relax}{%
    \pgfmathsetmacro{\xxx}{2*(\i)+2}%
    \qubitf{\xxx-\dlonger}{-\dlonger}%
    \qubitf{\xxx+\dlonger}{-\dlonger}%
  }%
  \pgfmathsetmacro{\myystart}{2*\tval-1+\dlonger}%
  \pgfmathtruncatemacro{\xfinalint}{2*\totalL}%
  \foreach\x in \xlistodd{%
    \ifnum\x=1\else
      \ifnum\x=\xfinalminus\else
        \pgfmathsetmacro{\dotX}{\x+1+2+\dlonger}%
        \qubitf{\dotX}{\myystart}%
      \fi
    \fi
  }%
  \foreach\x in \xlisteven{%
    \ifnum\x=\xfinalint\else
      \pgfmathsetmacro{\dotX}{\x+2-1-\dlonger}%
      \qubitf{\dotX}{\myystart}%
    \fi
  }%
  \pgfmathsetmacro{\dotX}{1+\dlonger}%
  \qubitf{\dotX}{\myystart}%
  \pgfmathsetmacro{\dotXright}{2*\totalL+1-\dlonger}%
  \qubitf{\dotXright}{\myystart}%
\end{tikzpicture} =
    \begin{tikzpicture}[baseline={([yshift=-0.6ex]current bounding box.center)},scale=0.5]
  \pgfmathparse{8}\let\tval\pgfmathresult
  \pgfmathparse{3}\let\totalL\pgfmathresult
   \pgfmathparse{1.5}\let\ellpos\pgfmathresult
    \pgfmathparse{3}\let\elllength\pgfmathresult
    \pgfmathparse{4}\let\period\pgfmathresult
    \pgfmathparse{1*\period}\let\ttone\pgfmathresult

  \localcircuitNOTARGETnogatef{\tval}{\totalL}
  \pgfmathsetmacro{\dlonger}{0.65}%
  \pgfmathtruncatemacro{\yfinalminus}{2*\tval-2}%
  \pgfmathsetmacro{\dotXleft}{1-\dlonger}%
  \foreach \y in {0,2,...,\yfinalminus}{%
    \pgfmathsetmacro{\dotYdown}{\y+1-\dlonger}%
    \pgfmathsetmacro{\dotYup}{\y+1+\dlonger}%
    \ifnum\y=0\else
      \dotf{\dotXleft}{\dotYdown}%
    \fi
    \dotf{\dotXleft}{\dotYup}%
  }
  \pgfmathsetmacro{\qbitX}{1-\dlonger}%
  \pgfmathsetmacro{\qbitY}{1-\dlonger}%
    \dotf{1-\dlonger}{1-\dlonger}
\rightfpnogatevf{2*\totalL}{1}{\tval}

\pgfmathtruncatemacro{\newtval}{2*\tval-1}%
\foreach \tt in {1,...,\numexpr\newtval\relax}{%
  \hlinequbitf[\staropf]{\ellpos-0.5}{\tt}{5}%
}%
  \btrianglequbitf{1}{1}{6}
  \pgfmathsetmacro{\linelength}{2*\elllength-1}
  \ttrianglequbitf{1}{\newtval}{5}
  \pgfmathsetmacro{\dlonger}{0.65}%
  \foreach \i in {0,...,\numexpr\totalL-1\relax}{%
    \pgfmathsetmacro{\xxx}{2*(\i)+2}%
    \qubitf{\xxx-\dlonger}{-\dlonger}%
    \qubitf{\xxx+\dlonger}{-\dlonger}%
  }%
  \pgfmathsetmacro{\myystart}{2*\tval-1+\dlonger}%
  \pgfmathtruncatemacro{\xfinalint}{2*\totalL}%
  \foreach\x in \xlistodd{%
    \ifnum\x=1\else
      \ifnum\x=\xfinalminus\else
        \pgfmathsetmacro{\dotX}{\x+1+2+\dlonger}%
        \qubitf{\dotX}{\myystart}%
      \fi
    \fi
  }%
  \foreach\x in \xlisteven{%
    \ifnum\x=\xfinalint\else
      \pgfmathsetmacro{\dotX}{\x+2-1-\dlonger}%
      \qubitf{\dotX}{\myystart}%
    \fi
  }%
  \pgfmathsetmacro{\dotX}{1+\dlonger}%
  \qubitf{\dotX}{\myystart}%
  \pgfmathsetmacro{\dotXright}{2*\totalL+1-\dlonger}%
  \qubitf{\dotXright}{\myystart}%
\end{tikzpicture}=
   \begin{tikzpicture}[baseline={([yshift=-0.6ex]current bounding box.center)},scale=0.5]
  \pgfmathparse{8}\let\tval\pgfmathresult
  \pgfmathparse{3}\let\totalL\pgfmathresult
   \pgfmathparse{1.5}\let\ellpos\pgfmathresult
    \pgfmathparse{3}\let\elllength\pgfmathresult
    \pgfmathparse{4}\let\period\pgfmathresult
    \pgfmathparse{1*\period}\let\ttone\pgfmathresult

  \localcircuitNOTARGETnogateatallf{\tval}{\totalL}
  \pgfmathsetmacro{\dlonger}{0.65}%
  \pgfmathtruncatemacro{\yfinalminus}{2*\tval-2}%
  \pgfmathsetmacro{\dotXleft}{1-\dlonger}%
  \foreach \y in {0,2,...,\yfinalminus}{%
    \pgfmathsetmacro{\dotYdown}{\y+1-\dlonger}%
    \pgfmathsetmacro{\dotYup}{\y+1+\dlonger}%
    \ifnum\y=0\else
      \dotf{\dotXleft}{\dotYdown}%
    \fi
    \dotf{\dotXleft}{\dotYup}%
  }
  \pgfmathsetmacro{\qbitX}{1-\dlonger}%
  \pgfmathsetmacro{\qbitY}{1-\dlonger}%
    \dotf{1-\dlonger}{1-\dlonger}

  \foreach \x in {4,6}{%
    \pgfmathsetmacro{\xshift}{\x-1}%
    \extensionf{\xshift}{2*\tval-1}\gatef{\xshift}{2*\tval-1}%
  }
  \dotf{5.65}{2*\tval-1.65} 
  \foreach \x in {0,2}{%
    \pgfmathsetmacro{\xshift}{\x+2}%
    \extensionf{\xshift}{2*\tval-2}\gatef{\xshift}{2*\tval-2}%
  }%
  \dotf{4.65}{2*\tval-2.65} 
  \foreach \x in {4}{%
    \pgfmathsetmacro{\xshift}{\x-1}%
    \extensionf{\xshift}{2*\tval-3}\gatef{\xshift}{2*\tval-3}%
  }
  \foreach \x in {0}{%
    \pgfmathsetmacro{\xshift}{\x+2}%
    \extensionf{\xshift}{2*\tval-4}\gatef{\xshift}{2*\tval-4}%
  }%
  \dotf{3.65}{2*\tval-3.65} 
  \dotf{2.65}{2*\tval-4.65} 
  \dotf{1.65}{2*\tval-5.65} 

  \nogaternolegsf[\staropf]{1}{7}
  \dotf{1.65}{7.65}
  \dotf{1.65}{7.0-0.65}
  
  \nogaternolegsf[\staropf]{1}{9}
  \dotf{1.65}{9.65}
  \dotf{1.65}{9.0-0.65}
  
  \foreach \x in {0,2,4}{%
    \pgfmathsetmacro{\xshift}{\x+2}%
    \extensionf{\xshift}{0}\gatef{\xshift}{0}%
  }%
  \dotf{6.65}{0.65} 
  \foreach \x in {4,6}{%
    \pgfmathsetmacro{\xshift}{\x-1}%
    \extensionf{\xshift}{1}\gatef{\xshift}{1}%
  }%
  \dotf{5.65}{1.65} 
  \foreach \x in {0,2}{%
    \pgfmathsetmacro{\xshift}{\x+2}%
    \extensionf{\xshift}{2}\gatef{\xshift}{2}%
  }%
  \dotf{4.65}{2.65} 
  \foreach \x in {4}{%
    \pgfmathsetmacro{\xshift}{\x-1}%
    \extensionf{\xshift}{3}\gatef{\xshift}{3}%
  }%
  \dotf{3.65}{3.65} 

  \foreach \x in {0}{%
    \pgfmathsetmacro{\xshift}{\x+2}%
    \extensionf{\xshift}{4}\gatef{\xshift}{4}%
  }%
  \dotf{2.65}{4.65} 
  \dotf{1.65}{5.65} 

\pgfmathtruncatemacro{\newtval}{2*\tval-1}%
  \btrianglequbitf[\staropf]{1}{1}{6}
  \btrianglequbitf{1}{1}{6}
  \pgfmathsetmacro{\linelength}{2*\elllength-1}
  \ttrianglequbitf[\staropf]{1}{\newtval}{5}
  \ttrianglequbitf{1}{\newtval}{5}
  \pgfmathsetmacro{\dlonger}{0.65}%
  \foreach \i in {0,...,\numexpr\totalL-1\relax}{%
    \pgfmathsetmacro{\xxx}{2*(\i)+2}%
    \qubitf{\xxx-\dlonger}{-\dlonger}%
    \qubitf{\xxx+\dlonger}{-\dlonger}%
  }%
  \pgfmathsetmacro{\myystart}{2*\tval-1+\dlonger}%
  \pgfmathtruncatemacro{\xfinalint}{2*\totalL}%
  \foreach\x in \xlistodd{%
    \ifnum\x=1\else
      \ifnum\x=\xfinalminus\else
        \pgfmathsetmacro{\dotX}{\x+1+2+\dlonger}%
        \qubitf{\dotX}{\myystart}%
      \fi
    \fi
  }%
  \foreach\x in \xlisteven{%
    \ifnum\x=\xfinalint\else
      \pgfmathsetmacro{\dotX}{\x+2-1-\dlonger}%
      \qubitf{\dotX}{\myystart}%
    \fi
  }%
  \pgfmathsetmacro{\dotX}{1+\dlonger}%
  \qubitf{\dotX}{\myystart}%
  \pgfmathsetmacro{\dotXright}{2*\totalL+1-\dlonger}%
\nogatelvf{\dotXright+0.5}{\myystart-0.5}
\qubitf{\dotXright}{\myystart}%
\pgfmathsetmacro{\dotXtopright}{\dotXright+1}%
\dotf{\dotXtopright}{\myystart}%
\end{tikzpicture}
\end{equation}
In the equation above, we use  Eq.\ttilde\eqref{eq:staropfgatefqubitf} to propagate the operators in  $ \begin{tikzpicture}[baseline={([yshift=-0.6ex]current bounding box.center)},scale=0.35]
    \extensionf{0}{0}\gatef{0}{0}
    \qubitf{-0.5}{0.5} 
    \staropf{-0.5}{-0.5} 
  \end{tikzpicture}$ and $ \begin{tikzpicture}[baseline={([yshift=-0.6ex]current bounding box.center)},scale=0.35]
    \extensionf{0}{0}\gatef{0}{0}
    \staropf{-0.5}{0.5} 
    \qubitf{-0.5}{-0.5} 
  \end{tikzpicture}$ diagonally to the right. Iterating this step yields the first equality. In the second equality, we use the last relation in Eq.\ttilde\eqref{eq:staropfgatef} to contract the rightmost column of the circuit; iterating this last step in turn gives the last diagram. 
  As in the previous case, using the identities in Eqs.\ttilde\eqref{eq:fs} and\ttilde\eqref{eq:normstatezero}--\eqref{eq:greencirclegate} we can finally compute the corresponding Loschmidt echo for $m\tau > 2\ell$ as:
\begin{equation}
\label{eq:Lmtl}
 \mathcal{L}(m) = 2^{-2m\tau}
        \begin{tikzpicture}[baseline={([yshift=-0.6ex]current bounding box.center)},scale=0.5]
  \pgfmathparse{8}\let\tval\pgfmathresult
  \pgfmathparse{3}\let\totalL\pgfmathresult
   \pgfmathparse{1.5}\let\ellpos\pgfmathresult
    \pgfmathparse{3}\let\elllength\pgfmathresult
    \pgfmathparse{4}\let\period\pgfmathresult
    \pgfmathparse{1*\period}\let\ttone\pgfmathresult

  \localcircuitNOTARGETnogateatallf{\tval}{\totalL}
  \pgfmathsetmacro{\dlonger}{0.65}%
  \pgfmathtruncatemacro{\yfinalminus}{2*\tval-2}%
  \pgfmathsetmacro{\dotXleft}{1-\dlonger}%
  \foreach \y in {0,2,...,\yfinalminus}{%
    \pgfmathsetmacro{\dotYdown}{\y+1-\dlonger}%
    \pgfmathsetmacro{\dotYup}{\y+1+\dlonger}%
    \ifnum\y=0\else
      \dotf{\dotXleft}{\dotYdown}%
    \fi
    \dotf{\dotXleft}{\dotYup}%
  }
  \pgfmathsetmacro{\qbitX}{1-\dlonger}%
  \pgfmathsetmacro{\qbitY}{1-\dlonger}%
    \dotf{1-\dlonger}{1-\dlonger}

  \foreach \x in {4,6}{%
    \pgfmathsetmacro{\xshift}{\x-1}%
    \extensionf{\xshift}{2*\tval-1}\nogatef{\xshift}{2*\tval-1}%
  }
  \qubitf{5.65}{2*\tval-1.65} 
  \foreach \x in {0,2}{%
    \pgfmathsetmacro{\xshift}{\x+2}%
    \extensionf{\xshift}{2*\tval-2}\nogatef{\xshift}{2*\tval-2}%
  }%
  \qubitf{4.65}{2*\tval-2.65} 
  \foreach \x in {4}{%
    \pgfmathsetmacro{\xshift}{\x-1}%
    \extensionf{\xshift}{2*\tval-3}\nogatef{\xshift}{2*\tval-3}%
  }
  \foreach \x in {0}{%
    \pgfmathsetmacro{\xshift}{\x+2}%
    \extensionf{\xshift}{2*\tval-4}\nogatef{\xshift}{2*\tval-4}%
  }%
  \qubitf{3.65}{2*\tval-3.65} 
  \qubitf{2.65}{2*\tval-4.65} 
  \qubitf{1.65}{2*\tval-5.65} 

  \nogaternolegsf{1}{7}
  \dotf{1.65}{7.65}
  \dotf{1.65}{7.0-0.65}
  
  \nogaternolegsf{1}{9}
  \dotf{1.65}{9.65}
  \dotf{1.65}{9.0-0.65}
  
  \foreach \x in {0,2,4}{%
    \pgfmathsetmacro{\xshift}{\x+2}%
    \extensionf{\xshift}{0}\nogatef{\xshift}{0}%
  }%
  \qubitf{6.65}{0.65} 
  \foreach \x in {4,6}{%
    \pgfmathsetmacro{\xshift}{\x-1}%
    \extensionf{\xshift}{1}\nogatef{\xshift}{1}%
  }%
  \qubitf{5.65}{1.65} 
  \foreach \x in {0,2}{%
    \pgfmathsetmacro{\xshift}{\x+2}%
    \extensionf{\xshift}{2}\nogatef{\xshift}{2}%
  }%
  \qubitf{4.65}{2.65} 
  \foreach \x in {4}{%
    \pgfmathsetmacro{\xshift}{\x-1}%
    \extensionf{\xshift}{3}\nogatef{\xshift}{3}%
  }%
  \qubitf{3.65}{3.65} 

  \foreach \x in {0}{%
    \pgfmathsetmacro{\xshift}{\x+2}%
    \extensionf{\xshift}{4}\nogatef{\xshift}{4}%
  }%
  \qubitf{2.65}{4.65} 
  \qubitf{1.65}{5.65} 

\pgfmathtruncatemacro{\newtval}{2*\tval-1}%
  \pgfmathsetmacro{\linelength}{2*\elllength-1}
  \pgfmathsetmacro{\dlonger}{0.65}%
  \foreach \i in {0,...,\numexpr\totalL-1\relax}{%
    \pgfmathsetmacro{\xxx}{2*(\i)+2}%
    \qubitf{\xxx-\dlonger}{-\dlonger}%
    \qubitf{\xxx+\dlonger}{-\dlonger}%
  }%
  \pgfmathsetmacro{\myystart}{2*\tval-1+\dlonger}%
  \pgfmathtruncatemacro{\xfinalint}{2*\totalL}%
  \foreach\x in \xlistodd{%
    \ifnum\x=1\else
      \ifnum\x=\xfinalminus\else
        \pgfmathsetmacro{\dotX}{\x+1+2+\dlonger}%
        \qubitf{\dotX}{\myystart}%
      \fi
    \fi
  }%
  \foreach\x in \xlisteven{%
    \ifnum\x=\xfinalint\else
      \pgfmathsetmacro{\dotX}{\x+2-1-\dlonger}%
      \qubitf{\dotX}{\myystart}%
    \fi
  }%
  \pgfmathsetmacro{\dotX}{1+\dlonger}%
  \qubitf{\dotX}{\myystart}%
  \pgfmathsetmacro{\dotXright}{2*\totalL+1-\dlonger}%
\nogatelvf{\dotXright+0.5}{\myystart-0.5}
\qubitf{\dotXright}{\myystart}%
\pgfmathsetmacro{\dotXtopright}{\dotXright+1}%
\dotf{\dotXtopright}{\myystart}%
  \pgfmathsetmacro{\dlonger}{0.65}%
  \pgfmathsetmacro{\myystart}{2*\tval-1+\dlonger}%
  \pgfmathsetmacro{\dxarrow}{1.5}%
  \pgfmathsetmacro{\arrowX}{2*\totalL+1+\dxarrow}%
  \pgfmathsetmacro{\arrowYbot}{-\dlonger}%
  \pgfmathsetmacro{\arrowYtop}{5+2*\dlonger}
  \pgfmathsetmacro{\arrowYmid}{0.5*(\arrowYbot+\arrowYtop)}%
  \draw[<->, thick] (\arrowX, \arrowYbot) -- (\arrowX, \arrowYtop);
  \node[anchor=south, rotate=90] at (\arrowX-0.1, \arrowYmid) {$\ell$};
  \pgfmathsetmacro{\arrowYbot}{5+2*\dlonger}%
  \pgfmathsetmacro{\arrowYtop}{\arrowYbot+4-\dlonger}
  \pgfmathsetmacro{\arrowYmid}{0.5*(\arrowYbot+\arrowYtop)}%
  \draw[<->, thick] (\arrowX, \arrowYbot) -- (\arrowX, \arrowYtop);
  \node[anchor=south, rotate=90] at (\arrowX-0.1, \arrowYmid) {$m\tau-2\ell$};
  \pgfmathsetmacro{\arrowYbot}{\arrowYbot+4-\dlonger}
  \pgfmathsetmacro{\arrowYtop}{\myystart}
  \pgfmathsetmacro{\arrowYmid}{0.5*(\arrowYbot+\arrowYtop)}%
  \draw[<->, thick] (\arrowX, \arrowYbot) -- (\arrowX, \arrowYtop);
  \node[anchor=south, rotate=90] at (\arrowX-0.1, \arrowYmid) {$\ell$};
\end{tikzpicture}=
2^{-2m\tau} \left( \; \begin{tikzpicture}[baseline={([yshift=-0.6ex]current bounding box.center)},scale=0.5]
    \newcommand{\dist}{0}
      \SinglePropf{0}{0}{2.0*\hprop+\dist}
      \pgfmathsetmacro{\Ypos}{(0.5)*(2*\hprop+\dist)}
     \dotf{0}{2*\Ypos}
     \dotf{0}{0}
      \end{tikzpicture}\; \right)^{2m\tau-2\ell} = 2^{-2\ell}\,.
\end{equation}
Therefore, combining the results in Eqs.\ttilde\eqref{eq:Lmts} and\ttilde\eqref{eq:Lmtl}, we reproduce
Eq.~(7). We finally note that the two branches match at $m\tau = 2\ell$, where both give $\mathcal{L}(m) = 2^{-2\ell}$.

\section{Subsystem Loschmidt echo decomposition}
In this section we present the derivation of an expansion of the first-detection probability in terms of the Loschmidt echo computed in the previous section. We begin, in Subsec.~\ref{subsec:IVA}, with the simplest case, detection at $k=1$, for which the expansion reduces to a single Loschmidt echo. The derivation for general $k$ follows in Subsec.~\ref{subsec:IVB}, and yields Eq.~(6) of main text. 
\subsection{Success at the first attempt}
\label{subsec:IVA}
For 
$k=1$, no unsuccessful measurement outcome occurs, so that the first-detection probability in Eq.~\eqref{eq:fdpfolded} involves only projectors onto the target state. We thus begin by observing that, for arbitrary total observation time $m\tau$, the following diagram
also reproduces the Loschmidt echo:
\begin{equation}
\label{eq:Lamplitudebottom}
 2^{-2m\tau+1}
    \begin{tikzpicture}[baseline={([yshift=-0.6ex]current bounding box.center)},scale=0.5]
  \pgfmathparse{4}\let\tval\pgfmathresult
  \pgfmathparse{3}\let\totalL\pgfmathresult
   \pgfmathparse{1.5}\let\ellpos\pgfmathresult
    \pgfmathparse{3}\let\elllength\pgfmathresult
    \pgfmathparse{2}\let\period\pgfmathresult
    \pgfmathparse{1*\period}\let\ttone\pgfmathresult

  \localcircuitf[0]{\tval}{\totalL}
  \pgfmathsetmacro{\dlonger}{0.65}%
  \pgfmathtruncatemacro{\yfinalminus}{2*\tval-2}%
  \pgfmathsetmacro{\dotXleft}{1-\dlonger}%
  \foreach \y in {0,2,...,\yfinalminus}{%
    \pgfmathsetmacro{\dotYdown}{\y+1-\dlonger}%
    \pgfmathsetmacro{\dotYup}{\y+1+\dlonger}%
    \ifnum\y=0\else
      \dotf{\dotXleft}{\dotYdown}%
    \fi
    \dotf{\dotXleft}{\dotYup}%
  }
  \pgfmathsetmacro{\qbitX}{1-\dlonger}%
  \pgfmathsetmacro{\qbitY}{1-\dlonger}%
  \qubittwodf{\qbitX+0.15}{\qbitY+0.15-1}
\rightfpf[0]{2*\totalL}{1}{\tval}
  \pgfmathsetmacro{\myystart}{2*\tval-1+\dlonger}%
  \pgfmathtruncatemacro{\xfinalint}{2*\totalL}%
  \foreach\x in \xlistodd{%
    \ifnum\x=1\else
      \ifnum\x=\xfinalminus\else
        \pgfmathsetmacro{\dotX}{\x+1+2+\dlonger}%
        \qubitf{\dotX}{\myystart}%
      \fi
    \fi
  }%
  \foreach\x in \xlisteven{%
    \ifnum\x=\xfinalint\else
      \pgfmathsetmacro{\dotX}{\x+2-1-\dlonger}%
      \qubitf{\dotX}{\myystart}%
    \fi
  }%
  \pgfmathsetmacro{\dotX}{1+\dlonger}%
  \qubitf{\dotX}{\myystart}%
  \pgfmathsetmacro{\dotXright}{2*\totalL+1-\dlonger}%
  \qubitf{\dotXright}{\myystart}%
  \nogaterf{0}{0}
  

  \pgfmathsetmacro{\dlonger}{0.65}%
  \pgfmathsetmacro{\myystart}{2*\tval-1+\dlonger}%
  \pgfmathsetmacro{\dxarrow}{1.5}%
  \pgfmathsetmacro{\arrowX}{2*\totalL+1+\dxarrow}%
  \pgfmathsetmacro{\arrowYbot}{-\dlonger}%
  \pgfmathsetmacro{\arrowYtop}{\myystart}%
  \pgfmathsetmacro{\arrowYmid}{0.5*(\arrowYbot+\arrowYtop)}%
  \draw[<->, thick] (\arrowX, \arrowYbot) -- (\arrowX, \arrowYtop);
  \node[anchor=west] at (\arrowX+0.2, \arrowYmid) {$m\tau$};
\end{tikzpicture}= \mathcal{L}(m)
\,.
\end{equation}
The diagram above differs from that used in the definition of the Loschmidt echo in Eq.\ttilde\eqref{eq:Amplitudecircuit} by the folded-space state defined in Eq.\ttilde\eqref{eq:qubittwod}, located at the bottom left. This state has unit norm, rather than norm $2$ as for the state defined in Eq.\ttilde\eqref{eq:fs}, however, the additional factor of $2$ in $2^{-2m\tau+1}$ in the equation above compensates for this difference, giving exactly the same result. This can be seen, for instance, by considering $m\tau <2\ell$:
\begin{equation}
2^{-2m\tau+1}
    \begin{tikzpicture}[baseline={([yshift=-0.6ex]current bounding box.center)},scale=0.5]
  \pgfmathparse{4}\let\tval\pgfmathresult
  \pgfmathparse{3}\let\totalL\pgfmathresult
   \pgfmathparse{1.5}\let\ellpos\pgfmathresult
    \pgfmathparse{3}\let\elllength\pgfmathresult
    \pgfmathparse{2}\let\period\pgfmathresult
  \localcircuitnogatef[0]{\tval}{\totalL}
  \pgfmathsetmacro{\dlonger}{0.65}%
  \pgfmathtruncatemacro{\yfinalminus}{2*\tval-2}%
  \pgfmathsetmacro{\dotXleft}{1-\dlonger}%
  \foreach \y in {0,2,...,\yfinalminus}{%
    \pgfmathsetmacro{\dotYdown}{\y+1-\dlonger}%
    \pgfmathsetmacro{\dotYup}{\y+1+\dlonger}%
    \ifnum\y=0\else
      \dotf{\dotXleft}{\dotYdown}%
    \fi
    \dotf{\dotXleft}{\dotYup}%
  }
  \pgfmathsetmacro{\qbitX}{1-\dlonger}%
  \pgfmathsetmacro{\qbitY}{1-\dlonger}%
    \nogaterf{0}{0}
  \qubittwodf{\qbitX+0.15}{\qbitY+0.15-1}
\rightfpnogatef[\qubitf]{2*\totalL}{1}{\tval}
  \pgfmathsetmacro{\myystart}{2*\tval-1+\dlonger}%
  \pgfmathtruncatemacro{\xfinalint}{2*\totalL}%
  \foreach\x in \xlistodd{%
    \ifnum\x=1\else
      \ifnum\x=\xfinalminus\else
        \pgfmathsetmacro{\dotX}{\x+1+2+\dlonger}%
        \qubitf{\dotX}{\myystart}%
      \fi
    \fi
  }%
  \foreach\x in \xlisteven{%
    \ifnum\x=\xfinalint\else
      \pgfmathsetmacro{\dotX}{\x+2-1-\dlonger}%
      \qubitf{\dotX}{\myystart}%
    \fi
  }%
  \pgfmathsetmacro{\dotX}{1+\dlonger}%
  \qubitf{\dotX}{\myystart}%
  \pgfmathsetmacro{\dotXright}{2*\totalL+1-\dlonger}%
  \qubitf{\dotXright}{\myystart}%
  \pgfmathsetmacro{\dotX}{1+\dlonger}%
  \qubitf{\dotX}{\myystart}%
\end{tikzpicture} =  2^{-2m\tau+1} \left( \; \begin{tikzpicture}[baseline={([yshift=-0.6ex]current bounding box.center)},scale=0.5]
    \newcommand{\dist}{0}
      \SinglePropf{0}{0}{2.0*\hprop+\dist}
      \pgfmathsetmacro{\Ypos}{(0.5)*(2*\hprop+\dist)}
     \dotf{0}{2*\Ypos}
     \dotf{0}{0}
      \end{tikzpicture}\; \right)^{m\tau-1} =2^{-m\tau}\,.
\end{equation} 
The equalities follow from the same arguments used in Section~\ref{sec:compLamplitude} to obtain Eqs.~\eqref{eq:Lmts} and~\eqref{eq:Lmtl}. Similar arguments show that the quantity in Eq.~\eqref{eq:Lamplitudebottom} gives $2^{-2\ell}$ for $m\tau>2\ell$.

Turning more specifically to the computation of 
 $F(1)$, Eq.\ttilde\eqref{eq:Lamplitudebottom} implies that
\begin{equation}
\label{eq:F1}
    F(1)=\mathcal{L}(1) = \begin{cases}
2^{-\tau}, & \tau \leq 2\ell, \\
2^{-2\ell}, & \tau > 2\ell\,.
\end{cases}
\end{equation}
This result is consistent with the expansion in Eq.~(6) evaluated at $k=1$. Moreover, as anticipated in the main text, $F(1)$ is the largest among the first-detection probabilities, so that detection is most likely to occur at the first attempt. We derive the expansion for general $k$ in the next subsection.

\subsection{Subsystem Loschmidt echo expansion}
\label{subsec:IVB}
For a number of attempts $k>1$, the first-detection probability 
involves unsuccessful outcomes as well, which we need to take into account in the computation. In particular, the expansion in Eq.~(6) is obtained by decomposing each unsuccessful measurement outcome appearing in the diagram of Eq.\ttilde\eqref{eq:localFDP} in terms of projectors onto the zero state. Each decomposition yields four terms
\begin{align}
\label{eq:projno}
    & \begin{tikzpicture}[baseline={([yshift=-0.6ex]current bounding box.center)},scale=0.5]
  \pgfmathparse{0.5}\let\dl\pgfmathresult
  \pgfmathsetmacro{\xstep}{2.0}
  \pgfmathsetmacro{\sep}{0.4}
  \foreach \k in {0,1,2}{
    \pgfmathsetmacro{\xoffset}{\k*(\xstep+\sep)}
    \begin{scope}[shift={(\xoffset,0)}]
      \gridLinef{(0.0)-\dl}{-\dl}{(0.0)}{(0.0)}
      \begin{scope}[shift={(0.5,0)}]
        \gridLinef{(0.0)}{(0.0)}{+\dl}{(0.0)-\dl} 
      \end{scope}
    \end{scope}
  }
  \pgfmathsetmacro{\elllength}{3*(2*\dl+0.5)+(3-1)*\sep-2}
  \getNof{1.5*\dl}{0.25*\dl}{\elllength}
\end{tikzpicture}= \nonumber \\
&\begin{tikzpicture}[baseline={([yshift=-0.6ex]current bounding box.center)},scale=0.5]
  \pgfmathparse{0.5}\let\dl\pgfmathresult
  \pgfmathsetmacro{\xstep}{2.0}
  \pgfmathsetmacro{\sep}{0.4}
  \foreach \k in {0,1,2}{
    \pgfmathsetmacro{\xoffset}{\k*(\xstep+\sep)}
    \begin{scope}[shift={(\xoffset,0)}]
      \gridLinef{(0.0)-\dl}{-\dl}{(0.0)}{(0.0)}
      \begin{scope}[shift={(0.5,0)}]
        \gridLinef{(0.0)}{(0.0)}{+\dl}{(0.0)-\dl} 
      \end{scope}
    \end{scope}
  }
\end{tikzpicture} 
 \quad\!\!+\quad\!\!
\begin{tikzpicture}[baseline={([yshift=-0.6ex]current bounding box.center)},scale=0.5]
  \pgfmathparse{0.5}\let\dl\pgfmathresult
  \pgfmathsetmacro{\xstep}{2.0}
  \pgfmathsetmacro{\sep}{0.4}
  \foreach \k in {0,1,2}{
    \pgfmathsetmacro{\xoffset}{\k*(\xstep+\sep)}
    \begin{scope}[shift={(\xoffset,0)}]
      \gridLinef{(0.0)-\dl}{-\dl}{(0.0)}{(0.0)}
     \qubitf{0.0-0.5*\dl}{0.0-0.5*\dl} 
      \begin{scope}[shift={(0.5,0)}]
        \gridLinef{(0.0)}{(0.0)}{+\dl}{(0.0)-\dl} 
        \qubitf{0.0+0.5*\dl}{0.0-0.5*\dl} 
      \end{scope}
    \end{scope}
  }
\end{tikzpicture} \quad\!\!-\quad\!\!
\begin{tikzpicture}[baseline={([yshift=-0.6ex]current bounding box.center)},scale=0.5]
  \pgfmathparse{0.5}\let\dl\pgfmathresult
  \pgfmathsetmacro{\xstep}{2.0}
  \pgfmathsetmacro{\sep}{0.4}
  \foreach \k in {0,1,2}{
    \pgfmathsetmacro{\xoffset}{\k*(\xstep+\sep)}
    \begin{scope}[shift={(\xoffset,0)}]
      \gridLinef{(0.0)-\dl}{-\dl}{(0.0)}{(0.0)}
     \frontqubitf{0.0-0.5*\dl}{0.0-0.5*\dl} 
      \begin{scope}[shift={(0.5,0)}]
        \gridLinef{(0.0)}{(0.0)}{+\dl}{(0.0)-\dl} 
        \frontqubitf{0.0+0.5*\dl}{0.0-0.5*\dl} 
      \end{scope}
    \end{scope}
  }
\end{tikzpicture} \quad\!\!-\quad\!\!
\begin{tikzpicture}[baseline={([yshift=-0.6ex]current bounding box.center)},scale=0.5]
  \pgfmathparse{0.5}\let\dl\pgfmathresult
  \pgfmathsetmacro{\xstep}{2.0}
  \pgfmathsetmacro{\sep}{0.4}
  \foreach \k in {0,1,2}{
    \pgfmathsetmacro{\xoffset}{\k*(\xstep+\sep)}
    \begin{scope}[shift={(\xoffset,0)}]
      \gridLinef{(0.0)-\dl}{-\dl}{(0.0)}{(0.0)}
     \backqubitf{0.0-0.5*\dl}{0.0-0.5*\dl} 
      \begin{scope}[shift={(0.5,0)}]
        \gridLinef{(0.0)}{(0.0)}{+\dl}{(0.0)-\dl} 
        \backqubitf{0.0+0.5*\dl}{0.0-0.5*\dl} 
      \end{scope}
    \end{scope}
  }
\end{tikzpicture} \,,
\end{align}
where the projectors $\projectorf[\frontqubitf]$ and $\projectorf[\backqubitf]$  in the last two terms act only on a single layer of the folded space and are defined in Eq.\ttilde\eqref{eq:onelayer}.
However, under the effect of the local dynamics (encoded in the operators $\projectorf[\staropf]$) or upon the action of the projectors $\projectorf$, these single-layer terms are forced to extend to both layers of the folded space, which is what ultimately produces Eq.~(6) of the main text. We now demonstrate this property in detail, for any arbitrary period $\tau$, length of the monitored region $\ell$ and number of attempts $k$. \\

We first consider  $k \tau > 2 \ell$, for which the diagram $D(k)= 2^{2k\tau-1} F(k)$ entering Eq.\ttilde\eqref{eq:localFDP} reads 
\begin{equation}
\label{eq:Dklarget}
D(k)= 
        \begin{tikzpicture}[baseline={([yshift=-0.6ex]current bounding box.center)},scale=0.5]
        \pgfmathparse{2}\let\period\pgfmathresult
    \pgfmathparse{3*\period}\let\ttthree\pgfmathresult
    \pgfmathparse{1*\period}\let\ttone\pgfmathresult
    \pgfmathparse{2*\period}\let\tttwo\pgfmathresult
  \pgfmathparse{8}\let\tval\pgfmathresult
  \pgfmathparse{3}\let\totalL\pgfmathresult
   \pgfmathparse{1.5}\let\ellpos\pgfmathresult
    \pgfmathparse{3}\let\elllength\pgfmathresult
    \pgfmathparse{2}\let\period\pgfmathresult

  \pgfmathsetmacro{\dlonger}{0.65}%
  \pgfmathsetmacro{\myystart}{2*\tval-1+\dlonger}%
  \pgfmathsetmacro{\dxarrow}{0.1}%
  \pgfmathsetmacro{\arrowX}{0 - \dxarrow}%
  \pgfmathsetmacro{\arrowYbot}{-\dlonger}%
  \pgfmathsetmacro{\arrowYtop}{\myystart}%
  \pgfmathsetmacro{\arrowYmid}{0.5*(\arrowYbot+\arrowYtop)}%
  \draw[<->, thick] (\arrowX, \arrowYbot) -- (\arrowX, \arrowYtop);
  \node[anchor=south, rotate=90] at (\arrowX-0.0, \arrowYmid) {$k\tau > 2\ell$};
  \localcircuitf[0]{\tval}{\totalL}
  \pgfmathsetmacro{\dlonger}{0.65}%
  \pgfmathtruncatemacro{\yfinalminus}{2*\tval-2}%
  \pgfmathsetmacro{\dotXleft}{1-\dlonger}%
  \foreach \y in {0,2,...,\yfinalminus}{%
    \pgfmathsetmacro{\dotYdown}{\y+1-\dlonger}%
    \pgfmathsetmacro{\dotYup}{\y+1+\dlonger}%
    \ifnum\y=0\else
      \dotf{\dotXleft}{\dotYdown}%
    \fi
    \dotf{\dotXleft}{\dotYup}%
  }
  \pgfmathsetmacro{\qbitX}{1-\dlonger}%
  \pgfmathsetmacro{\qbitY}{1-\dlonger}%
  \qubittwodf{\qbitX+0.15}{\qbitY+0.15-1}
  \nogaterf{0}{0}
\rightfpf[0]{2*\totalL}{1}{\tval}
  \pgfmathsetmacro{\myystart}{2*\tval-1+\dlonger}%
  \pgfmathtruncatemacro{\xfinalint}{2*\totalL}%
  \foreach\x in \xlistodd{%
    \ifnum\x=1\else
      \ifnum\x=\xfinalminus\else
        \pgfmathsetmacro{\dotX}{\x+1+2+\dlonger}%
        \qubitf{\dotX}{\myystart}%
      \fi
    \fi
  }%
  \foreach\x in \xlisteven{%
    \ifnum\x=\xfinalint\else
      \pgfmathsetmacro{\dotX}{\x+2-1-\dlonger}%
      \qubitf{\dotX}{\myystart}%
    \fi
  }%
  \pgfmathsetmacro{\dotX}{1+\dlonger}%
  \qubitf{\dotX}{\myystart}%
  \pgfmathsetmacro{\dotXright}{2*\totalL+1-\dlonger}%
  \qubitf{\dotXright}{\myystart}%
  \pgfmathsetmacro{\dotX}{1+\dlonger}%
  \qubitf{\dotX}{\myystart}%
  \getNof{\ellpos}{\ttone}{\elllength}
    \getNof{\ellpos}{\tttwo}{\elllength}
     \getNof{\ellpos}{\ttthree}{\elllength}
\end{tikzpicture}=  
\begin{tikzpicture}[baseline={([yshift=-0.6ex]current bounding box.center)},scale=0.5]
  \pgfmathparse{8}\let\tval\pgfmathresult
  \pgfmathparse{3}\let\totalL\pgfmathresult
   \pgfmathparse{1.5}\let\ellpos\pgfmathresult
    \pgfmathparse{3}\let\elllength\pgfmathresult
    \pgfmathparse{4}\let\period\pgfmathresult
    \pgfmathparse{1*\period}\let\ttone\pgfmathresult

  \localcircuitNOTARGETnogatef{\tval}{\totalL}
  \pgfmathsetmacro{\dlonger}{0.65}%
  \pgfmathtruncatemacro{\yfinalminus}{2*\tval-2}%
  \pgfmathsetmacro{\dotXleft}{1-\dlonger}%
  \foreach \y in {0,2,...,\yfinalminus}{%
    \pgfmathsetmacro{\dotYdown}{\y+1-\dlonger}%
    \pgfmathsetmacro{\dotYup}{\y+1+\dlonger}%
    \ifnum\y=0\else
      \dotf{\dotXleft}{\dotYdown}%
    \fi
    \dotf{\dotXleft}{\dotYup}%
  }
  \pgfmathsetmacro{\qbitX}{1-\dlonger}%
  \pgfmathsetmacro{\qbitY}{1-\dlonger}%
  \qubittwodf{\qbitX+0.15}{\qbitY+0.15-1}
  \nogaterf{0}{0}
\rightfpf[0]{2*\totalL}{1}{\tval}

\pgfmathparse{2}\let\period\pgfmathresult
\pgfmathparse{3*\period}\let\ttthree\pgfmathresult
\pgfmathparse{1*\period}\let\ttone\pgfmathresult
\pgfmathparse{2*\period}\let\tttwo\pgfmathresult
  \getNof{\ellpos}{\ttone}{\elllength}
    \getNof{\ellpos}{\tttwo}{\elllength}
     \getNof{\ellpos}{\ttthree}{\elllength}
\pgfmathtruncatemacro{\newtval}{2*\tval-1}%
\foreach \tt in {1,...,\numexpr\newtval\relax}{%
  \hlinequbitf[\staropf]{\ellpos-0.5}{\tt}{6}%
}%
  \btrianglequbitf{1}{1}{6}
  \pgfmathsetmacro{\linelength}{2*\elllength-1}
  \ttrianglequbitf{1}{\newtval}{5}
  \pgfmathsetmacro{\dlonger}{0.65}%
  \foreach \i in {0,...,\numexpr\totalL-1\relax}{%
    \pgfmathsetmacro{\xxx}{2*(\i)+2}%
    \qubitf{\xxx-\dlonger}{-\dlonger}%
    \qubitf{\xxx+\dlonger}{-\dlonger}%
  }%
  \pgfmathsetmacro{\myystart}{2*\tval-1+\dlonger}%
  \pgfmathtruncatemacro{\xfinalint}{2*\totalL}%
  \foreach\x in \xlistodd{%
    \ifnum\x=1\else
      \ifnum\x=\xfinalminus\else
        \pgfmathsetmacro{\dotX}{\x+1+2+\dlonger}%
        \qubitf{\dotX}{\myystart}%
      \fi
    \fi
  }%
  \foreach\x in \xlisteven{%
    \ifnum\x=\xfinalint\else
      \pgfmathsetmacro{\dotX}{\x+2-1-\dlonger}%
      \qubitf{\dotX}{\myystart}%
    \fi
  }%
  \pgfmathsetmacro{\dotX}{1+\dlonger}%
  \qubitf{\dotX}{\myystart}%
  \pgfmathsetmacro{\dotXright}{2*\totalL+1-\dlonger}%
  \qubitf{\dotXright}{\myystart}%
\end{tikzpicture}=  
       \begin{tikzpicture}[baseline={([yshift=-0.6ex]current bounding box.center)},scale=0.5]
        \pgfmathparse{2}\let\period\pgfmathresult
    \pgfmathparse{3*\period}\let\ttthree\pgfmathresult
    \pgfmathparse{1*\period}\let\ttone\pgfmathresult
    \pgfmathparse{2*\period}\let\tttwo\pgfmathresult
  \pgfmathparse{8}\let\tval\pgfmathresult
  \pgfmathparse{3}\let\totalL\pgfmathresult
   \pgfmathparse{1.5}\let\ellpos\pgfmathresult
    \pgfmathparse{3}\let\elllength\pgfmathresult
    \pgfmathparse{2}\let\period\pgfmathresult
  \localcircuitf[0]{\tval}{\totalL}
  \pgfmathsetmacro{\dlonger}{0.65}%
  \pgfmathtruncatemacro{\yfinalminus}{2*\tval-2}%
  \pgfmathsetmacro{\dotXleft}{1-\dlonger}%
  \foreach \y in {0,2,...,\yfinalminus}{%
    \pgfmathsetmacro{\dotYdown}{\y+1-\dlonger}%
    \pgfmathsetmacro{\dotYup}{\y+1+\dlonger}%
    \ifnum\y=0\else
      \dotf{\dotXleft}{\dotYdown}%
    \fi
    \dotf{\dotXleft}{\dotYup}%
  }
  \pgfmathsetmacro{\qbitX}{1-\dlonger}%
  \pgfmathsetmacro{\qbitY}{1-\dlonger}%
  \qubittwodf{\qbitX+0.15}{\qbitY+0.15-1}
  \nogaterf{0}{0}
\rightfpf[0]{2*\totalL}{1}{\tval}
  \pgfmathsetmacro{\myystart}{2*\tval-1+\dlonger}%
  \pgfmathtruncatemacro{\xfinalint}{2*\totalL}%
  \foreach\x in \xlistodd{%
    \ifnum\x=1\else
      \ifnum\x=\xfinalminus\else
        \pgfmathsetmacro{\dotX}{\x+1+2+\dlonger}%
        \qubitf{\dotX}{\myystart}%
      \fi
    \fi
  }%
  \foreach\x in \xlisteven{%
    \ifnum\x=\xfinalint\else
      \pgfmathsetmacro{\dotX}{\x+2-1-\dlonger}%
      \qubitf{\dotX}{\myystart}%
    \fi
  }%
  \pgfmathsetmacro{\dotX}{1+\dlonger}%
  \qubitf{\dotX}{\myystart}%
  \pgfmathsetmacro{\dotXright}{2*\totalL+1-\dlonger}%
  \qubitf{\dotXright}{\myystart}%
  \pgfmathsetmacro{\dotX}{1+\dlonger}%
  \qubitf{\dotX}{\myystart}%
  \getNof{\ellpos}{\ttone}{\elllength}
    \getNof{\ellpos}{\tttwo}{\elllength}
     \getNof{\ellpos}{\ttthree}{\elllength}
     \hlinequbitf[\staropf]{1}{2*\ttone}{2*\elllength}
    \hlinequbitf[\staropf]{1}{2*\tttwo}{2*\elllength}
     \hlinequbitf[\staropf]{1}{2*\ttthree}{2*\elllength}
\end{tikzpicture}
\end{equation}
The first equality follows by the same route as in Eq.\ttilde\eqref{eq:Lamplitudelargetau}: we iterate Eq.\ttilde\eqref{eq:relstar} to introduce the operators $\projectorf[\staropf]$, and then use the first relation in Eq.\ttilde\eqref{eq:staropfgatef} together with Eqs.\ttilde \eqref{eq:greencircleprop} and\ttilde\eqref{eq:staropfgatefqubitf} to propagate them rightwards. Here, we also use that each operator $\projectorf[\staropf]$ commutes with the red operator in Eq.\ttilde\eqref{eq:projno} irrespective of  the qubit $j$ on which the latter acts:
\begin{equation}
\label{eq:commstar}
\begin{tikzpicture}[baseline={([yshift=-0.6ex]current bounding box.center)},scale=0.5]
  \pgfmathparse{1.1}\let\dl\pgfmathresult
  \pgfmathsetmacro{\xstep}{2.0}
  \pgfmathsetmacro{\sep}{1.0}
  \foreach \k in {0,1,2}{
    \pgfmathsetmacro{\xoffset}{\k*(\xstep+\sep)}
    \begin{scope}[shift={(\xoffset,0)}]
      \gridLinef{(0.0)-\dl}{-\dl}{(0.0)}{(0.0)}
      \begin{scope}[shift={(0.5,0)}]
        \gridLinef{(0.0)}{(0.0)}{+\dl}{(0.0)-\dl} 
      \end{scope}
    \end{scope}
  }
  \pgfmathsetmacro{\totalwidth}{3*(\xstep+\sep)}%
  \pgfmathsetmacro{\getNoflen}{0.5*\totalwidth+0.1}%
  \pgfmathsetmacro{\getNofstart}{0.35}
  \pgfmathsetmacro{\getNofY}{0.25*(1-\dl)}%
  \getNof{\getNofstart}{\getNofY}{\getNoflen}%
  
  \pgfmathsetmacro{\shiftt}{0.35}
  \foreach \k in{1}{ 
    \pgfmathsetmacro{\xoffset}{\k*(\xstep+\sep)}
    \begin{scope}[shift={(\xoffset,0)}]
     \staropf{0.0-0.5*\dl-\shiftt}{0.0-0.5*\dl-\shiftt} 
 \pgfmathsetmacro{\starX}{0.0-0.5*\dl-\shiftt}%
\pgfmathsetmacro{\starY}{0.0-0.5*\dl-\shiftt}%
\node at (\starX, \starY-0.7) {$j$};%
    \end{scope}
  }
\end{tikzpicture} =
\begin{tikzpicture}[baseline={([yshift=-0.6ex]current bounding box.center)},scale=0.5]
  \pgfmathparse{1.1}\let\dl\pgfmathresult
  \pgfmathsetmacro{\xstep}{2.0}
  \pgfmathsetmacro{\sep}{1.0}
  \foreach \k in {0,1,2}{
    \pgfmathsetmacro{\xoffset}{\k*(\xstep+\sep)}
    \begin{scope}[shift={(\xoffset,0)}]
      \gridLinef{(0.0)-\dl}{-\dl}{(0.0)}{(0.0)}
      \begin{scope}[shift={(0.5,0)}]
        \gridLinef{(0.0)}{(0.0)}{+\dl}{-\dl} 
      \end{scope}
    \end{scope}
  }
  \pgfmathsetmacro{\totalwidth}{3*(\xstep+\sep)}%
  \pgfmathsetmacro{\getNoflen}{0.5*\totalwidth+0.1}%
  \pgfmathsetmacro{\getNofstart}{0.35}
  \pgfmathsetmacro{\getNofY}{0.25*(1-\dl)}%
  \getNof{\getNofstart}{\getNofY}{\getNoflen}%
  
  \pgfmathsetmacro{\shiftt}{0.35}
  \foreach \k in{1}{ 
    \pgfmathsetmacro{\xoffset}{\k*(\xstep+\sep)}
    \begin{scope}[shift={(\xoffset,0)}]
     \staropf{0.0-0.5*\dl+\shiftt}{0.0-0.5*\dl+\shiftt} 
\pgfmathsetmacro{\starX}{0.0-0.5*\dl-\shiftt}%
\pgfmathsetmacro{\starY}{0.0-0.5*\dl-\shiftt}%
\node at (\starX, \starY-0.7) {$j$};%
    \end{scope}
  }
\end{tikzpicture}=
\begin{tikzpicture}[baseline={([yshift=-0.6ex]current bounding box.center)},scale=0.5]
  \pgfmathparse{0.70}\let\dl\pgfmathresult
  \pgfmathsetmacro{\xstep}{2.0}
  \pgfmathsetmacro{\sep}{0.4}
  \foreach \k in {0,1,2}{
    \pgfmathsetmacro{\xoffset}{\k*(\xstep+\sep)}
    \begin{scope}[shift={(\xoffset,0)}]
      \gridLinef{(0.0)-\dl}{-\dl}{(0.0)}{(0.0)}
      \begin{scope}[shift={(0.5,0)}]
        \gridLinef{(0.0)}{(0.0)}{+\dl}{(0.0)-\dl} 
      \end{scope}
    \end{scope}
  }
  \pgfmathsetmacro{\elllength}{3*(2*\dl+0.5)+(3-1)*\sep-2}
   \pgfmathsetmacro{\totalwidth}{3*(\xstep+\sep)}%
  \pgfmathsetmacro{\getNofstart}{0.5*\dl+0.25}%
  \pgfmathsetmacro{\getNoflen}{0.5*\totalwidth+\dl-0.5}%
  \pgfmathsetmacro{\getNofY}{0.25*(1-\dl)}%
  \getNof{\getNofstart}{\getNofY}{\getNoflen}%
  \foreach \k in{1}{ 
    \pgfmathsetmacro{\xoffset}{\k*(\xstep+\sep)}
    \begin{scope}[shift={(\xoffset,0)}]
     \staropf{0.0-0.5*\dl}{0.0-0.5*\dl} 
     \pgfmathsetmacro{\shiftt}{0.35}
    \pgfmathsetmacro{\starX}{0.0-0.5*\dl-\shiftt}%
\pgfmathsetmacro{\starY}{0.0-0.5*\dl-\shiftt}%
\node at (\starX, \starY-0.7) {$j$};%
    \end{scope}
 }
\end{tikzpicture}
\,.
\end{equation}
This relation is, in turn, a consequence of  the commutation relations in Eqs.\ttilde\eqref{eq:actionstarop} and\ttilde\eqref{eq:projgaterelations}. Furthermore, the second equality in Eq.\ttilde\eqref{eq:Dklarget} follows from  Eq.\ttilde\eqref{eq:staropfgateftwo} along with the relations above applied in reverse, used to reabsorb all the remaining operators $\projectorf[\staropf]$ into the left boundary. Crucially, the action of the operators $\projectorf[\staropf]$ on the whole support of the operator in Eq.\ttilde\eqref{eq:projno} leaves only two terms 
\begin{equation}
\label{eq:redandstar}
\begin{tikzpicture}[baseline={([yshift=-0.6ex]current bounding box.center)},scale=0.5]
  \pgfmathparse{0.70}\let\dl\pgfmathresult
  \pgfmathsetmacro{\xstep}{2.0}
  \pgfmathsetmacro{\sep}{0.4}
  \foreach \k in {0,1,2}{
    \pgfmathsetmacro{\xoffset}{\k*(\xstep+\sep)}
    \begin{scope}[shift={(\xoffset,0)}]
      \gridLinef{(0.0)-\dl}{-\dl}{(0.0)}{(0.0)}
      \begin{scope}[shift={(0.5,0)}]
        \gridLinef{(0.0)}{(0.0)}{+\dl}{(0.0)-\dl} 
      \end{scope}
    \end{scope}
  }
  \pgfmathsetmacro{\elllength}{3*(2*\dl+0.5)+(3-1)*\sep-2}
   \pgfmathsetmacro{\totalwidth}{3*(\xstep+\sep)}%
  \pgfmathsetmacro{\getNofstart}{0.5*\dl+0.25}%
  \pgfmathsetmacro{\getNoflen}{0.5*\totalwidth+\dl-0.5}%
  \pgfmathsetmacro{\getNofY}{0.25*(1-\dl)}%
  \getNof{\getNofstart}{\getNofY}{\getNoflen}%
  \foreach \k in {0,1,2}{
    \pgfmathsetmacro{\xoffset}{\k*(\xstep+\sep)}
    \begin{scope}[shift={(\xoffset,0)}]
     \staropf{0.0-0.5*\dl}{0.0-0.5*\dl} 
      \begin{scope}[shift={(0.5,0)}]
        \staropf{0.0+0.5*\dl}{0.0-0.5*\dl} 
      \end{scope}
    \end{scope}
  }
\end{tikzpicture} =
    \begin{tikzpicture}[baseline={([yshift=-0.6ex]current bounding box.center)},scale=0.5]
  \pgfmathparse{0.70}\let\dl\pgfmathresult
  \pgfmathsetmacro{\xstep}{2.0}
  \pgfmathsetmacro{\sep}{0.4}
  \foreach \k in {0,1,2}{
    \pgfmathsetmacro{\xoffset}{\k*(\xstep+\sep)}
    \begin{scope}[shift={(\xoffset,0)}]
      \gridLinef{(0.0)-\dl}{-\dl}{(0.0)}{(0.0)}
      \begin{scope}[shift={(0.5,0)}]
        \gridLinef{(0.0)}{(0.0)}{+\dl}{(0.0)-\dl} 
      \end{scope}
    \end{scope}
  }
\end{tikzpicture}-
\begin{tikzpicture}[baseline={([yshift=-0.6ex]current bounding box.center)},scale=0.5]
  \pgfmathparse{1.0}\let\dl\pgfmathresult
  \pgfmathsetmacro{\xstep}{2.0}%
  \pgfmathsetmacro{\sep}{0.4}%
  \pgfmathsetmacro{\proplen}{1.414*\dl}%
  \foreach \k in {0,1,2}{%
    \pgfmathsetmacro{\xoffset}{\k*(\xstep+\sep)}%
    \begin{scope}[shift={(\xoffset,0)}]%
      \begin{scope}[shift={(-\dl,-\dl)},rotate=-45]%
        \doublepropf{0}{0}{\proplen}%
      \end{scope}%
      \begin{scope}[shift={(\dl,-\dl)},rotate=45]%
        \doublepropf{0}{0}{\proplen}%
      \end{scope}%
    \end{scope}%
  }%
\end{tikzpicture}\,,
\end{equation}
where we also express the two-layer projectors in terms of the folded-space state, as in the first relation of Eq.\ttilde\eqref{eq:projrel1f}. By plugging the above relation into the first unsuccessful outcome, we obtain 
\begin{equation}
    \label{eq:expansionresult}
    F(k)=2^{-2k\tau+1}
     \begin{tikzpicture}[baseline={([yshift=-0.6ex]current bounding box.center)},scale=0.5]
        \pgfmathparse{2}\let\period\pgfmathresult
    \pgfmathparse{3*\period}\let\ttthree\pgfmathresult
    \pgfmathparse{1*\period}\let\ttone\pgfmathresult
    \pgfmathparse{2*\period}\let\tttwo\pgfmathresult
  \pgfmathparse{8}\let\tval\pgfmathresult
  \pgfmathparse{3}\let\totalL\pgfmathresult
   \pgfmathparse{1.5}\let\ellpos\pgfmathresult
    \pgfmathparse{3}\let\elllength\pgfmathresult
    \pgfmathparse{2}\let\period\pgfmathresult
  \localcircuitf[0]{\tval}{\totalL}
  \pgfmathsetmacro{\dlonger}{0.65}%
  \pgfmathtruncatemacro{\yfinalminus}{2*\tval-2}%
  \pgfmathsetmacro{\dotXleft}{1-\dlonger}%
  \foreach \y in {0,2,...,\yfinalminus}{%
    \pgfmathsetmacro{\dotYdown}{\y+1-\dlonger}%
    \pgfmathsetmacro{\dotYup}{\y+1+\dlonger}%
    \ifnum\y=0\else
      \dotf{\dotXleft}{\dotYdown}%
    \fi
    \dotf{\dotXleft}{\dotYup}%
  }
  \pgfmathsetmacro{\qbitX}{1-\dlonger}%
  \pgfmathsetmacro{\qbitY}{1-\dlonger}%
  \qubittwodf{\qbitX+0.15}{\qbitY+0.15-1}
  \nogaterf{0}{0}
\rightfpf[0]{2*\totalL}{1}{\tval}
  \pgfmathsetmacro{\myystart}{2*\tval-1+\dlonger}%
  \pgfmathtruncatemacro{\xfinalint}{2*\totalL}%
  \foreach\x in \xlistodd{%
    \ifnum\x=1\else
      \ifnum\x=\xfinalminus\else
        \pgfmathsetmacro{\dotX}{\x+1+2+\dlonger}%
        \qubitf{\dotX}{\myystart}%
      \fi
    \fi
  }%
  \foreach\x in \xlisteven{%
    \ifnum\x=\xfinalint\else
      \pgfmathsetmacro{\dotX}{\x+2-1-\dlonger}%
      \qubitf{\dotX}{\myystart}%
    \fi
  }%
  \pgfmathsetmacro{\dotX}{1+\dlonger}%
  \qubitf{\dotX}{\myystart}%
  \pgfmathsetmacro{\dotXright}{2*\totalL+1-\dlonger}%
  \qubitf{\dotXright}{\myystart}%
  \pgfmathsetmacro{\dotX}{1+\dlonger}%
  \qubitf{\dotX}{\myystart}%
  \getNof{\ellpos}{\ttone}{\elllength}
    \getNof{\ellpos}{\tttwo}{\elllength}
     \getNof{\ellpos}{\ttthree}{\elllength}
     \hlinequbitf[\staropf]{1}{2*\ttone}{2*\elllength}
    \hlinequbitf[\staropf]{1}{2*\tttwo}{2*\elllength}
     \hlinequbitf[\staropf]{1}{2*\ttthree}{2*\elllength}
\end{tikzpicture}= 
2^{-2k\tau+1}
     \begin{tikzpicture}[baseline={([yshift=-0.6ex]current bounding box.center)},scale=0.5]
        \pgfmathparse{2}\let\period\pgfmathresult
    \pgfmathparse{3*\period}\let\ttthree\pgfmathresult
    \pgfmathparse{1*\period}\let\ttone\pgfmathresult
    \pgfmathparse{2*\period}\let\tttwo\pgfmathresult
  \pgfmathparse{8}\let\tval\pgfmathresult
  \pgfmathparse{3}\let\totalL\pgfmathresult
   \pgfmathparse{1.5}\let\ellpos\pgfmathresult
    \pgfmathparse{3}\let\elllength\pgfmathresult
    \pgfmathparse{2}\let\period\pgfmathresult
  \localcircuitf[0]{\tval}{\totalL}
  \pgfmathsetmacro{\dlonger}{0.65}%
  \pgfmathtruncatemacro{\yfinalminus}{2*\tval-2}%
  \pgfmathsetmacro{\dotXleft}{1-\dlonger}%
  \foreach \y in {0,2,...,\yfinalminus}{%
    \pgfmathsetmacro{\dotYdown}{\y+1-\dlonger}%
    \pgfmathsetmacro{\dotYup}{\y+1+\dlonger}%
    \ifnum\y=0\else
      \dotf{\dotXleft}{\dotYdown}%
    \fi
    \dotf{\dotXleft}{\dotYup}%
  }
  \pgfmathsetmacro{\qbitX}{1-\dlonger}%
  \pgfmathsetmacro{\qbitY}{1-\dlonger}%
  \qubittwodf{\qbitX+0.15}{\qbitY+0.15-1}
  \nogaterf{0}{0}
\rightfpf[0]{2*\totalL}{1}{\tval}
  \pgfmathsetmacro{\myystart}{2*\tval-1+\dlonger}%
  \pgfmathtruncatemacro{\xfinalint}{2*\totalL}%
  \foreach\x in \xlistodd{%
    \ifnum\x=1\else
      \ifnum\x=\xfinalminus\else
        \pgfmathsetmacro{\dotX}{\x+1+2+\dlonger}%
        \qubitf{\dotX}{\myystart}%
      \fi
    \fi
  }%
  \foreach\x in \xlisteven{%
    \ifnum\x=\xfinalint\else
      \pgfmathsetmacro{\dotX}{\x+2-1-\dlonger}%
      \qubitf{\dotX}{\myystart}%
    \fi
  }%
  \pgfmathsetmacro{\dotX}{1+\dlonger}%
  \qubitf{\dotX}{\myystart}%
  \pgfmathsetmacro{\dotXright}{2*\totalL+1-\dlonger}%
  \qubitf{\dotXright}{\myystart}%
  \pgfmathsetmacro{\dotX}{1+\dlonger}%
  \qubitf{\dotX}{\myystart}%
    \getNof{\ellpos}{\tttwo}{\elllength}
     \getNof{\ellpos}{\ttthree}{\elllength}
    \hlinequbitf[\staropf]{1}{2*\tttwo}{2*\elllength}
     \hlinequbitf[\staropf]{1}{2*\ttthree}{2*\elllength}
\end{tikzpicture}- \frac{\mathcal{L}(1)}{2^{2(k-1)\tau}}
     \begin{tikzpicture}[baseline={([yshift=-0.6ex]current bounding box.center)},scale=0.5]
        \pgfmathparse{2}\let\period\pgfmathresult
    \pgfmathparse{3*\period}\let\ttthree\pgfmathresult
    \pgfmathparse{1*\period}\let\ttone\pgfmathresult
    \pgfmathparse{2*\period}\let\tttwo\pgfmathresult
  \pgfmathparse{6}\let\tval\pgfmathresult
  \pgfmathparse{3}\let\totalL\pgfmathresult
   \pgfmathparse{1.5}\let\ellpos\pgfmathresult
    \pgfmathparse{3}\let\elllength\pgfmathresult
    \pgfmathparse{2}\let\period\pgfmathresult
  \localcircuitf[0]{\tval}{\totalL}
  \pgfmathsetmacro{\dlonger}{0.65}%
  \pgfmathtruncatemacro{\yfinalminus}{2*\tval-2}%
  \pgfmathsetmacro{\dotXleft}{1-\dlonger}%
  \foreach \y in {0,2,...,\yfinalminus}{%
    \pgfmathsetmacro{\dotYdown}{\y+1-\dlonger}%
    \pgfmathsetmacro{\dotYup}{\y+1+\dlonger}%
    \ifnum\y=0\else
      \dotf{\dotXleft}{\dotYdown}%
    \fi
    \dotf{\dotXleft}{\dotYup}%
  }
  \pgfmathsetmacro{\qbitX}{1-\dlonger}%
  \pgfmathsetmacro{\qbitY}{1-\dlonger}%
  \dotf{\qbitX}{\qbitY}%
\rightfpf[0]{2*\totalL}{1}{\tval}
  \pgfmathsetmacro{\myystart}{2*\tval-1+\dlonger}%
  \pgfmathtruncatemacro{\xfinalint}{2*\totalL}%
  \foreach\x in \xlistodd{%
    \ifnum\x=1\else
      \ifnum\x=\xfinalminus\else
        \pgfmathsetmacro{\dotX}{\x+1+2+\dlonger}%
        \qubitf{\dotX}{\myystart}%
      \fi
    \fi
  }%
  \foreach\x in \xlisteven{%
    \ifnum\x=\xfinalint\else
      \pgfmathsetmacro{\dotX}{\x+2-1-\dlonger}%
      \qubitf{\dotX}{\myystart}%
    \fi
  }%
  \pgfmathsetmacro{\dotX}{1+\dlonger}%
  \qubitf{\dotX}{\myystart}%
  \pgfmathsetmacro{\dotXright}{2*\totalL+1-\dlonger}%
  \qubitf{\dotXright}{\myystart}%
  \pgfmathsetmacro{\dotX}{1+\dlonger}%
  \qubitf{\dotX}{\myystart}%
  \getNof{\ellpos}{\ttone}{\elllength}
    \getNof{\ellpos}{\tttwo}{\elllength}
     \hlinequbitf[\staropf]{1}{2*\ttone}{2*\elllength}
    \hlinequbitf[\staropf]{1}{2*\tttwo}{2*\elllength}
\end{tikzpicture}\,,
\end{equation}
where the Loschmidt echo $\mathcal{L}(1)$ originates from the second term in Eq.~\eqref{eq:redandstar}. Iterating the same decomposition on each of the remaining red operators in Eq.\ttilde\eqref{eq:expansionresult}, we arrive at the Loschmidt-echo expansion of Eq.~(6) of main text.

We now consider $k\tau\leq 2 \ell$. In this case, it follows from Eq.\ttilde\eqref{eq:projgaterelations} that the action of the operator in Eq.\ttilde\eqref{eq:projno} on a single projector $\projectorf$ acting on qubit $j$ gives:
\begin{equation}
\label{eq:commgreen}
\begin{tikzpicture}[baseline={([yshift=-0.6ex]current bounding box.center)},scale=0.5]
  \pgfmathparse{1.1}\let\dl\pgfmathresult
  \pgfmathsetmacro{\xstep}{2.0}
  \pgfmathsetmacro{\sep}{1.0}
  \foreach \k in {0,1,2}{
    \pgfmathsetmacro{\xoffset}{\k*(\xstep+\sep)}
    \begin{scope}[shift={(\xoffset,0)}]
      \gridLinef{(0.0)-\dl}{-\dl}{(0.0)}{(0.0)}
      \begin{scope}[shift={(0.5,0)}]
        \gridLinef{(0.0)}{(0.0)}{+\dl}{(0.0)-\dl} 
      \end{scope}
    \end{scope}
  }
  \pgfmathsetmacro{\totalwidth}{3*(\xstep+\sep)}%
  \pgfmathsetmacro{\getNoflen}{0.5*\totalwidth+0.1}%
  \pgfmathsetmacro{\getNofstart}{0.35}
  \pgfmathsetmacro{\getNofY}{0.25*(1-\dl)}%
  \getNof{\getNofstart}{\getNofY}{\getNoflen}%
  
  \pgfmathsetmacro{\shiftt}{0.35}
  \foreach \k in{1}{ 
    \pgfmathsetmacro{\xoffset}{\k*(\xstep+\sep)}
    \begin{scope}[shift={(\xoffset,0)}]
     \qubitf{0.0-0.5*\dl-\shiftt}{0.0-0.5*\dl-\shiftt} 
 \pgfmathsetmacro{\starX}{0.0-0.5*\dl-\shiftt}%
\pgfmathsetmacro{\starY}{0.0-0.5*\dl-\shiftt}%
\node at (\starX, \starY-0.7) {$j$};%
    \end{scope}
  }
\end{tikzpicture} =
\begin{tikzpicture}[baseline={([yshift=-0.6ex]current bounding box.center)},scale=0.5]
  \pgfmathparse{1.1}\let\dl\pgfmathresult
  \pgfmathsetmacro{\xstep}{2.0}
  \pgfmathsetmacro{\sep}{1.0}
  \foreach \k in {0,1,2}{
    \pgfmathsetmacro{\xoffset}{\k*(\xstep+\sep)}
    \begin{scope}[shift={(\xoffset,0)}]
      \gridLinef{(0.0)-\dl}{-\dl}{(0.0)}{(0.0)}
      \begin{scope}[shift={(0.5,0)}]
        \gridLinef{(0.0)}{(0.0)}{+\dl}{-\dl} 
      \end{scope}
    \end{scope}
  }
  \pgfmathsetmacro{\totalwidth}{3*(\xstep+\sep)}%
  \pgfmathsetmacro{\getNoflen}{0.5*\totalwidth+0.1}%
  \pgfmathsetmacro{\getNofstart}{0.35}
  \pgfmathsetmacro{\getNofY}{0.25*(1-\dl)}%
  \getNof{\getNofstart}{\getNofY}{\getNoflen}%
  
  \pgfmathsetmacro{\shiftt}{0.35}
  \foreach \k in{1}{ 
    \pgfmathsetmacro{\xoffset}{\k*(\xstep+\sep)}
    \begin{scope}[shift={(\xoffset,0)}]
     \qubitf{0.0-0.5*\dl+\shiftt}{0.0-0.5*\dl+\shiftt} 
\pgfmathsetmacro{\starX}{0.0-0.5*\dl-\shiftt}%
\pgfmathsetmacro{\starY}{0.0-0.5*\dl-\shiftt}%
\node at (\starX, \starY-0.7) {$j$};%
    \end{scope}
  }
\end{tikzpicture}=
\begin{tikzpicture}[baseline={([yshift=-0.6ex]current bounding box.center)},scale=0.5]
  \pgfmathparse{0.70}\let\dl\pgfmathresult
  \pgfmathsetmacro{\xstep}{2.0}
  \pgfmathsetmacro{\sep}{0.4}
  \foreach \k in {0,1,2}{
    \pgfmathsetmacro{\xoffset}{\k*(\xstep+\sep)}
    \begin{scope}[shift={(\xoffset,0)}]
      \gridLinef{(0.0)-\dl}{-\dl}{(0.0)}{(0.0)}
      \begin{scope}[shift={(0.5,0)}]
        \gridLinef{(0.0)}{(0.0)}{+\dl}{(0.0)-\dl} 
      \end{scope}
    \end{scope}
  }
  \pgfmathsetmacro{\elllength}{3*(2*\dl+0.5)+(3-1)*\sep-2}
   \pgfmathsetmacro{\totalwidth}{3*(\xstep+\sep)}%
  \pgfmathsetmacro{\getNofstart}{0.5*\dl+0.25}%
  \pgfmathsetmacro{\getNoflen}{0.5*\totalwidth+\dl-0.5}%
  \pgfmathsetmacro{\getNofY}{0.25*(1-\dl)}%
  \getNof{\getNofstart}{\getNofY}{\getNoflen}%
  \foreach \k in{1}{ 
    \pgfmathsetmacro{\xoffset}{\k*(\xstep+\sep)}
    \begin{scope}[shift={(\xoffset,0)}]
     \qubitf{0.0-0.5*\dl}{0.0-0.5*\dl} 
     \pgfmathsetmacro{\shiftt}{0.35}
    \pgfmathsetmacro{\starX}{0.0-0.5*\dl-\shiftt}%
\pgfmathsetmacro{\starY}{0.0-0.5*\dl-\shiftt}%
\node at (\starX, \starY-0.7) {$j$};%
    \end{scope}
 }
\end{tikzpicture}
\,,
\end{equation}
for any qubit. Iterating the relations in Eqs.\ttilde\eqref{eq:greencircleprop} and\ttilde\eqref{eq:greencirclegate}, together with the above property, we show that the following diagram gives zero:
\begin{equation} 
    \label{eq:Dksmallt}
    D(k)=  2^{2k\tau-1} F(k) =
        \begin{tikzpicture}[baseline={([yshift=-0.6ex]current bounding box.center)},scale=0.5]
        \pgfmathparse{1}\let\period\pgfmathresult
    \pgfmathparse{4*\period}\let\tfour\pgfmathresult
    \pgfmathparse{3*\period}\let\ttthree\pgfmathresult
    \pgfmathparse{1*\period}\let\ttone\pgfmathresult
    \pgfmathparse{2*\period}\let\tttwo\pgfmathresult
  \pgfmathparse{5}\let\tval\pgfmathresult
  \pgfmathparse{3}\let\totalL\pgfmathresult
   \pgfmathparse{1.5}\let\ellpos\pgfmathresult
    \pgfmathparse{3}\let\elllength\pgfmathresult

  \pgfmathsetmacro{\dlonger}{0.65}%
  \pgfmathsetmacro{\myystart}{2*\tval-1+\dlonger}%
  \pgfmathsetmacro{\dxarrow}{0.1}%
  \pgfmathsetmacro{\arrowX}{0 - \dxarrow}%
  \pgfmathsetmacro{\arrowYbot}{-\dlonger}%
  \pgfmathsetmacro{\arrowYtop}{\myystart}%
  \pgfmathsetmacro{\arrowYmid}{0.5*(\arrowYbot+\arrowYtop)}%
  \draw[<->, thick] (\arrowX, \arrowYbot) -- (\arrowX, \arrowYtop);
  \node[anchor=south, rotate=90] at (\arrowX-0.0, \arrowYmid) {$k\tau \leq 2\ell$};
  \localcircuitf[0]{\tval}{\totalL}
  \pgfmathsetmacro{\dlonger}{0.65}%
  \pgfmathtruncatemacro{\yfinalminus}{2*\tval-2}%
  \pgfmathsetmacro{\dotXleft}{1-\dlonger}%
  \foreach \y in {0,2,...,\yfinalminus}{%
    \pgfmathsetmacro{\dotYdown}{\y+1-\dlonger}%
    \pgfmathsetmacro{\dotYup}{\y+1+\dlonger}%
    \ifnum\y=0\else
      \dotf{\dotXleft}{\dotYdown}%
    \fi
    \dotf{\dotXleft}{\dotYup}%
  }
  \pgfmathsetmacro{\qbitX}{1-\dlonger}%
  \pgfmathsetmacro{\qbitY}{1-\dlonger}%
  \qubittwodf{\qbitX+0.15}{\qbitY+0.15-1}
  \nogaterf{0}{0}
\rightfpf[0]{2*\totalL}{1}{\tval}
  \pgfmathsetmacro{\myystart}{2*\tval-1+\dlonger}%
  \pgfmathtruncatemacro{\xfinalint}{2*\totalL}%
  \foreach\x in \xlistodd{%
    \ifnum\x=1\else
      \ifnum\x=\xfinalminus\else
        \pgfmathsetmacro{\dotX}{\x+1+2+\dlonger}%
        \qubitf{\dotX}{\myystart}%
      \fi
    \fi
  }%
  \foreach\x in \xlisteven{%
    \ifnum\x=\xfinalint\else
      \pgfmathsetmacro{\dotX}{\x+2-1-\dlonger}%
      \qubitf{\dotX}{\myystart}%
    \fi
  }%
  \pgfmathsetmacro{\dotX}{1+\dlonger}%
  \qubitf{\dotX}{\myystart}%
  \pgfmathsetmacro{\dotXright}{2*\totalL+1-\dlonger}%
  \qubitf{\dotXright}{\myystart}%
  \pgfmathsetmacro{\dotX}{1+\dlonger}%
  \qubitf{\dotX}{\myystart}%
  \getNof{\ellpos}{\ttone}{\elllength}
    \getNof{\ellpos}{\tttwo}{\elllength}
     \getNof{\ellpos}{\ttthree}{\elllength}
     \getNof{\ellpos}{\tfour}{\elllength}
\end{tikzpicture}= 
    \begin{tikzpicture}[baseline={([yshift=-0.6ex]current bounding box.center)},scale=0.5]
        \pgfmathparse{1}\let\period\pgfmathresult
    \pgfmathparse{4*\period}\let\tfour\pgfmathresult
    \pgfmathparse{3*\period}\let\ttthree\pgfmathresult
    \pgfmathparse{1*\period}\let\ttone\pgfmathresult
    \pgfmathparse{2*\period}\let\tttwo\pgfmathresult
  \pgfmathparse{5}\let\tval\pgfmathresult
  \pgfmathparse{3}\let\totalL\pgfmathresult
   \pgfmathparse{1.5}\let\ellpos\pgfmathresult
    \pgfmathparse{3}\let\elllength\pgfmathresult
   
  \localcircuitf[0]{\tval}{\totalL}
  \pgfmathsetmacro{\dlonger}{0.65}%
  \pgfmathtruncatemacro{\yfinalminus}{2*\tval-2}%
  \pgfmathsetmacro{\dotXleft}{1-\dlonger}%
  \foreach \y in {0,2,...,\yfinalminus}{%
    \pgfmathsetmacro{\dotYdown}{\y+1-\dlonger}%
    \pgfmathsetmacro{\dotYup}{\y+1+\dlonger}%
    \ifnum\y=0\else
      \dotf{\dotXleft}{\dotYdown}%
    \fi
    \dotf{\dotXleft}{\dotYup}%
  }
  \pgfmathsetmacro{\qbitX}{1-\dlonger}%
  \pgfmathsetmacro{\qbitY}{1-\dlonger}%
  \qubittwodf{\qbitX+0.15}{\qbitY+0.15-1}
  \nogaterf{0}{0}
\rightfpf[0]{2*\totalL}{1}{\tval}
  \pgfmathsetmacro{\myystart}{2*\tval-1+\dlonger}%
  \pgfmathtruncatemacro{\xfinalint}{2*\totalL}%
  \foreach\x in \xlistodd{%
    \ifnum\x=1\else
      \ifnum\x=\xfinalminus\else
        \pgfmathsetmacro{\dotX}{\x+1+2+\dlonger}%
        \qubitf{\dotX}{\myystart}%
      \fi
    \fi
  }%
  \foreach\x in \xlisteven{%
    \ifnum\x=\xfinalint\else
      \pgfmathsetmacro{\dotX}{\x+2-1-\dlonger}%
      \qubitf{\dotX}{\myystart}%
    \fi
  }%
  \pgfmathsetmacro{\dotX}{1+\dlonger}%
  \qubitf{\dotX}{\myystart}%
  \pgfmathsetmacro{\dotXright}{2*\totalL+1-\dlonger}%
  \qubitf{\dotXright}{\myystart}%
  \pgfmathsetmacro{\dotX}{1+\dlonger}%
  \qubitf{\dotX}{\myystart}%
  \getNof{\ellpos}{\ttone}{\elllength}
    \getNof{\ellpos}{\tttwo}{\elllength}
     \getNof{\ellpos}{\ttthree}{\elllength}
     \getNof{\ellpos}{\tfour}{\elllength}
\pgfmathtruncatemacro{\ycooord}{2*\tval-1}%
\pgfmathtruncatemacro{\nummm}{2*\totalL-1}%
\pgfmathtruncatemacro{\totalLtwo}{2*\totalL}%
\btrianglequbitf{1}{1}{\totalLtwo}%
\ttrianglequbitf{1}{\ycooord}{\nummm}%
\end{tikzpicture}=
\begin{tikzpicture}[baseline={([yshift=-0.6ex]current bounding box.center)},scale=0.5]
        \pgfmathparse{1}\let\period\pgfmathresult
    \pgfmathparse{4*\period}\let\tfour\pgfmathresult
    \pgfmathparse{3*\period}\let\ttthree\pgfmathresult
    \pgfmathparse{1*\period}\let\ttone\pgfmathresult
    \pgfmathparse{2*\period}\let\tttwo\pgfmathresult
  \pgfmathparse{5}\let\tval\pgfmathresult
  \pgfmathparse{3}\let\totalL\pgfmathresult
   \pgfmathparse{1.5}\let\ellpos\pgfmathresult
    \pgfmathparse{3}\let\elllength\pgfmathresult
   
  \localcircuitf[0]{\tval}{\totalL}
  \pgfmathsetmacro{\dlonger}{0.65}%
  \pgfmathtruncatemacro{\yfinalminus}{2*\tval-2}%
  \pgfmathsetmacro{\dotXleft}{1-\dlonger}%
  \foreach \y in {0,2,...,\yfinalminus}{%
    \pgfmathsetmacro{\dotYdown}{\y+1-\dlonger}%
    \pgfmathsetmacro{\dotYup}{\y+1+\dlonger}%
    \ifnum\y=0\else
      \dotf{\dotXleft}{\dotYdown}%
    \fi
    \dotf{\dotXleft}{\dotYup}%
  }
  \pgfmathsetmacro{\qbitX}{1-\dlonger}%
  \pgfmathsetmacro{\qbitY}{1-\dlonger}%
  \qubittwodf{\qbitX+0.15}{\qbitY+0.15-1}
  \nogaterf{0}{0}
\rightfpf[0]{2*\totalL}{1}{\tval}
  \pgfmathsetmacro{\myystart}{2*\tval-1+\dlonger}%
  \pgfmathtruncatemacro{\xfinalint}{2*\totalL}%
  \foreach\x in \xlistodd{%
    \ifnum\x=1\else
      \ifnum\x=\xfinalminus\else
        \pgfmathsetmacro{\dotX}{\x+1+2+\dlonger}%
        \qubitf{\dotX}{\myystart}%
      \fi
    \fi
  }%
  \foreach\x in \xlisteven{%
    \ifnum\x=\xfinalint\else
      \pgfmathsetmacro{\dotX}{\x+2-1-\dlonger}%
      \qubitf{\dotX}{\myystart}%
    \fi
  }%
  \pgfmathsetmacro{\dotX}{1+\dlonger}%
  \qubitf{\dotX}{\myystart}%
  \pgfmathsetmacro{\dotXright}{2*\totalL+1-\dlonger}%
  \qubitf{\dotXright}{\myystart}%
  \pgfmathsetmacro{\dotX}{1+\dlonger}%
  \qubitf{\dotX}{\myystart}%
  \getNof{\ellpos}{\ttone}{\elllength}
    \getNof{\ellpos}{\tttwo}{\elllength}
     \getNof{\ellpos}{\ttthree}{\elllength}
     \getNof{\ellpos}{\tfour}{\elllength}
\pgfmathtruncatemacro{\ycooord}{2*\tval-1}%
\pgfmathtruncatemacro{\nummm}{2*\totalL}%
\pgfmathtruncatemacro{\totalLtwo}{2*\totalL}%
\foreach \i in {1,...,\ycooord}{%
  \pgfmathtruncatemacro{\ii}{\i}%
  \hlinequbitf{1}{\ii}{\nummm}%
}%
\end{tikzpicture}=0\;,
\end{equation}
where the last equality follows from Eq.\ttilde\eqref{eq:projno}:
\begin{equation}
\label{eq:redandgreen}
 \begin{tikzpicture}[baseline={([yshift=-0.6ex]current bounding box.center)},scale=0.5]
  \pgfmathparse{0.70}\let\dl\pgfmathresult
  \pgfmathsetmacro{\xstep}{2.0}
  \pgfmathsetmacro{\sep}{0.4}
  \foreach \k in {0,1,2}{
    \pgfmathsetmacro{\xoffset}{\k*(\xstep+\sep)}
    \begin{scope}[shift={(\xoffset,0)}]
      \gridLinef{(0.0)-\dl}{-\dl}{(0.0)}{(0.0)}
      \begin{scope}[shift={(0.5,0)}]
        \gridLinef{(0.0)}{(0.0)}{+\dl}{(0.0)-\dl} 
      \end{scope}
    \end{scope}
  }
  \pgfmathsetmacro{\elllength}{3*(2*\dl+0.5)+(3-1)*\sep-2}
   \pgfmathsetmacro{\totalwidth}{3*(\xstep+\sep)}%
  \pgfmathsetmacro{\getNofstart}{0.5*\dl+0.25}%
  \pgfmathsetmacro{\getNoflen}{0.5*\totalwidth+\dl-0.5}%
  \pgfmathsetmacro{\getNofY}{0.25*(1-\dl)}%
  \getNof{\getNofstart}{\getNofY}{\getNoflen}%
  \foreach \k in {0,1,2}{
    \pgfmathsetmacro{\xoffset}{\k*(\xstep+\sep)}
    \begin{scope}[shift={(\xoffset,0)}]
     \qubitf{0.0-0.5*\dl}{0.0-0.5*\dl} 
      \begin{scope}[shift={(0.5,0)}]
        \qubitf{0.0+0.5*\dl}{0.0-0.5*\dl} 
      \end{scope}
    \end{scope}
  }
\end{tikzpicture}=
\begin{tikzpicture}[baseline={([yshift=-0.6ex]current bounding box.center)},scale=0.5]
  \pgfmathparse{1.1}\let\dl\pgfmathresult
  \pgfmathsetmacro{\xstep}{2.0}
  \pgfmathsetmacro{\sep}{1.0}
  \foreach \k in {0,1,2}{
    \pgfmathsetmacro{\xoffset}{\k*(\xstep+\sep)}
    \begin{scope}[shift={(\xoffset,0)}]
      \gridLinef{(0.0)-\dl}{-\dl}{(0.0)}{(0.0)}
      \begin{scope}[shift={(0.5,0)}]
        \gridLinef{(0.0)}{(0.0)}{+\dl}{(0.0)-\dl} 
      \end{scope}
    \end{scope}
  }
  \pgfmathsetmacro{\totalwidth}{3*(\xstep+\sep)}%
  \pgfmathsetmacro{\getNoflen}{0.5*\totalwidth+0.1}%
  \pgfmathsetmacro{\getNofstart}{0.35}
  \pgfmathsetmacro{\getNofY}{0.25*(1-\dl)}%
  \getNof{\getNofstart}{\getNofY}{\getNoflen}%
  
  \pgfmathsetmacro{\shiftt}{0.35}
  \foreach \k in {0,1,2}{
    \pgfmathsetmacro{\xoffset}{\k*(\xstep+\sep)}
    \begin{scope}[shift={(\xoffset,0)}]
     \qubitf{0.0-0.5*\dl-\shiftt}{0.0-0.5*\dl-\shiftt} 
      \begin{scope}[shift={(0.5,0)}]
        \qubitf{0.0+0.5*\dl+\shiftt}{0.0-0.5*\dl-\shiftt} 
      \end{scope}
    \end{scope}
  }
\end{tikzpicture} =
\begin{tikzpicture}[baseline={([yshift=-0.6ex]current bounding box.center)},scale=0.5]
  \pgfmathparse{1.1}\let\dl\pgfmathresult
  \pgfmathsetmacro{\xstep}{2.0}
  \pgfmathsetmacro{\sep}{1.0}
  \foreach \k in {0,1,2}{
    \pgfmathsetmacro{\xoffset}{\k*(\xstep+\sep)}
    \begin{scope}[shift={(\xoffset,0)}]
      \gridLinef{(0.0)-\dl}{-\dl}{(0.0)}{(0.0)}
      \begin{scope}[shift={(0.5,0)}]
        \gridLinef{(0.0)}{(0.0)}{+\dl}{-\dl} 
      \end{scope}
    \end{scope}
  }
  \pgfmathsetmacro{\totalwidth}{3*(\xstep+\sep)}%
  \pgfmathsetmacro{\getNoflen}{0.5*\totalwidth+0.1}%
  \pgfmathsetmacro{\getNofstart}{0.35}
  \pgfmathsetmacro{\getNofY}{0.25*(1-\dl)}%
  \getNof{\getNofstart}{\getNofY}{\getNoflen}%
  
  \pgfmathsetmacro{\shiftt}{0.35}
  \foreach \k in {0,1,2}{
    \pgfmathsetmacro{\xoffset}{\k*(\xstep+\sep)}
    \begin{scope}[shift={(\xoffset,0)}]
     \qubitf{0.0-0.5*\dl+\shiftt}{0.0-0.5*\dl+\shiftt} 
      \begin{scope}[shift={(0.5,0)}]
        \qubitf{0.0+0.5*\dl-\shiftt}{0.0-0.5*\dl+\shiftt} 
      \end{scope}
    \end{scope}
  }
\end{tikzpicture} = 0\,.
\end{equation} 
Therefore, we obtain
\begin{equation}
\label{eq:resulttausmall}
    F(k) = 0\,, \quad\quad\text{for}\; k\tau\leq 2 \ell\,,\; k>1\,.
\end{equation}
We now verify that the expansion in Eq.\ttilde(6) reproduces this same result. Since $k\tau\leq 2 \ell$
 Eq.~(7) implies  that $\mathcal{L}(m_j) = 2^{-m_j\tau}$ for any $m_j$, so that for fixed $r$ and any ordered partition $(m_1,m_2, \dots, m_r)$ with  $\sum_{m_j}=k$, the product $\prod_j \mathcal{L}(m_j)$ depends only on $k\tau$ not on the individual $m_j$. The corresponding sum in  Eq.~(6) therefore becomes
\begin{equation}
 \sum_{m_1,\dots,m_r=1}^\infty \left( \prod_{j=1}^r \mathcal{L} (m_j) \right) \delta_{\sum_i m_i, k} = 2^{-k\tau} N_r (k)
\end{equation}
with $N_r (k)$ being the number of ordered partitions of $k$ into $r$ integer parts,
\begin{equation}
    N_r (k) = \sum_{m_1,\dots,m_r=1}^\infty  \delta_{\sum_i m_i, k} = \binom{k-1}{r-1}\,.
\end{equation}
We can then write
\begin{equation}
 F(k)= 2^{-k\tau} \sum_{r=1}^\infty  \binom{k-1}{r-1} (-1)^{r+1} =2^{-k\tau} \sum_{s=0}^\infty  \binom{k-1}{s} (-1)^{s} (1)^{k-1-s} = 2^{-k\tau} (1 - 1)^{k-1} \,,
 \end{equation}
 where we used the substitution $s= r-1$ and the binomial theorem. For $k=1$ the last step gives $2^{-\tau}$, consistent with the result in Eq.\ttilde\eqref{eq:F1}. For any $k>1$, instead, the factor $(1 - 1)^{k-1}$ vanishes identically, confirming Eq.\ttilde\eqref{eq:resulttausmall}. \\

We have thus shown that the expansion in Eq.\ttilde\eqref{eq:expansion} holds  for arbitrary $\tau$ and $\ell$. The expansion can be evaluated directly in the early-detection regime, where $k$ --- and hence the number of terms in the expansion --- is small.
 For instance, for $k=2$ one finds
   \begin{equation}
 F(2) = \mathcal{L}(2) -  \mathcal{L}(1)^2\,= \begin{cases}
0, & \tau \leq \ell, \\
2^{-2\ell}-2^{-2\tau}, & \ell < \tau < 2\ell \\
2^{-2\ell}-2^{-4\ell}, & \tau \geq 2\ell .
\end{cases}
\end{equation}
As discussed in the main text, the different contributions to the expansion depend on   how $\tau$ compares to $\ell$ and can be understood through a geometrical argument, illustrated in Fig.~3(a,b) for $k=3$. We also remark that Eq.~\eqref{eq:expansionresult} is a crucial and nontrivial result since it expresses $F(k)$ only in terms of two-layer projectors $\projectorf$. 

This is the property that leads to Eq.~(6) and, crucially, to the renewal equation in Eq.~(8) of the main text. This expression is fundamentally different from the renewal equations from the field of quantum walks \cite{Barkai2016,Friedman2017QuantumProblem,Dhar2015QuantumModel,Dhar2015b,Barkai2018,Barkai2020,Thiel_spectral,Walter2025,roy_causality_2025,roy2026phase}, which instead involve the first-detection amplitude, whose modulus squared gives $F(k)$. Moreover, the non-negativity of the first-detection probability $F(k)$ follows from Eq.~(8) and the log-convexity of the Loschmidt echo in Eq.~(7), that is, $\log\mathcal{L}(m)-\log\mathcal{L}(m-1)\leq\log\mathcal{L}(m+1)-\log\mathcal{L}(m)$ for all $m\geq1$~\cite{Kaluza1928}. We stress that we have used Eq.\ttilde(8) to recursively obtain the first-detection probability for arbitrary $k$ and to generate the results shown in Fig.~3(c,d) of the main text.

\section{Exact first-detection probability}

In this section we present an alternative method to compute the first-detection probability, as the higher-order derivative of a generating function $\hat{\mathcal{G}}(z)$ in Eq.~(10) of the main text, for which an exact expression can be obtained. This approach is particularly convenient at large $k$, where the number of terms in the expansion~(6) becomes large and the latter is no longer efficient to compute. In Subsec.\ttilde\ref{subsec:VA}, we derive a closed expression for the first-detection probability and we report the analytical result for $\hat{\mathcal{G}}(z)$ in Eq.~(10). In Subsec.\ttilde\ref{subsec:VB} we discuss in detail the exponential decay at large numbers of attempts $k$.

\subsection{Derivation via the generating function}
\label{subsec:VA}
The starting point is the renewal equation\ttilde(8) of the main text, whose right-hand side is a convolution of $\mathcal{L}$ and $F$. One way to simplify that equation is to map the problem onto the complex plane, $z\in\mathbb{C}$, via the discrete Laplace transform ($Z$-transform):
\begin{equation}
\label{eq:Fz}
   \hat{ \mathcal{F}} (z) = \sum_{k=1}^{\infty} F (k)\, z^k = \frac{\hat{\mathcal{L}}(z)}{1+\hat{\mathcal{L}}(z)}\,,
\end{equation}
where we define the discrete Laplace transform of the Loschmidt echo as
\begin{equation}
\label{eq:Lcomputation}
    \hat{\mathcal{L}}(z)=\sum_{k=1}^{\infty} \mathcal{L} ({k}) z^k\, = \sum_{k=1}^{n-1} \left( 2^{-\tau} z \right)^k + 2^{-2\ell} \sum_{k=n}^{\infty}   z^k\,.
\end{equation}
Here, we used Eq.~(7) and introduced a threshold $n$, fixed by the measurement period $\tau$ and the size  $\ell$ of the monitored subregion, namely,
\begin{equation}
\label{eq:n}
    n= n(\tau, \ell) =\lceil 2\ell / \tau \rceil,
\end{equation}
where $\lceil x \rceil$ for a real number $x$ denotes the ceiling function, i.e., the smallest integer that is greater than or equal to $x$. This number measures how far the monitored system is from the infrequent regime $n =1$. The two sums in Eq.~\eqref{eq:Lcomputation} can be rewritten as finite and infinite geometric series, respectively. For the convergence domain $| z | < 1$, we have that 
\begin{align}
 \hat{\mathcal{L}}(z)&= \sum_{k=0}^{n-1} \left( 2^{-\tau} z \right)^k -1 + 2^{-2\ell} \sum_{k=0}^{\infty} z^k -2^{-2\ell} \sum_{k=0}^{n-1}   z^k \nonumber  \\
 &= \frac{1-  \left( 2^{-\tau} z \right)^n}{1- 2^{-\tau} z }-1  + \frac{2^{-2\ell}}{1-z}- \frac{2^{-2\ell} (1-z^n)}{1-z} \nonumber 
 \\
 &= \frac{(2^{-\tau n} -2^{-\tau}\, 2^{-2\ell})z^{n+1}+(2^{-2\ell}-2^{-\tau n}) z^n - 2^{-\tau} z^2 + 2^{-\tau} z}{(1-2^{-\tau}z) \, (1-z)} \,, \label{eq:Lz3} 
\end{align}
where in the first line we shift the summation indices to start the sums from $k=0$, and subtract the $k=0$ contribution when necessary.  We now define the generating function $\hat{\mathcal{G}} (z)$ as
\begin{equation}
\label{eq:SMgen}
    \hat{\mathcal{G}} (z) = z^{-1} \,  \hat{ \mathcal{F}} (z) =  \frac{(2^{-\tau n} -2^{-\tau}\, 2^{-2\ell})z^{n}+(2^{-2\ell}-2^{-\tau n}) z^{n-1} - 2^{-\tau} z + 2^{-\tau}}{(2^{-\tau n} -2^{-\tau}\, 2^{-2\ell})z^{n+1}+(2^{-2\ell}-2^{-\tau n}) z^{n} -  z + 1} \,,
\end{equation}
where $ \hat{ \mathcal{F}} (z)$ is obtained by plugging Eq.~\eqref{eq:Lz3} into Eq.~\eqref{eq:Fz}. Eq.\ttilde\eqref{eq:SMgen} provides the expression of the generating function $\hat{\mathcal{G}}(z)$ reported in Eq.~(10) of the main text. The generating function is expressed as the ratio of two polynomials of degree $n$ and $n+1$, respectively. The only exception occurs in the infrequent regime ($n=1$), where Eq.~\eqref{eq:SMgen} simplifies to 
\begin{equation}
\label{eq:geninfrequent}
    \hat{\mathcal{G}} (z) = \frac{2^{-2\ell}}{1-z+2^{-2\ell}z} \,,\quad \quad \quad \quad \text{for}\; n=1\,.
\end{equation}
Note that the definition of the generating function is not unique, however, for the choice here presented, the first-detection probability $F(k)$ can be obtained from the inverse transformation,
\begin{equation}
    F(k) = \oint \frac{dz}{2\pi i} \, \frac{  \hat{\mathcal{G}}(z)}{z^k} =\frac{1}{(k-1)!} \frac{d^{k-1}}{dz^{k-1}} \left. \hat{\mathcal{G}} (z)\right|_{z=0}\,,
\end{equation}
where the contour encircles the origin and it is entirely within the domain of analyticity of $\hat{\mathcal{G}}(z)$, and we finally recover Eq.~\eqref{eq:FDPgen}.

\subsection{Exponential decay}
\label{subsec:VB}

\begin{figure}[t]
    \centering
    \includegraphics[width=0.70\textwidth]{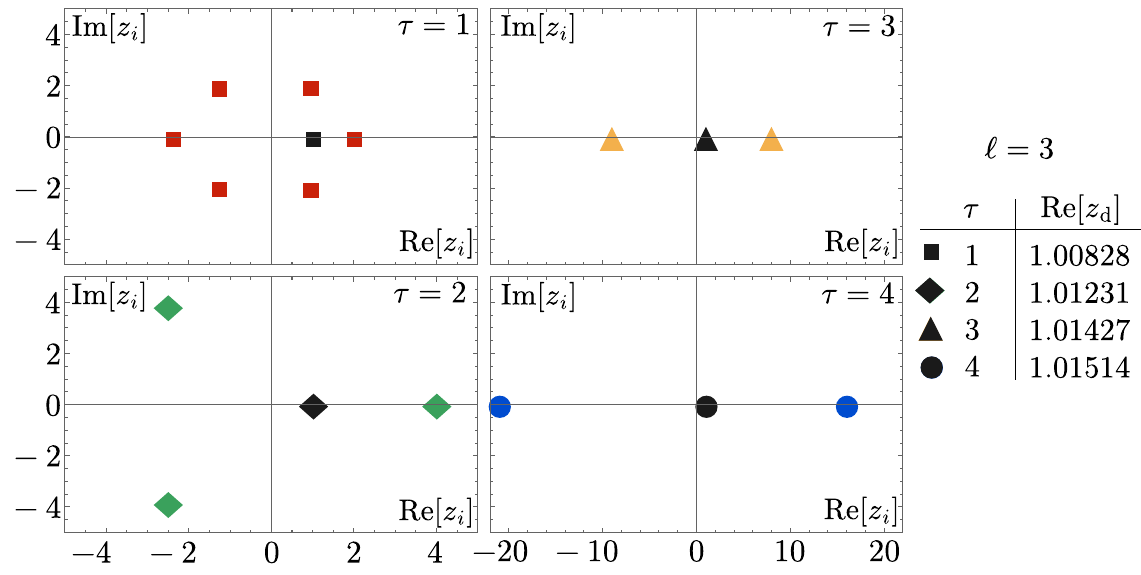}
    \caption{ \textbf{Poles of the generating function of the first-detection return probability.} Tiles display poles of the generating function~\eqref{eq:SMgen} for $\ell=3$ and several $\tau=1,2,3,4$. We have $n+1$ simple poles where $n=6,3,2,2$ in the present examples and is given by Eq.~\eqref{eq:n}. Poles with nonvanishing imaginary parts always appear in complex-conjugate pairs. Dominant poles $z_\textrm{d}$ (black markers) have $\textrm{Im} [z_\textrm{d}] = 0$ and $\textrm{Re} [z_\textrm{d}]$ near $1$ (see the legend for exact numerical value). 
    }
    \label{figSM:poles}
\end{figure}
We now focus on the frequent regime $n>1$, and rewrite Eq.~\eqref{eq:SMgen} as a partial fraction decomposition
\begin{equation}
\label{eq:pfd}
      \hat{\mathcal{G}} (z) =  \sum^{n+1}_{i=1} \frac{C_i}{(z-z_i)}
\end{equation}
where $z_i$ are roots of the denominator in Eq.~\eqref{eq:SMgen} and $C_i= \operatorname{Res}_{z=z_i} \hat{\mathcal{G}}(z)$ the associated residue. We stress that since the denominator is a high-order polynomial for $n> 1$, the poles must be determined numerically, and each coefficient $C_i$ is determined by the corresponding pole $z_i$. We have performed this computation by using standard functions in Mathematica.

In the main text, we proposed the existence of a dominant pole $z_\textrm{d}$ governing the asymptotics of the first-detection probability (see Eq.~(11)). This ansatz can be expressed in terms of the generating function, as 
\begin{equation}
\label{eq:ansatzG}
      \hat{\mathcal{G}} (z) \sim  \frac{C_\textrm{d}}{(z-z_\textrm{d})}, \quad\quad k\gg n.
\end{equation}
Here, we assumed that the generating function has only simple poles; this is also confirmed by our numerics. In particular, Fig.~\ref{figSM:poles} shows an example for fixed $\ell$ and several $\tau$. In each case, we observe $n+1$ distinct poles with $n$ given by Eq.~\eqref{eq:n}. We note that since both the polynomial in the numerator and that in the denominator of Eq.~\eqref{eq:SMgen} have real-valued coefficients, non-real poles of the generating function always appear in complex-conjugate pairs. This ensures that the asymptotics of $F(k)$ is real valued, as it must be. We find that the dominant pole corresponds to the pole of smallest modulus (real and slightly larger than $1$). As we showed in the main text, this ansatz accurately describes the large-$k$ behavior of the first-detection probability. 

While the expression in Eq.~\eqref{eq:ansatzG} is generally asymptotic, it becomes exact in the infrequent regime, where there is just one pole $z_\textrm{d}$ in Eq.~\eqref{eq:geninfrequent} with 
\begin{equation}
\label{eq:geninfrequentnew}
    z_\textrm{d}= \frac{1}{1-2^{-2\ell}}\,, \quad C_\textrm{d} = -\frac{2^{-2\ell}}{1-2^{-2\ell}}\quad\quad \textrm{for}\;\; n=1\,.
\end{equation}
This result implies that in the infrequent measurement regime, the first-detection probabilities take a geometric form,
\begin{equation}
\label{eq:geometric}
    F(k) = 2^{-2\ell} \, (1-  2^{-2\ell} )^{k-1} \quad \quad \textrm{for}\;\; n=1\,.
\end{equation}
This result matches Eq.~(9) of the main text. In the infrequent regime, the monitored subsystem relaxes in-between consecutive measurements and this renders independent measurement outcomes with probability of success $p=2^{-2\ell}$. The latter value has a clear interpretation by noting that the local stationary state of the system without measurements corresponds to an infinite temperature mixed state. 

\begin{figure}[t]
    \centering
    \includegraphics[width=0.45\textwidth]{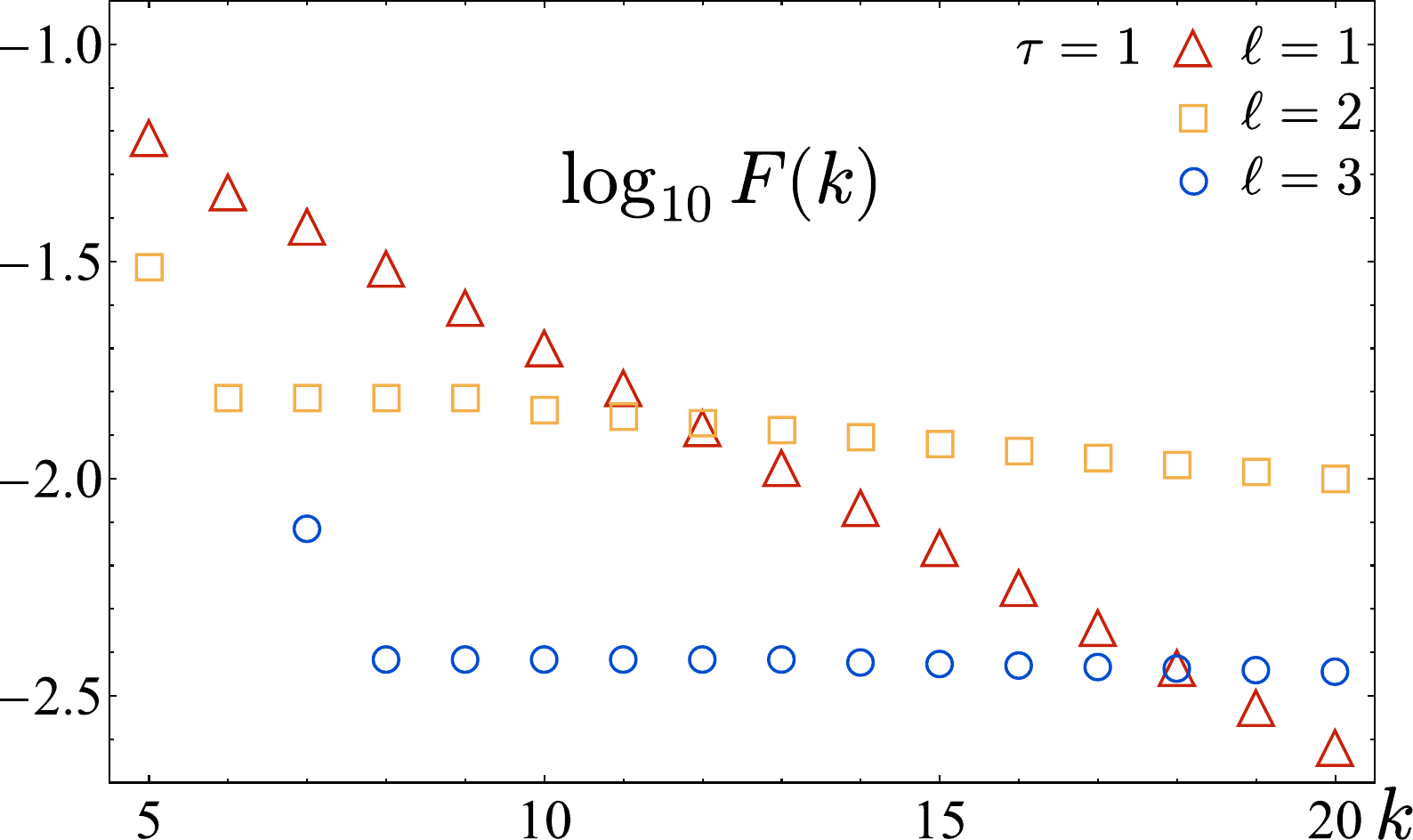}
 \caption{\textbf{Size dependence of the first-detection return probability.} We plot the logarithm of the first-detection return probability $\log_{10} F(k)$ as a function of the number of attempts $k$, for fixed measurement period $\tau$ and various lengths $\ell$ of the monitored subregion, all chosen in the frequent-measurement regime. At intermediate $k$ the first-detection return probability decreases as the size $\ell$ of the monitored subregion grows, while at larger $k$ its decay rate, which depends on $\ell$, becomes relevant. 
 }
    \label{figSM:elldep}
\end{figure}
We conclude this section by briefly commenting on the scaling of the probabilities $F(k)$ with the length of the monitored region $\ell$, fixing the measurement period $\tau$ for simplicity. As shown in Fig.~\ref{figSM:elldep}, the magnitude of the first-detection probability, measured by $\log_{10} F(k)$, strongly depends on $\ell$. In particular, increasing $\ell$ by four qubits can change $\log_{10} F(k)$ by one order of magnitude. For $\ell=1$, for instance, one has $F(8)\sim 10^{-1.5} \sim 3 \cdot 10^{-2}$, while for $\ell=3$ one obtains $F(8)\sim 10^{-2.5} \sim 3\cdot 10^{-3}$. This further confirms that $\ell$ provides an effective control parameter for the magnitude of the detection probabilities, allowing their magnitude to be adjusted by varying the size of the monitored region. This allows one to obtain finite values of the first-detection probability, which can thus be accessed experimentally, e.g., in digital quantum simulations \cite{midcircuit_1,midcircuit_2,midcircuit_3,midcircuit_4,midcircuit_5}. 

\section{Properties of the first-detection statistics}
In this final section of the Supplemental Material, we discuss the calculations and the numerical analysis behind the main properties of the first-detection statistics presented in the main text. In Subsec.~\ref{subsecVIA} we show that the local detection protocol for the state $\ket{\mathbf{{0}}_\ell}$ in the DFQEM is always recurrent. In Subsec.~\ref{subsecVIB}, we compute the coefficient of variation of $F(k)$.

\subsection{Recurrence of local first detection}
\label{subsecVIA}
In this subsection, we provide more details on the recurrent nature of the first-detection probability for the local detection protocol of the manuscript. 

Recalling Eq.\ttilde\eqref{eq:Lcomputation} for the discrete Laplace transform of the Loschmidt echo, $\hat{\mathcal{L}}(z)$  can be written in terms of a geometric series:
\begin{equation}
\label{eq:Lznew}
\hat{\mathcal{L}}(z)=\sum_{m=1}^{n-1}2^{-m\tau}z^m+2^{-2\ell} \,z^n\sum_{j=0}^{\infty}z^j=\sum_{m=1}^{n-1}2^{-m\tau}z^m+2^{-2\ell}\,\frac{z^{n}}{1-z}\,, \quad  \quad |z| < 1\,,
\end{equation}
where the threshold $n$ is given by Eq.\ttilde\eqref{eq:n}. For $n=1$ (infrequent-measurement regime) the first sum gives zero and $\hat{\mathcal{L}}(z)$ reduces to a geometric series, while for $n>1$ (frequent-measurement regime) the first sum is generally finite and non-zero. In both cases, due to the second term, $\hat{\mathcal{L}}(z)$ diverges as $z\to 1^-$. Therefore, the total detection probability is given by the limit $z\to1^-$ of $\hat{\mathcal{F}}(z)$, and Eq.\ttilde\eqref{eq:Fz} yields
\begin{equation}
\label{eq:sumFk}
\sum^\infty_{k=1} {F}(k)= \lim_{z\to1^-} \hat{\mathcal{F}}(z)=\lim_{z\to1^-}\frac{1}{1+1/\hat{\mathcal{L}}(z)}= 1\,.
\end{equation}
Recurrence is therefore a consequence of the saturation of the Loschmidt echo $\mathcal{L}(m)$ to the finite value $2^{-2\ell}$ for $m\geq n$. More generally, we stress that, for first-detection probabilities fulfilling the renewal equation\ttilde(8) of the main text, whether the process is recurrent can be read off directly from the behavior of $\hat{\mathcal{L}}(z)$ close to $z=1$. Indeed, in contrast to our case, a finite limit $\hat{\mathcal{L}}(z)\to S$ as $z\to1^-$ would give $\sum_{k}F(k)=S/(1+S)<1$, i.e., a transient dynamics.

In the following, we confirm this result numerically, which also provides a benchmark for the recursive solution of Eq.\ttilde(8) and for the asymptotic form\ttilde(11) used in the main text. To do so, we rewrite the sum by replacing the large-$k$
terms with their late-detection form given by Eq.~(11) of the main text:
\begin{equation}
\label{eq:split}
    \sum_{k=1}^\infty F(k) =  \sum_{k=1}^{k^*} F(k) \;-  \sum_{k=k^*+1}^\infty C_\textrm{d} \, z_\textrm{d}^{-k} \,.
\end{equation}
Similarly to  Eq.\ttilde\eqref{eq:Lznew}, the second term can be completed to obtaining a geometric series, as a result, Eq.~\eqref{eq:split} can be rewritten as 
\begin{equation}
\label{eq:cumFnum}
    \sum_{k=1}^\infty F(k) = -C_\textrm{d} \frac{ z_\textrm{d}^{-1}}{1-z_\textrm{d}^{-1}} + \Delta\,,
\end{equation}
where the remainder $\Delta$ contains the small-$k$ contributions where the probability $F(k)$ differs from the exponential asymptotics:
\begin{equation}
\label{eq:delta}
    \Delta = \sum^{k^*}_{k=1} \left( F(k) +C_\textrm{d} z_\textrm{d}^{-k} \right)\,.
\end{equation}
Note that $\Delta = 0$ in the infrequent measurement regime $n=1$, where $F(k)$ is geometric for all $k$. Hence $\Delta$ quantifies the deviation of the first-detection probability from a geometric distribution. Furthermore, for $n=1$, Eq.~\eqref{eq:geninfrequentnew} gives $-C_\textrm{d}\,z_\textrm{d}^{-1}/(1-z_\textrm{d}^{-1})=1$, so that the right-hand side of Eq.\ttilde\eqref{eq:cumFnum} equals $1$.

\begin{figure}[t]
    \centering
    \includegraphics[width=0.9\textwidth]{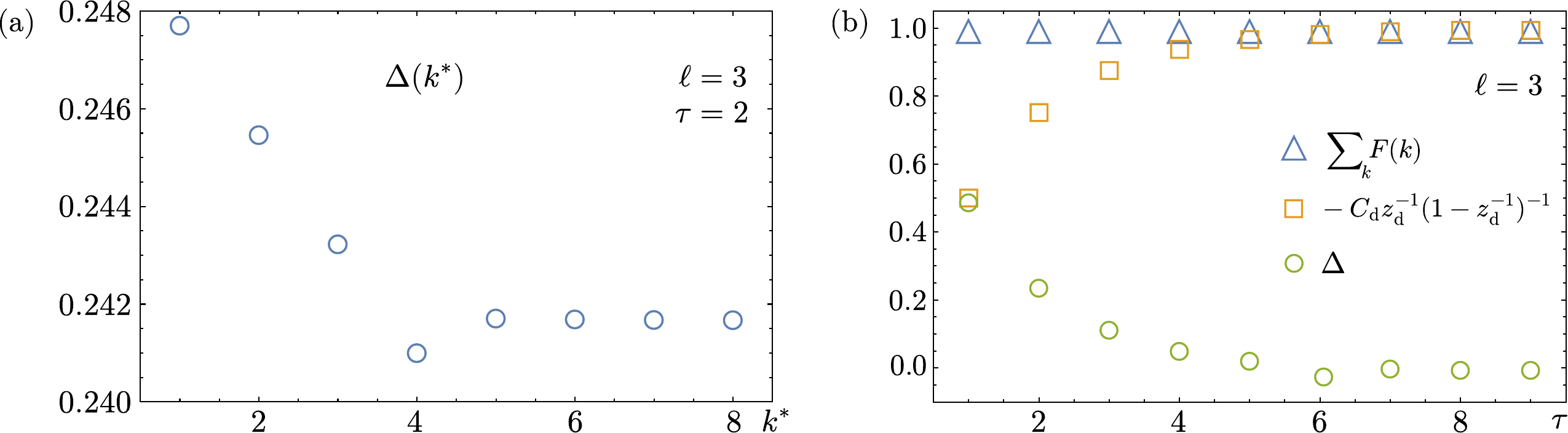}
    \caption{ \textbf{Total detection probability for various values of the probing time.} (a) The remainder $\Delta$ defined in Eq.~\eqref{eq:delta} as a function of the cut-off $k^*$ for a given choice of $\tau$ and $\ell$. The remainder $\Delta$ is independent of $k^{*}$ for sufficiently large $k^{*}$. (b) Numerical evaluation for fixed $\ell$ and various $\tau$ values of the two terms on the right-hand side of Eq.~\eqref{eq:cumFnum} (orange squares and green circles, respectively) and of the total detection probability $\sum_{k\geq 1}F(k)$ (blue triangle). This result confirms that $\sum_{k\geq 1}F(k) =1 $ for every $\tau$. The local detection protocol is thus always recurrent.}
    \label{figSM:delta}
\end{figure}

In the frequent regime $n>1$, we do not need to know the value $k^*$ at which the asymptotic (late-detection) behavior is reached in order to estimate $\Delta$. Indeed, as shown in Fig.~\ref{figSM:delta}(a), $ \Delta$ is  $k^*$-independent for 
 large $k^*$. This is because $F(k)$ converges asymptotically to $-C_\textrm{d} z_\textrm{d}^{-k}$, so that the two terms in the summand of Eq.\ttilde\eqref{eq:delta} cancel for sufficiently large $k^*$. As shown in Fig~\ref{figSM:delta}(b), by evaluating numerically the right-hand side of Eq.~\eqref{eq:cumFnum}, we find that the sum of the two contributions equals $1$ for any choice of $\ell$ and $\tau$. This reproduces the result in Eq.~\eqref{eq:sumFk} and once again confirms the asymptotic form in Eq.~(11). 
 
\subsection{Coefficient of variation of the first-detection probability}
\label{subsecVIB}
We conclude the Supplemental Material by computing the coefficient of variation $\textrm{CV}$,
defined in Eq.~(12) of the main text. In the infrequent regime, this quantity is simply predicted from the geometric distribution~\eqref{eq:geometric} and gives
\begin{equation}
    \textrm{CV}_\textrm{geo} = \sqrt{1-2^{-2\ell}} \,, \quad \quad \textrm{for}\,\, n=1\,.
\end{equation}
In the frequent regime $n>1$, we can obtain an exact expression for $\textrm{CV}$ by computing the mean number of attempts $\langle k \rangle$ and its second moment $\langle k^2 \rangle$. We can express these two quantities in terms of derivatives of the generating function $\hat{\mathcal{G}} (z)$ evaluated at $z=1$. In particular, $\langle k \rangle$  
can be expressed as
\begin{equation}
    \langle k \rangle =  \sum_{k=1}^\infty k\, F(k) = \frac{d}{dz} \left. \hat{\mathcal{F}} (z)\right|_{z=1} = \frac{d}{dz}  \left[ z\, \hat{\mathcal{G}} (z)\right]_{z=1} = \hat{\mathcal{G}} (1)  + \left. \frac{d}{dz} \hat{\mathcal{G}} (z) \right|_{z=1} \,,
\end{equation}
where the function $\hat{\mathcal{F}} (z) =z\, \hat{\mathcal{G}} (z)$ is defined in Eq.~\eqref{eq:Fz}. Similarly,  $\langle k^2\rangle$ can be derived from the second-order derivative of Eq.~\eqref{eq:Fz}: 
\begin{equation}
    \langle k^2 \rangle = \sum_{k=1}^\infty k^2 \, F(k) =   \frac{d^2}{dz^2}  \left[ z\, \hat{\mathcal{G}} (z)\right]_{z=1} +  \langle k \rangle\,,
\end{equation}
where the above expression is obtained from 
\begin{equation}
  \frac{d^2}{dz^2}  \left[ z\, \hat{\mathcal{G}} (z)\right]_{z=1} = \frac{d^2}{dz^2}  \left.  \hat{\mathcal{F}} (z)  \right|_{z=1}  =  \sum_{k=1}^\infty k \, (k-1 )\, F(k) = \langle k^2 \rangle -\langle k \rangle  \,. 
\end{equation}
Evaluating these derivatives from  Eq.~\eqref{eq:SMgen} for $n>1$, we obtain
\begin{align}
\label{eq:mean_detection}
 \langle k \rangle & =  2^{2\ell} \,, \\
\langle k^2 \rangle &= \frac{  2-2^{-n\tau+1} }{2^{-4\ell}(1-2^{-\tau})} - \frac{2n}{2^{-2\ell}} + 2^{2\ell} \,.
\end{align}
Therefore, the coefficient of variation $\textrm{CV}$ can be  computed exactly and reads
\begin{equation}
    \textrm{CV} = 
        \sqrt{
    \frac{  2-2^{-n\tau+1} }{1-2^{-\tau}} - 2n\, 2^{-2\ell} - 1 + 2^{-2\ell}}
    \,.
\end{equation}
For $n>1$ (frequent-measurement regime), the mean number of attempts \eqref{eq:mean_detection} matches the mean predicted by the geometric distribution~\eqref{eq:geometric}: $\langle k \rangle_\textrm{geo}=2^{2\ell}$. In contrast, the standard deviation, entering the numerator of the coefficient of variation, does depend on $\tau$ and differs from its geometric value $\sigma_\textrm{geo} =2^{2\ell}\sqrt{1-2^{-2\ell}}$. This leads to the results for the coefficient of variation shown in Fig.~3(d) of the main text. Its deviation from the geometric value reveals correlations between measurement outcomes, which emerge when the probing time $\tau$ is shorter than the relaxation time $2\ell$.
\end{document}